\documentclass{emulateapj}
\usepackage{graphicx}
\usepackage{amssymb,amsfonts,amsmath,amstext,amsgen,amsopn,amsxtra,indentfirst,times}

\begin{document}

\title{Atmospheric Retrieval Analysis of the Directly Imaged Exoplanet HR 8799b}
\author{Jae-Min Lee\altaffilmark{1,2}}
\author{Kevin Heng\altaffilmark{2}}
\author{Patrick G.J. Irwin\altaffilmark{3}}
\altaffiltext{1}{University of Z\"{u}rich, Institute for Theoretical Physics, Winterthurerstrasse 190, CH-8057, Z\"{u}rich, Switzerland.  Email: lee@physik.uzh.ch}
\altaffiltext{2}{University of Bern, Center for Space and Habitability, Sidlerstrasse 5, CH-3012, Bern, Switzerland.  Email: kevin.heng@csh.unibe.ch}
\altaffiltext{3}{University of Oxford, Atmospheric, Oceanic and Planetary Physics, Clarendon Laboratory, Parks Road, Oxford OX1 3PU, U.K.  Email: irwin@atm.ox.ac.uk}

\begin{abstract}
Directly-imaged exoplanets are unexplored laboratories for the application of the spectral and temperature retrieval method, where the chemistry and composition of their atmospheres are inferred from inverse modeling of the available data.  As a pilot study, we focus on the extrasolar gas giant HR 8799b for which more than 50 data points are available.  We upgrade our non-linear optimal estimation retrieval method to include a  phenomenological model of clouds that requires the cloud optical depth and monodisperse particle size to be specified.  Previous studies have focused on forward models with assumed values of the exoplanetary properties; there is no consensus on the best-fit values of the radius, mass, surface gravity and effective temperature of HR 8799b.  We show that cloudfree models produce reasonable fits to the data if the atmosphere is of super-solar metallicity and non-solar elemental abundances.  Intermediately cloudy models with moderate values of the cloud optical depth and micron-sized particles provide an equally reasonable fit to the data and require a lower mean molecular weight.  We report our best-fit values for the radius, mass, surface gravity and effective temperature of HR 8799b.  The mean molecular weight is about 3.8, while the carbon-to-oxygen ratio is about unity due to the prevalence of carbon monoxide.  Our study emphasizes the need for robust claims about the nature of an exoplanetary atmosphere to be based on analyses involving both photometry and spectroscopy and inferred from beyond a few photometric data points, such as are typically reported for hot Jupiters.
\end{abstract}

\keywords{planets and satellites: atmospheres}

\section{Introduction}

\subsection{Background and Motivation}

One of the biggest surprises of the global exoplanet hunt was the discovery of four self-luminous, sub-stellar objects orbiting the A star HR 8799 \citep{marois08,marois10}.  The stellar age---and hence the age of these objects themselves---was initially a source of controversy, since a $\sim 1$-Gyr age, as claimed via asteroseismology, would imply that the objects are brown dwarfs \citep{moya10}, but an age of $\sim 0.1$ Gyr is more compatible with stability analyses of the system \citep{gm09,fm10,mm10,currie11,sh12}.  Subsequently, an improved determination of the stellar parameters of HR 8799 using optical interferometry, in combination with the Yonsei-Yale evolutionary tracks, yielded an age of about 30--90 Myr \citep{baines12}, thus confirming the exoplanetary nature of these objects.  

Unlike for hot Jupiters, where the atmospheric properties are extracted via transit or eclipse measurements, the HR 8799 exoplanets provide prototypical examples of atmospheres that are photometrically distinct from their star.  The tradeoff is that their radii and masses cannot be directly measured.  Numerous observational studies have been performed on these gas-giant exoplanets, located about 40 pc away, mostly using photometry \citep{lafreniere09,fukagawa09,metchev09,hinz10,bergfors11,galicher11,currie11,currie12,skemer12,esposito13} and in some cases using spectroscopy \citep{bowler10,barman11,opp13}.  Consequently, this has enabled the construction of a spectral energy distribution (SED) of HR 8799b, which has inspired a series of theoretical interpretations of its atmospheric chemistry, composition and cloud/haze properties \citep{barman11,madhu11b,marley12}.  The clearly triangular shape of the spectra in the H band indicates the weakness of collision-induced absorption, which possibly implies a low surface gravity and youth, further hinting that HR 8799b is an exoplanet \citep{barman11}. 

All of the existing theoretical studies of the atmosphere of HR 8799b have so far focused on ``forward modeling": the calculation of synthetic spectra based on a set of preconceived ideas, often influenced by the study of brown dwarfs \citep{bowler10,currie11,madhu11b,barman11,marley12}.  General examples of forward models include the work of \cite{fortney06,fortney08,fortney10}, \cite{helling08b} and \cite{sb12}.  It remains unclear if brown dwarfs and directly imaged exoplanets share a common origin and therefore possess similar atmospheric properties.  A consensus is emerging that clouds and hazes are playing an integral role in the spectral appearance of HR 8799b.  Clouds and hazes have long been an important ingredient in the study of brown dwarfs and planetary-mass objects (e.g., \citealt{barman11b,burrows11}) and are emerging as a major theme in the studies of hot Jupiters \citep{barman01,burrows01,pont08,demory11,demory13,pont13,sing09,sing11,gibson12,heng12,evans13,hd13}.  When benchmarked against brown dwarfs, HR 8799b exhibits unusually red colors and a higher effective temperature than predicted \citep{barman11}.  

It is worth summarizing the other difficulties previously encountered.  Specifically, \cite{madhu11b} find that cloudfree models over-predict the fluxes at wavelengths of $\lambda \lesssim 2.2$ $\mu$m.  \cite{barman11} find it challenging to simultaneously match the H and K spectra of HR 8799b with the template spectra of typical brown dwarfs.  Only the very reddest L dwarfs resemble HR 8799b in color.  \cite{marley12} report that none of their solar-abundance models provide a satisfactory fit to all of the observational constraints and that their results are sensitive to whether the spectroscopic data is included in their analysis (alongside the photometric data).  There is also evidence for disequilibrium chemistry at work: carbon monoxide (CO) appears to be relatively more abundant than methane (CH$_4$); the CO/CH$_4$ ratio inferred from theoretical studies exceeds that expected from equilibrium chemistry \citep{madhu11b,barman11,marley12}.  Generally, the published theoretical studies do not agree on the inferred values of the radius, surface gravity and effective temperature of HR 8799b.  They also tend to assume solar abundances \citep{ag89}.  Moreover, the evolutionary cooling tracks have difficulty producing the radius and surface gravity of HR 8799b given its age and luminosity.

An alternative method of interpreting the spectra of any atmosphere is to employ an inverse modeling technique, pioneered by the Earth climate community, known as ``atmospheric retrieval" \citep{rodgers00}.  In essence, the method allows one to take a measured spectrum and ask: what are the atmospheric chemistry and composition consistent with the data?  The inferred results are only as good as the data are---conversely, the interpretation improves as the quality and quantity of the data increase.  Such a technique has been applied to hot Jupiters.  The various published studies differ in the details of their techniques, ranging from performing a brute-force sweep of parameter space \citep{madhu09} to employing a Markov chain Monte Carlo method (e.g., \citealt{bs12}) or a non-linear optimal estimation method \citep{lee12,line12,barstow13} for exploring parameter space.  See \cite{line13} for a comparision of these methods.  Most of these studies do not include a treatment of clouds or hazes, which is an obstacle towards studying the atmosphere of HR 8799b using spectral retrieval.  The directly-imaged exoplanet HR 8799b is an ideal target for atmospheric retrieval studies, because the number of data points available is $\sim$50, intermediate between typically $\sim 1$--10 for hot Jupiters and $\sim$100--1000 for brown dwarfs.

\begin{figure}
\hspace{-0.5cm}
\begin{minipage}{9.4cm}
\includegraphics[width=\columnwidth]{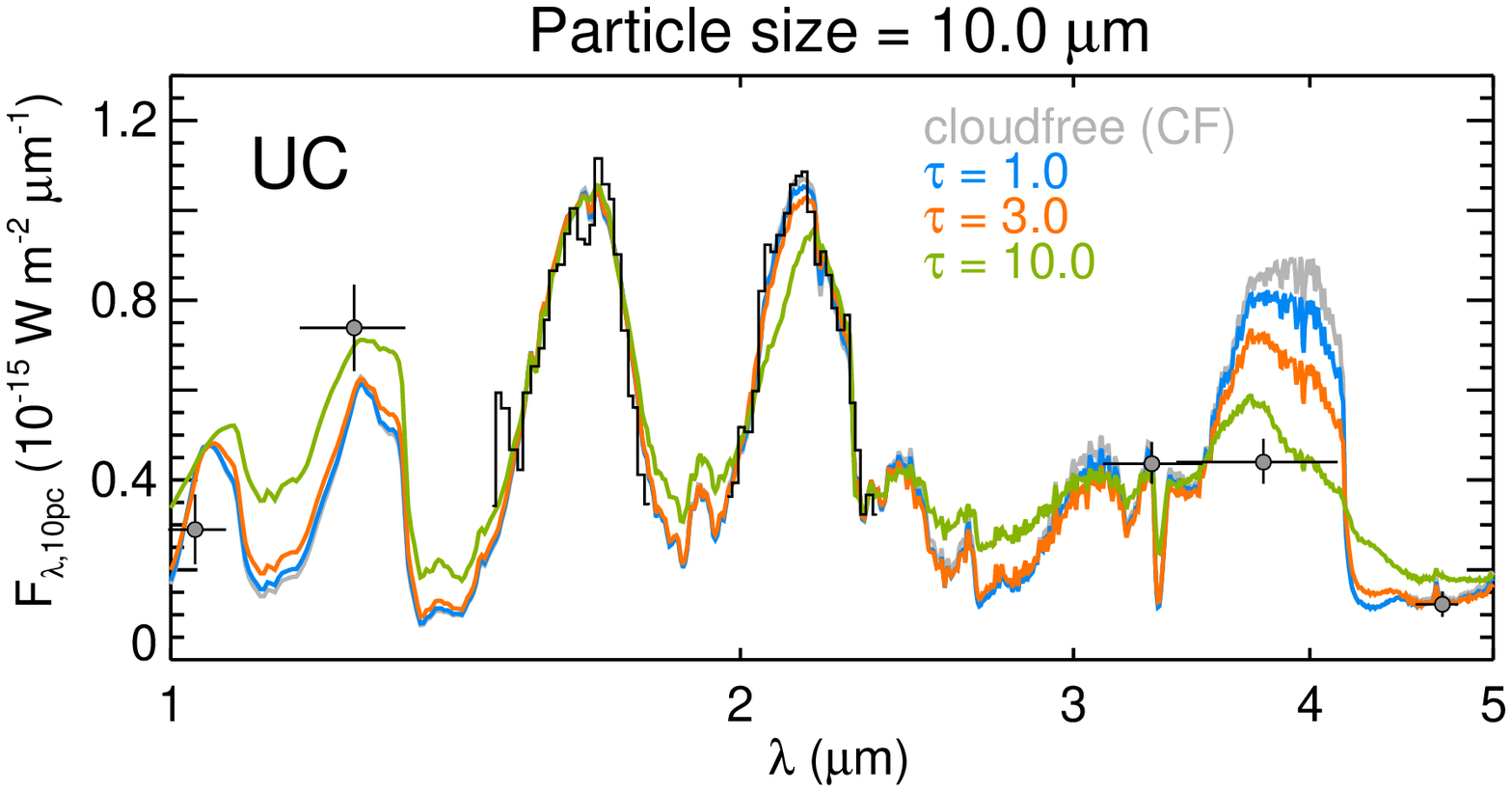} 
\includegraphics[width=\columnwidth]{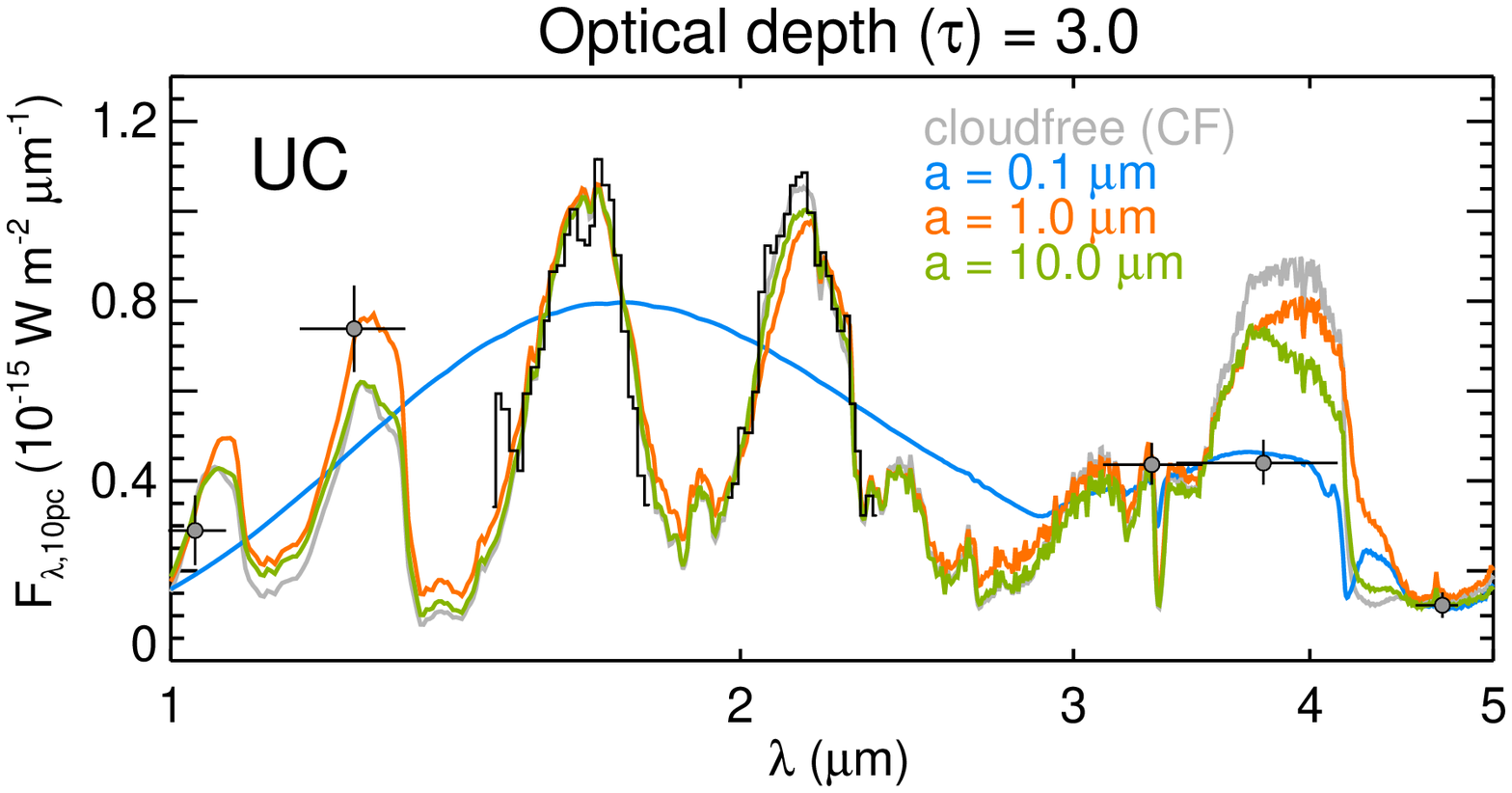}
\end{minipage}
\caption{Synthetic spectra of HR 8799b, as a function of wavelength, compared against the measured photometry (data points) and spectra (binned curve).  The top and bottom panels show the variation of the computed spectra with cloud opacity ($\tau$) and cloud particle size ($a$), respectively.  The quantity $F_{\lambda,{\rm 10pc}}$ denotes the flux at a distance of 10 pc.  All of the curves shown are best-fit models, with $R=R_{\rm J}$, from the CF and UC suites.  The top and bottom panels show models with $\log{g}=4.0$ and 3.2, respectively.  Each curve has its own set of elemental abundances and temperature profiles (see Figure \ref{fig:tp}).}
\label{fig:spectra}
\end{figure}

\begin{figure}
\centering
\includegraphics[width=\columnwidth]{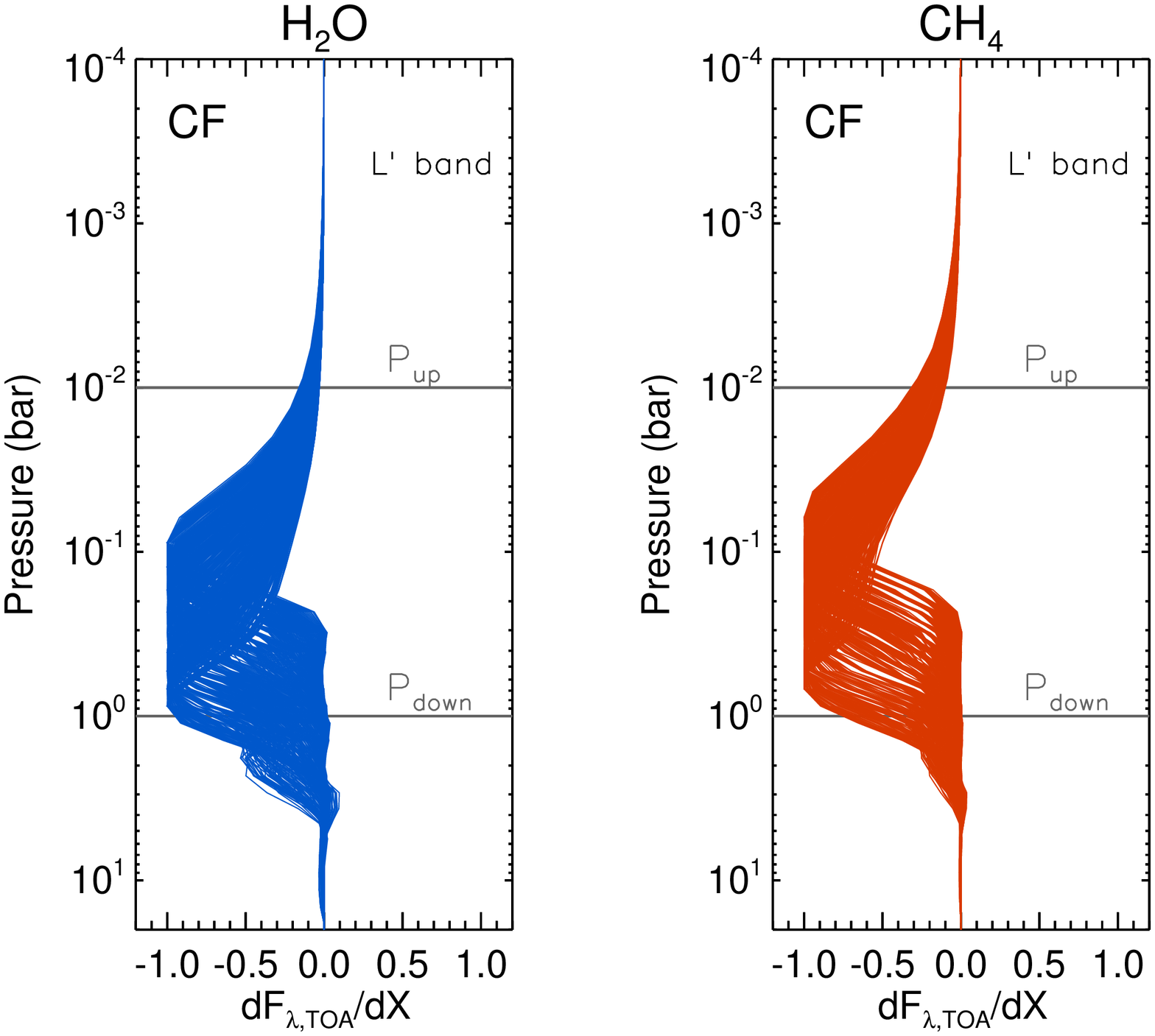}
\includegraphics[width=9cm]{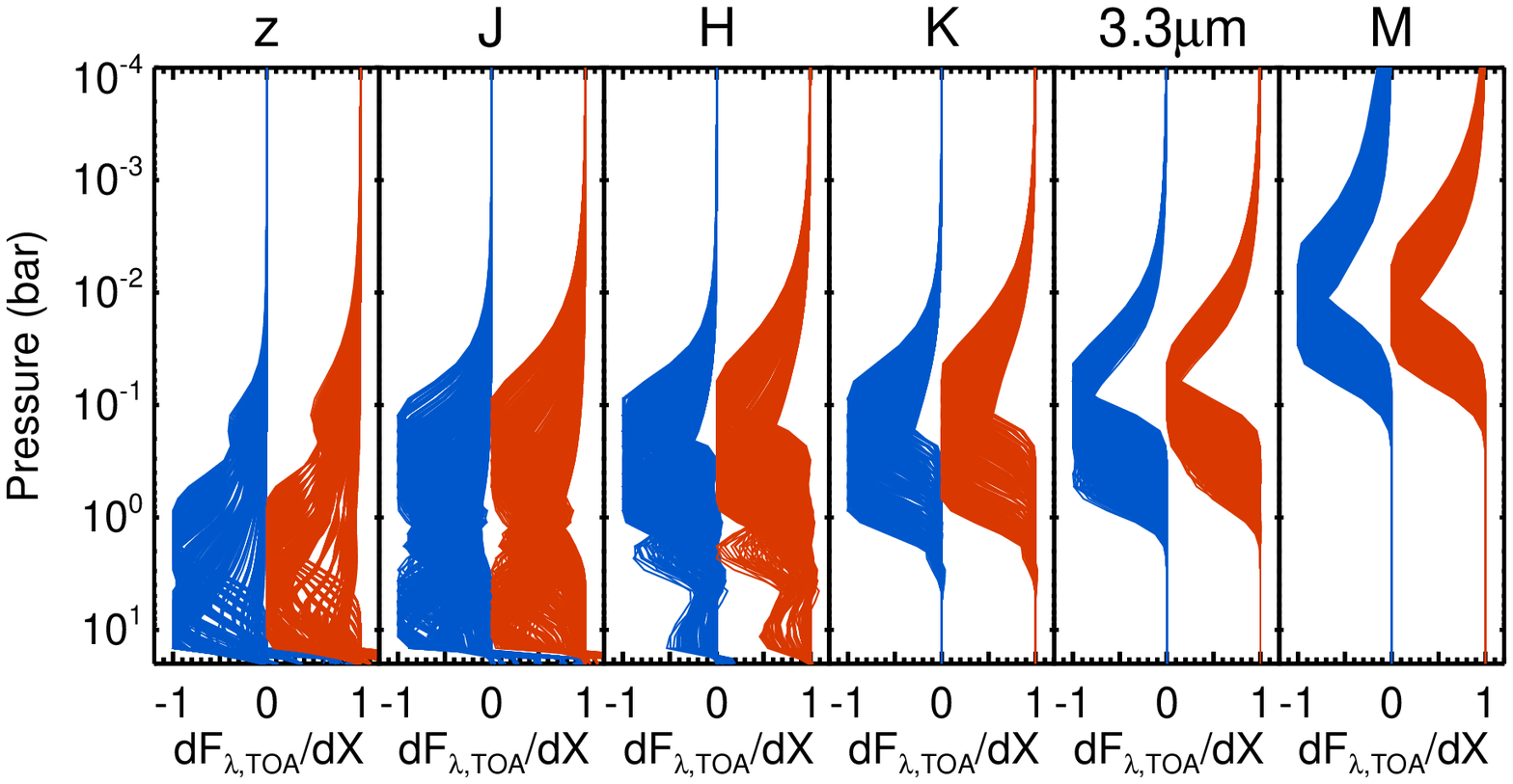}
\caption{Top panel: normalized spectrum sensitivity of the L$^\prime$-band flux to water and methane across pressure or altitude, computed using our suite of CF models.  Most of the sensitivity lies between $\sim 0.01$ and $\sim 1$ bar.  Bottom panel: for completeness, the sensitivity functions for the other wavebands are included.  The quantity $dF_{\lambda,{\rm TOA}}/dX$ is given in arbitrary units.}
\label{fig:L_band}
\end{figure}

\subsection{A More General Approach to Understanding HR 8799b}

The synthetic spectra shown in Figure \ref{fig:spectra} illustrate the difficulty of matching models with data, where cloudfree atmospheres produce L$^\prime$-band (3.78 $\mu$m) fluxes that are too high.  Consistent with previous studies \citep{madhu11b,barman11,marley12}, this mismatch points to the necessity of including clouds in the model atmosphere.  The effects of clouds are also shown in Figure \ref{fig:spectra}.  Increasing the cloud optical depth ($\tau$) or adopting the appropriate cloud particle radius ($a$) suppresses the L$^\prime$-band flux.  However, if the optical depth is too high or the particles are too large, then the water absorption features start to become muted and discrepant with the measured H- and K- band spectra.  The general expectation is that neither cloudfree nor overly cloudy models produce a good match to the data.

Thus, it is plausible to construct three types of models.
\begin{itemize}

\item \textbf{Cloudfree (CF):} Guided by Occam's Razor, a natural, rational starting point is to construct a cloudfree (or clear, ``blue sky") model, analogous to the \texttt{COND} model of \cite{baraffe02}.

\item \textbf{Uniformly Cloudy (UC):} The second simplest model to construct is one in which the entire atmosphere is permeated uniformly\footnote{By ``uniform", we mean both vertically and horizontally.} by spherical cloud particles of radius $a$.  Measured from the top of the model atmosphere, the cloud has an optical depth of $\tau$.  This model is analogous to the \texttt{DUSTY} model of \cite{baraffe02}.

\item \textbf{Intermediate (IN):} The third simplest model retains the assumption of a monodisperse population of cloud particles, but allows for the cloud deck to have a finite thickness ranging from $P_{\rm down}$ to $P_{\rm up}$ as has been explored by previous authors (e.g., \citealt{tn03,burrows11}).

\end{itemize}
To implement the IN models requires the specification of the cloud deck boundaries.  Based on the results in Figure \ref{fig:spectra}, we focus on the sensitivity of the L$^\prime$-band flux to absorption by water and methane molecules.  By computing a suite of CF models (Figure \ref{fig:L_band}), we estimate that the L$^\prime$-band flux mostly emanates from between $\sim 0.01$ and $\sim 1$ bar.  Therefore, we set $P_{\rm down} = 1$ bar and $P_{\rm up} = 0.01$ bar for all of our IN models.

In the present paper, we conduct the first ever study of the atmospheric retrieval of HR 8799b.  Our main goals are summarized as follows.
\begin{enumerate}

\item To determine if cloudy models are superior to cloudfree models for describing the atmosphere of HR 8799b.

\item To bracket the broad and degenerate range of cloud properties consistent with the observed SED of HR 8799b.

\item To formally constrain the radius ($R$), mass ($M$), surface gravity ($g$), effective temperature ($T_{\rm eff}$), mean molecular weight ($\mu$) and elemental abundances of HR 8799b by analyzing the currently available data (Table 2).  

\end{enumerate}
In \S\ref{sect:method}, we describe our methodology, including the data used and our modeling techniques.  Our results are presented in \S\ref{sect:results} and Table 1, while their implications are discussed in \S\ref{sect:discussion}.

\begin{table}
\label{tab:results}
\renewcommand{\arraystretch}{1.5}
\begin{center}
\caption{Best-fit values from super suite of models}
\begin{tabular}{cccc}
\hline
\hline
Quantity & CF & UC & IN \\
(Units) & (Cloudfree) & (Uniformly Cloudy) & (Intermediate) \\
\hline

$a$ ($\mu$m) & -- & 1.5 & 1.5 \\

$\tau$ & -- & 2 & 2 \\

$\chi^2/N$ & 1.36 & 1.41 & 1.36 \\

$R$ $(R_{\rm J})$ & 0.66$^{+0.04}_{-0.03}$ & 0.71$^{+0.09}_{-0.07}$ & 0.66$^{+0.07}_{-0.04}$ \\

$\log{g}$ (cgs) & $4.9 \pm 0.1$ & $5.0 \pm 0.1$ & 5.0$^{+0.1}_{-0.2}$ \\

$M$ $(M_{\rm J})$ & 13$^{+3 }_{-4 }$ & $21 \pm 8$ & 16$^{+5 }_{-4 }$ \\

$\rho$ (g/cm$^3$) & 57$^{+13}_{-16}$ & 75$^{+22}_{-25}$ & 68$^{+19}_{-14}$ \\

$T_{\rm eff}$ (K) & 880$^{+20 }_{-30 }$ & 880$^{+40 }_{-70 }$ & 900$^{+30 }_{-60 }$ \\

$\mu$ & 4.2$^{+1.5}_{-0.5}$ & 3.3$^{+1.1}_{-0.3}$ & 3.8$^{+1.5}_{-0.4}$ \\

C/O & 0.97$^{+0.00}_{-0.01}$ & 0.94$^{+0.02}_{-0.01}$ & $0.96 \pm 0.01$ \\

$X_{\rm CO}/X_{\rm CH_4}$ & 980$^{+890}_{-220}$ & 730$^{+1150}_{-260}$ & 880$^{+1010}_{-180}$ \\

$X_{\rm H_2O}/X_{\rm CO}$ & 0.036$^{+0.007}_{-0.008}$ & 0.062$^{+0.019}_{-0.021}$ & 0.044$^{+0.009}_{-0.012}$ \\

$X_{\rm CO}$ $(10^{-2})$ & 8$^{+5}_{-2}$ & 4$^{+4}_{-1}$ & 6$^{+5}_{-1}$\\

$X_{\rm H_2O}$ $(10^{-2})$ & $0.3 \pm 0.1$  & 0.2$^{+0.1}_{-0.0}$ & 0.3$^{+0.0}_{-0.1}$ \\

$X_{\rm CH_4}$ $(10^{-4})$ & 0.8$^{+0.7}_{-0.3}$ & 0.5$^{+0.7}_{-0.2}$ & 0.7$^{+0.8}_{-0.3}$ \\

$X_{\rm CO_2}$  $(10^{-4})$ & 0.2$^{+0.5}$  & 0.4$^{+0.5}$ & 0.2$^{+0.4}$ \\

$X_{\rm H_2}$ & 0.84$^{+0.02}_{-0.05}$ & 0.87$^{+0.01}_{-0.04}$ & 0.85$^{+0.02}_{-0.05}$ \\

$X_{\rm He}$ & 0.082$^{+0.001}_{-0.005}$ & 0.085$^{+0.001}_{-0.004}$ & 0.083$^{+0.001}_{-0.005}$  \\

\hline
\end{tabular}
\end{center}
Note: if a lower or upper limit is listed as zero, it means that the central value itself is the minimum or maximum value.  The uncertainties associated with $R$ and $\log{g}$ are computed by projecting the $\Delta \chi^2 = 2.30$ contour \citep{avni76} along the respective axis.  The uncertainties associated with their dependent quantities ($M$ and $\rho$) are obtained via quadrature.  The uncertainties associated with the chemical abundances (and their dependent quantities) are computed by exploring the sensitivity of the retrievals to the variation of the abundances of the other chemical species (Figure \ref{fig:abund_var}).  The abundances of carbon dioxide are unbounded from below, due to the relative insensitivity of the synthetic spectra to their presence.\\
\end{table}


\section{Methodology}
\label{sect:method}

\subsection{Selected Data and Filter Functions}

\begin{figure}
\hspace{-0.5cm}
\includegraphics[width=9.5cm]{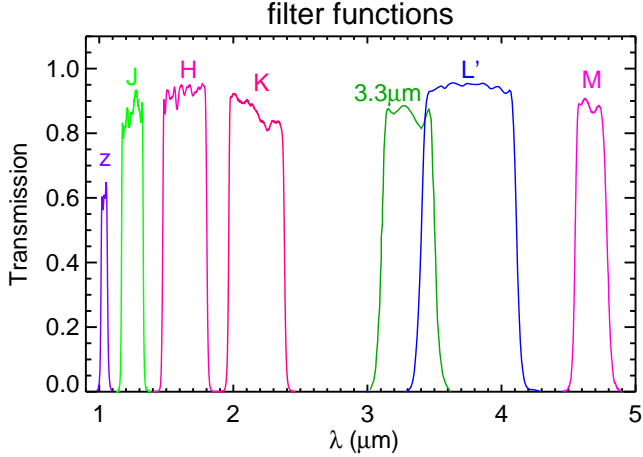}
\caption{Filter functions for the photometric (z, J, 3.3 $\mu$m, L$^\prime$, and M) and spectroscopic (OSIRIS H and K) wavebands of the various observations of HR 8799b.}
\label{fig:filters}
\end{figure}

No new data is presented in this study.  Instead, we cull the data from the published literature.  In the interest of scientific reproducibility, we list the data points in the same physical units alongside their published source in Table 2.  For the photometric points and OSIRIS H and K spectra, we obtain the filter functions from the appropriate observatory/telescope where the data were gathered (Figure \ref{fig:filters}).  A few features of the data are worth highlighting.  The data point at 3.3 $\mu$m is generally believed to be a sensitive indicator for the abundance of methane \citep{barman11}.  The K band is an indicator of water absorption, which appears to be evident in HR 8799b \citep{barman11}.  We have not included the data of \cite{opp13} for two reasons: their published spectra of HR 8799b are in arbitrary, rather than absolute, flux units; their H-band spectrum is consistent with the published results of \cite{barman11} (see their Figure 5).  We judge that the exclusion of the \cite{opp13} will not affect the broad conclusions of our study.

In Table 2, the published fluxes of HR 8799b are normalized at a distance of $d=10$ pc, whereas the required input for our models is the flux at the top of the atmosphere (TOA).  The conversion between the two fluxes follows from energy conservation (and assuming isotropy) and requires the radius $R$ to be stated,
\begin{equation}
F_{\lambda,{\rm TOA}} = \left(\frac{d}{R}\right)^2 F_{\lambda,{\rm 10pc}}.
\end{equation}
Thus, lower values of $R$ correspond to higher values of $F_{\lambda,{\rm TOA}}$.  In other words, if two¤ exoplanets are located at the same distance away from us, the smaller one needs to be intrinsically brighter to produce the same flux received at Earth.  Therefore, \emph{lower values of $R$ correspond to hotter exoplanets, which translate into subtle differences in the atmospheric chemistry as the different major molecules have opacities that possess different temperature-dependent sensitivities,} an issue we will explore in detail later.

\begin{figure}
\centering
\includegraphics[width=\columnwidth]{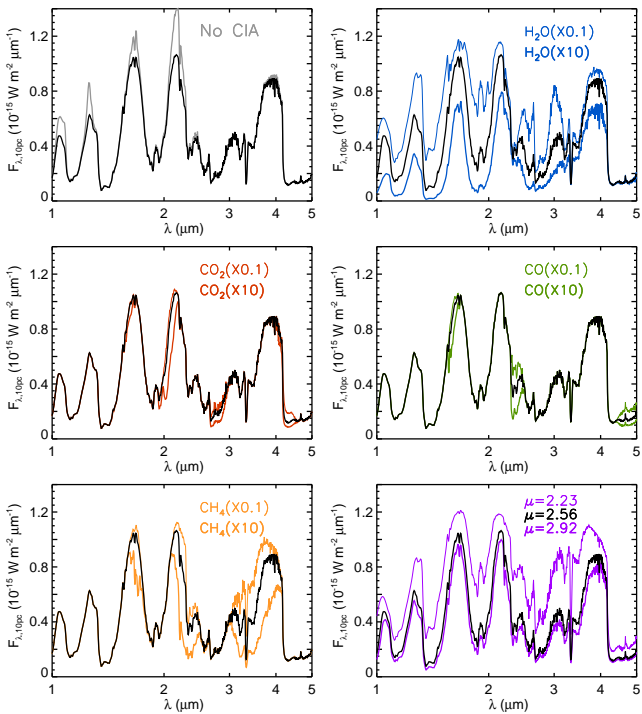}
\caption{Variations of the synthetic spectra as a function of wavelength by changing the effects of collision-induced absorption by molecular hydrogen and helium (top left panel), water abundance (top right panel), carbon dioxide abundance (middle left panel), carbon monoxide abundance (middle right panel), methane abundance (bottom left panel) and the mean molecular weight (bottom right panel).  The fiducial or baseline spectrum is our best-fit solution to the suite of CF models with $R=R_{\rm J}$ and $\log{g}=4.0$.}
\label{fig:spectra_variations}
\end{figure}

There is no consensus on the previously reported values of the radius, mass, surface gravity and effective temperature of HR 8799b.  \cite{madhu11b} report values of $T_{\rm eff} \approx 850$ K, $\log{g} \approx 4.3$ and $M \approx 12 ~M_{\rm J}$ for HR 8799b, which translates into $R \approx 1.2 ~R_{\rm J}$ with $M_{\rm J}$ and $R_{\rm J}$ being the mass and radius of Jupiter, respectively.  The best-fit model obtained by \cite{barman11} yields $T_{\rm eff} = 1100 \pm 100$ K, $\log{g} = 3.5 \pm 0.5$ and $R = 0.75^{+0.17}_{-0.12} ~R_{\rm J}$.  The derived mass of $M \approx 0.7 M_{\rm J}$ is consistent with the $M \lesssim 20 ~M_{\rm J}$ limit derived by \cite{fm10} based on orbital stability considerations.  The low mass expected is also the reason why $\log{g} > 5$ models are disfavored.  \cite{marley12} remark on how the solar-abundance models they explored do not satisfy all of the observational constraints.  When only the photometric data are considered (setting aside the H- and K-band spectra obtained by \citealt{barman11}), they find $T_{\rm eff} = 1000$ K and $M \approx 26 ~M_{\rm J}$; the inferred mass value is in conflict with the dynamical stability constraints of \cite{fm10}.  When the spectroscopic results are included, the mass obtained becomes more reasonable ($M \approx 3 ~M_{\rm J}$), but the low effective temperature of $T_{\rm eff} = 750$ K inferred is inconsistent with the observed luminosity.  In their Table 1, \cite{marley12} provide a summary of the diverse ranges of $\log{g}$, $R$ and $T_{\rm eff}$ inferred from various studies.  They favor $R = 1.11 ~R_{\rm J}$ and consider $R < R_{\rm J}$ values to be unphysical.  It is interesting to note that the study of \cite{madhu11b}, which only uses photometric data, produces $T_{\rm eff} < 1000$ K, in seeming contradiction to the results of \cite{marley12}.  From examining the studies of \cite{madhu11b}, \cite{barman11} and \cite{marley12}, we conclude that the radius, mass, surface gravity and effective temperature of HR 8799b remain poorly constrained, largely due to the challenges of interpretation presented by its unusual atmosphere.

\subsection{Retrieval Method}

Our retrieval method (\texttt{NEMESIS}) utilizes a non-linear optimal estimation scheme, in which the state of the fitting parameters are optimised by minimising the ``cost function", which finds the balance between the constraints from priors and measurements, described by a Gaussian probability density function (PDF) \citep{rodgers00,irwin08}.  Since the prior information for HD 8799b is unknown, we adopt an initial guess of the fitting parameters, with a large 1-$\sigma$ error associated with its Gaussian PDF, to be the starting point of our retrieval.  These initial values are applied to our 43-layer temperature-pressure profiles and to the abundances of the major molecules.  As a result, the optimal state is fully unbiased towards initial values that are given nominally.  Reassuringly, the initial temperature-pressure profile of our model atmosphere mimics the profile in Figure 2 of \cite{marley12} with $\log{g}=4.5$ and $T_{\rm eff}$=900 K.  We adopt the correlated-k approximation \citep{gy89,lo91} for performing fast spectral integration in the forward model by using the pre-tabulated k-distributions, known as ``k-coefficient tables" (or ``k-tables" for short).  Details of the k-tables and the references for the line lists used are given in \cite{lee12}, which also describes a previous application of our cloudfree retrieval method to the hot Jupiter HD 189733b.

While the advantage of the retrieval method is to infer the properties of the exoplanetary atmosphere without being tied down by numerous model assumptions, we do introduce some preconceived ideas into it.  

\begin{enumerate}

\item In addition to molecular hydrogen (H$_2$) and helium (He), we assume that the only major molecules present are water (H$_2$O), carbon dioxide (CO$_2$), carbon monoxide (CO) and methane (CH$_4$).  In other words, we neglect the possibility of nitrogen- and sulphur-carrying species.  For example, if S is substantially enhanced, then molecular species such as H$_2$S may contribute non-negligibly to the spectral appearance of HR 8799b.  There is also the possibility that species that are trace elements by mass may contribute disproportionately to the spectrum (e.g., phosphine or PH$_3$).

\item We assume that all of the atmospheric constituents are vertically well-mixed (except for the clouds in the case of the IN model).  It is expected that a larger space of solutions will be allowed if this assumption is relaxed, but the current quality and quantity of data available do not merit such an investigation.

\item In the present investigation, we have assumed our clouds to be composed of enstatite grains.  While the composition of cloud species in HR 8799b (if they are present) is generally unknown, such an investigation retains some generality because of the nearly universal shape of the extinction efficiency curve (see the Appendix), thus allowing our conclusions to apply to other refractory species.

\end{enumerate}

For a chemical species $S$, the normalized abundance---or ``volume mixing ratio" as it is more commonly termed in the atmospheric sciences---is denoted by $X_S$.  During the iterative fitting process, the molecular abundances retrieved instantaneously adjust the abundances of hydrogen and helium, which are the inert elements in our atmospheric model, such that the sum of the volume mixing ratios is always unity and the $X_{\rm He}/X_{\rm H_2}$ ratio remains the same ($\approx 10\%$). H$_{2}$ and He contribute via the action of collision-induced absorption (CIA) and largely determine the mean molecular weight ($\mu$), the latter of which sets the pressure scale height of the atmosphere.  Unlike for most of the previous studies, we do not assume solar abundances ($\mu=2.35$).  The carbon-to-oxygen (C/O) ratio is a \emph{consequence} of the retrieved molecular abundances and is not set to be the solar value (C/O $\approx 0.5$--0.6).  It is computed from
\begin{equation}
\mbox{C/O} = \frac{X_{\rm CO} + X_{\rm CH_4} + X_{\rm CO_2}}{X_{\rm CO} + X_{\rm H_2O} + 2X_{\rm CO_2}}.
\end{equation}
When carbon monoxide is the dominant species, we have C/O $\approx 1$.  (See \citealt{line13} for more discussion on the C/O ratio.)

To compute the uncertainties associated with the abundance of each chemical species ($\Delta X_S$), we explore the sensitivity of the retrievals to the variation of each individual species \citep{lee12}.  This procedure affects the dependent quantities as well ($\mu$ and C/O ratio).  The posterior distribution of each $X_S$ is computed and its uncertainty is determined using $\Delta \chi^2=1$, where $\Delta \chi^2$ is the change in the goodness of fit from its best-fit value (Figure \ref{fig:abund_var}).

Figure \ref{fig:spectra_variations} shows the effects of varying the strength of CIA, the mean molecular weight and the various molecular abundances.  CIA mostly affects the H- and K-band water features.  Water and methane generally have pronounced effects throughout the wavelength range considered (1--5 $\mu$m), as does the mean molecular weight.  Note that the invariance of methane just longward of 1 $\mu$m is due to the lack of these specific line lists in our database.  The effects of carbon dioxide and carbon monoxide are more subdued, but nonetheless important.

Finally, we note that our definition of the effective temperature ($T_{\rm eff}$) follows that of the classical Milne's solution for self-luminous objects: it is the temperature where the total optical depth of the atmosphere is $\tau_{\rm total} = 2/3$ \citep{mihalas}.  We compute $\tau_{\rm total}$ using the flux-weighted mean optical depth, where the wavelength-dependent optical depth is weighted by the retrieved SED of HR 8799b for a given model.  An alternative and equivalent method is to integrate the retrieved SED of HR 8799b across wavelength and equate this bolometric flux to $\sigma_{\rm SB} T^4_{\rm eff}$, where $\sigma_{\rm SB}$ is the Stefan-Boltzmann constant.

\subsection{Cloud Model}

\subsubsection{Summary of Previous Work}

Forward modelers have used a variety of sophisticated codes to model the condensation physics.  \cite{madhu11b} use the \texttt{COOLTLUSTY} code, which combines atmospheric and evolutionary calculations self-consistently, and a ``painted-on" cloud model (see \citealt{burrows11} and references therein).  It is assumed that a cloud deck forms at the intersection of a condensation curve with the temperature-pressure profile---see, e.g., \cite{helling08} for a counter-opinion---and that its density drops off as a power law in either direction.  Thus, for a given composition, the cloud model involves four free parameters: two power-law indices, the thickness of the cloud deck and the modal size of the condensates.  Furthermore, it is assumed that the size distribution of the cloud particles follows that of cumulus water clouds on Earth.  The broad range of model particle size (1--60 $\mu$m), cloud deck thickness and compositions (forsterite and iron) reported by \cite{madhu11b} to be consistent with the data on HR 8799b highlights the degeneracy inherent in the condensation physics.  \cite{madhu11b} remark how the thickness of the clouds is ``almost unconstrained by theory" and that thicker clouds may plausibly be a result of higher metallicity and stronger atmospheric mixing, but do not formally explore the interplay between these effects.

\cite{barman11} improved the \texttt{PHOENIX} code, beyond the legacy \texttt{COND} (cloudfree) and \texttt{DUSTY} (uniformly cloudy) modes of operation, to include an intermediate cloud model, the need of which was already discussed by \cite{marois08}.  The prescription of \cite{madhu11b} for clouds of intermediate thickness is similar to this approach.  Their calculations are not self-consistent in the sense that the atmospheric modeling is not tied to evolutionary cooling tracks---which is also the case in the present study---but the intention of their study is precisely to point out that evolutionary models have difficulty matching the observed properties of HR 8799b.  \cite{barman11} assume the lower base of their cloud to be fixed by the condensation curve of a given chemical species, specify the cloud deck thickness as a free parameter, and demand that the cloud density falls off exponentially above the upper base.  They assume a log-normal distribution for the size distribution of the cloud particles with the modal size being prescribed as a free parameter ($\approx 5$--10 $\mu$m).  The sizes of the condensates considered range from 1 to 100 $\mu$m.  \cite{barman11} remark that photochemistry may be a non-negligible effect, but that they did not consider it due to the inherent technical challenges.

\cite{marley12} use combined atmospheric and evolutionary models, which include a treatment of clouds that considers the sedimentation and lofting of the cloud particles.  Their models have been tested extensively against brown dwarfs and utilize a sedimentation parameter ($f_{\rm sed}$) that is traditionally tuned to match the L to T transition of the color-magnitude diagram of brown dwarfs \citep{sm08,burrows11}, but is used as a free parameter in the study of HR 8799b.

It should be noted that all of the published models are one-dimensional in nature and consider the effects of clouds only in the radial direction.  Thus, cloud patchiness, hints or signs of which have been observed in some brown dwarfs \citep{artigau09,buenzli12,radigan12}, cannot be modeled in a strict sense.

\subsubsection{A Simple, Phenomenological Approach}
\label{subsect:cloudmodel}

\begin{figure}
\hspace{-0.5cm}
\begin{minipage}{9.4cm}
\includegraphics[width=\columnwidth]{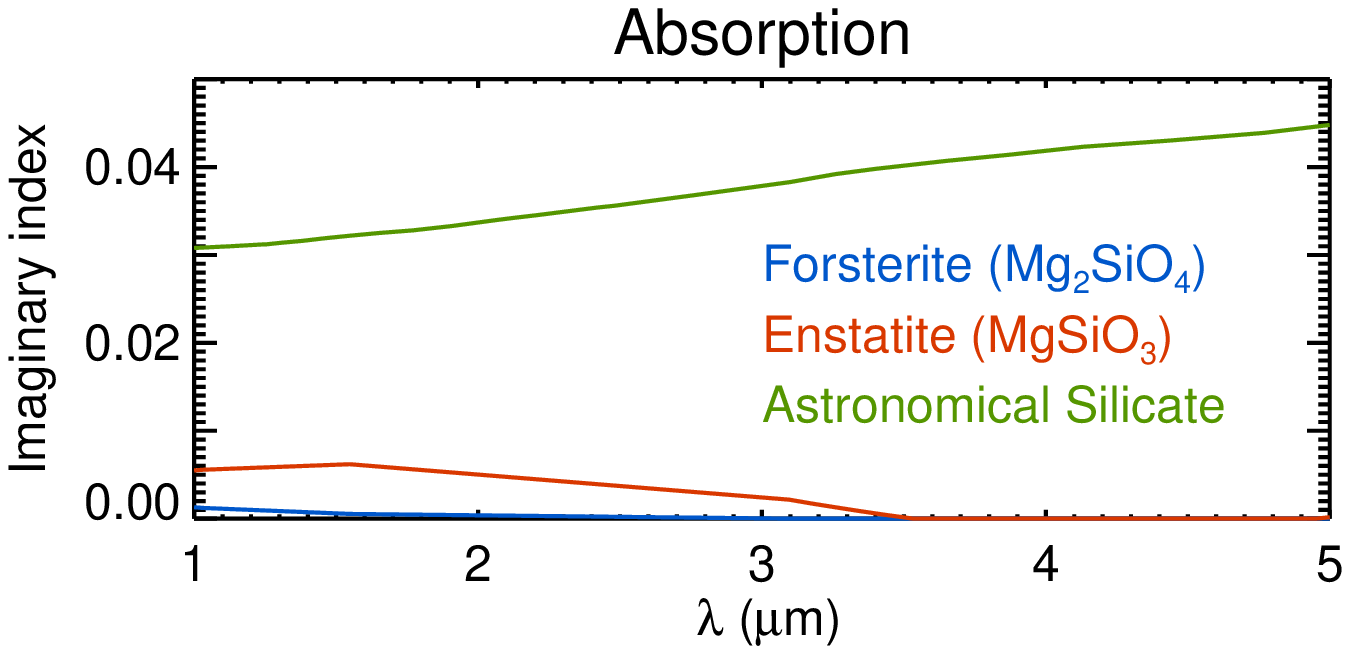}
\includegraphics[width=\columnwidth]{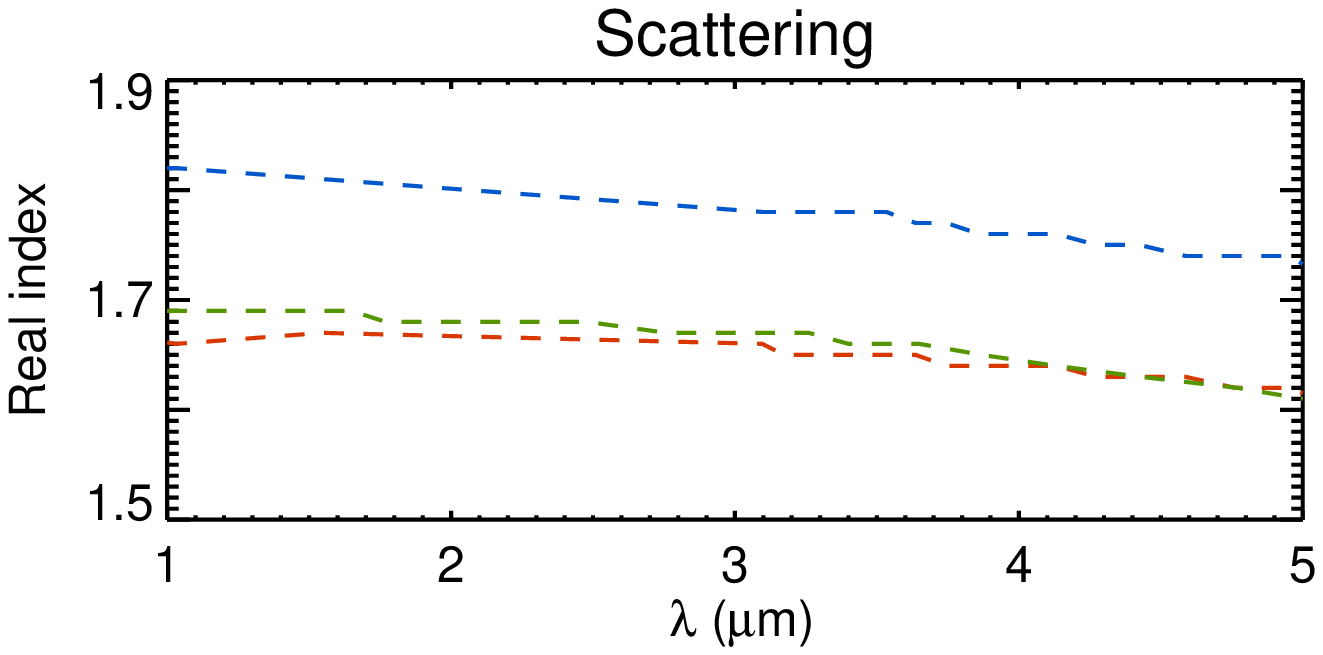}
\end{minipage}
\caption{Absorption (top panel) and scattering (bottom panel) refractive indices, as functions of wavelength, for various materials.  Only enstatite is considered in the present study.  The data for astronomical silicate are taken from \cite{draine84} and \cite{ld93}.}
\label{fig:refract}
\end{figure}

\begin{figure}
\hspace{-0.3cm}
\begin{minipage}{9cm}
\includegraphics[width=\columnwidth]{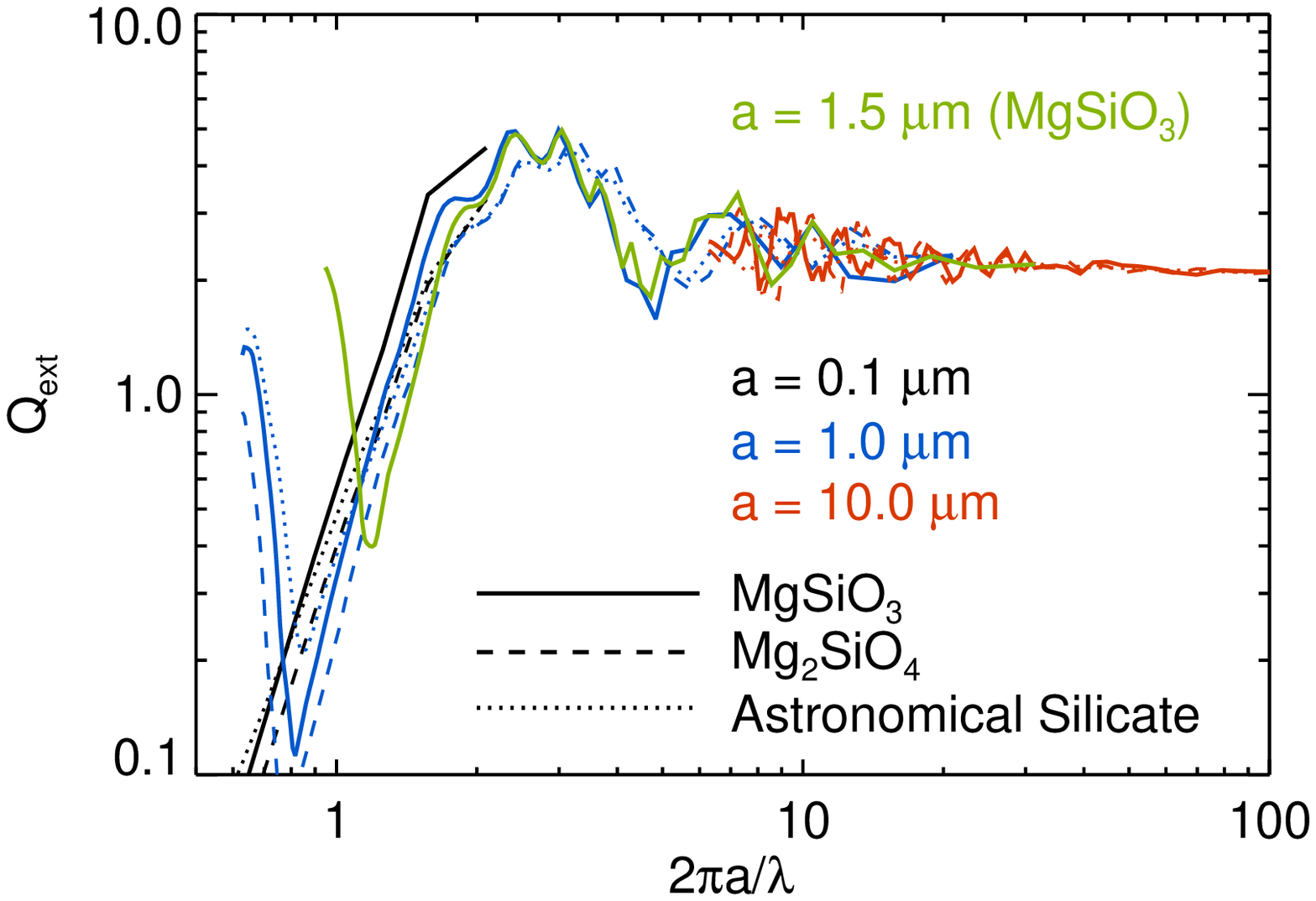}
\includegraphics[width=\columnwidth]{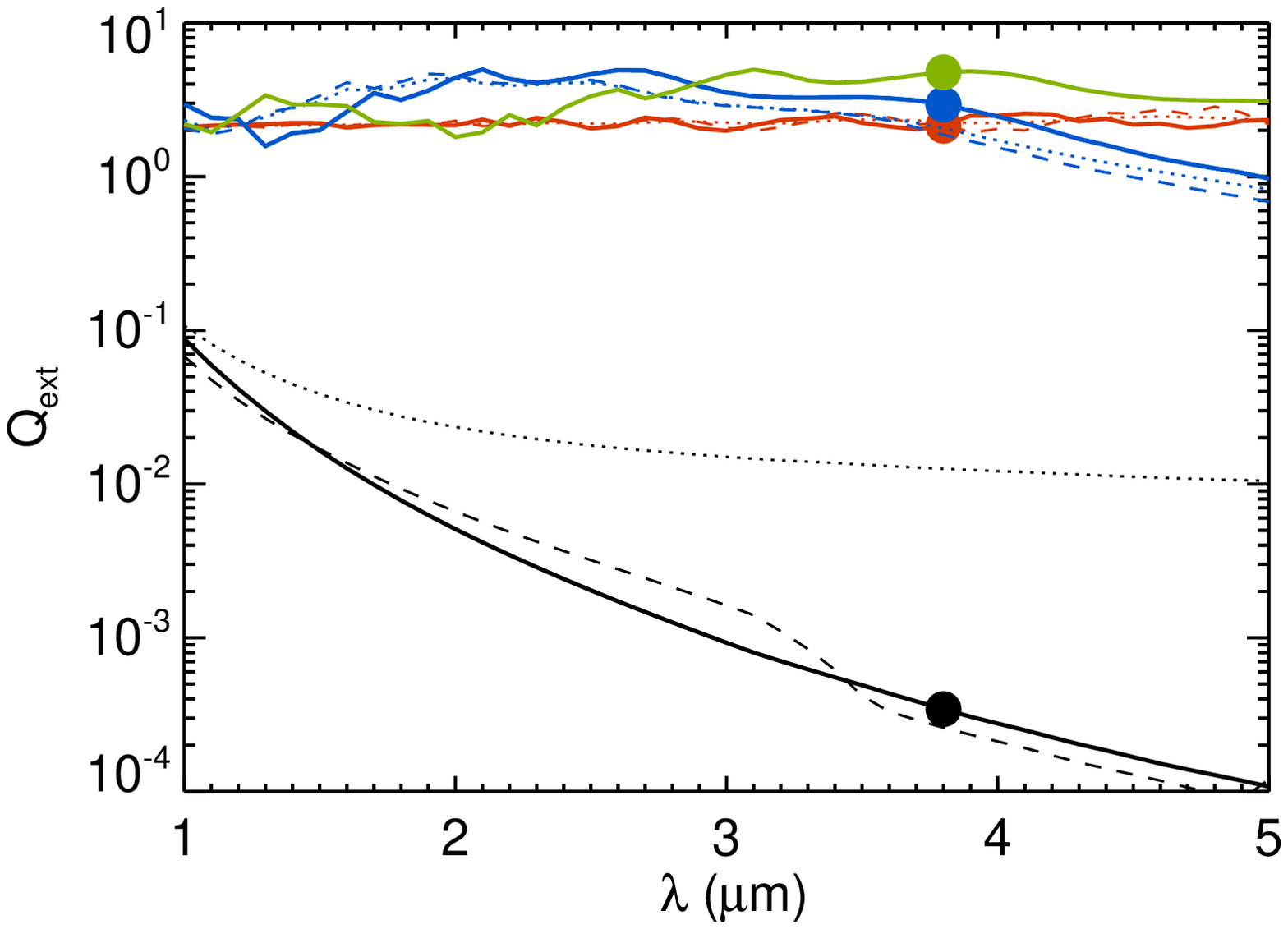}
\end{minipage}
\caption{Extinction efficiency of enstatite, as a function of the size parameter ($2\pi a/\lambda$; top panel) and wavelength ($\lambda$; bottom panel), for the three particle radii considered in our study.  The solid circles indicate the values of $Q_{\rm ext}$ used for the L$^\prime$ band.  We have considered only enstatite (solid curves) in the present study, but we show examples of other materials to illustrate the similarity of $Q_{\rm ext}$ when the particles are large ($2 \pi a/\lambda \gtrsim 1$).  The data for astronomical silicate are taken from \cite{draine84} and \cite{ld93}.}
\label{fig:qext}
\end{figure}

\begin{figure*}
\centering
\includegraphics[width=0.65\columnwidth]{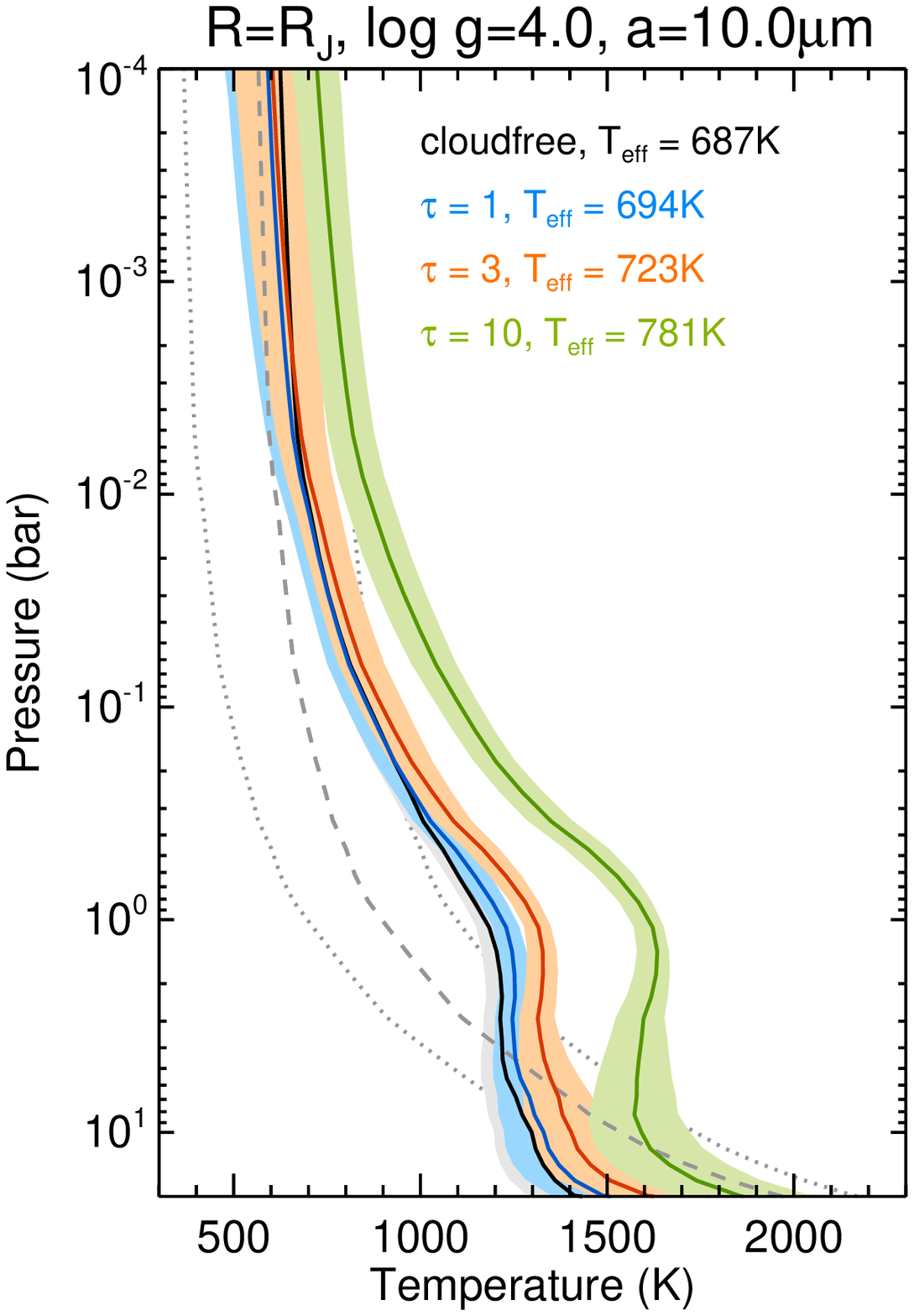}
\includegraphics[width=0.65\columnwidth]{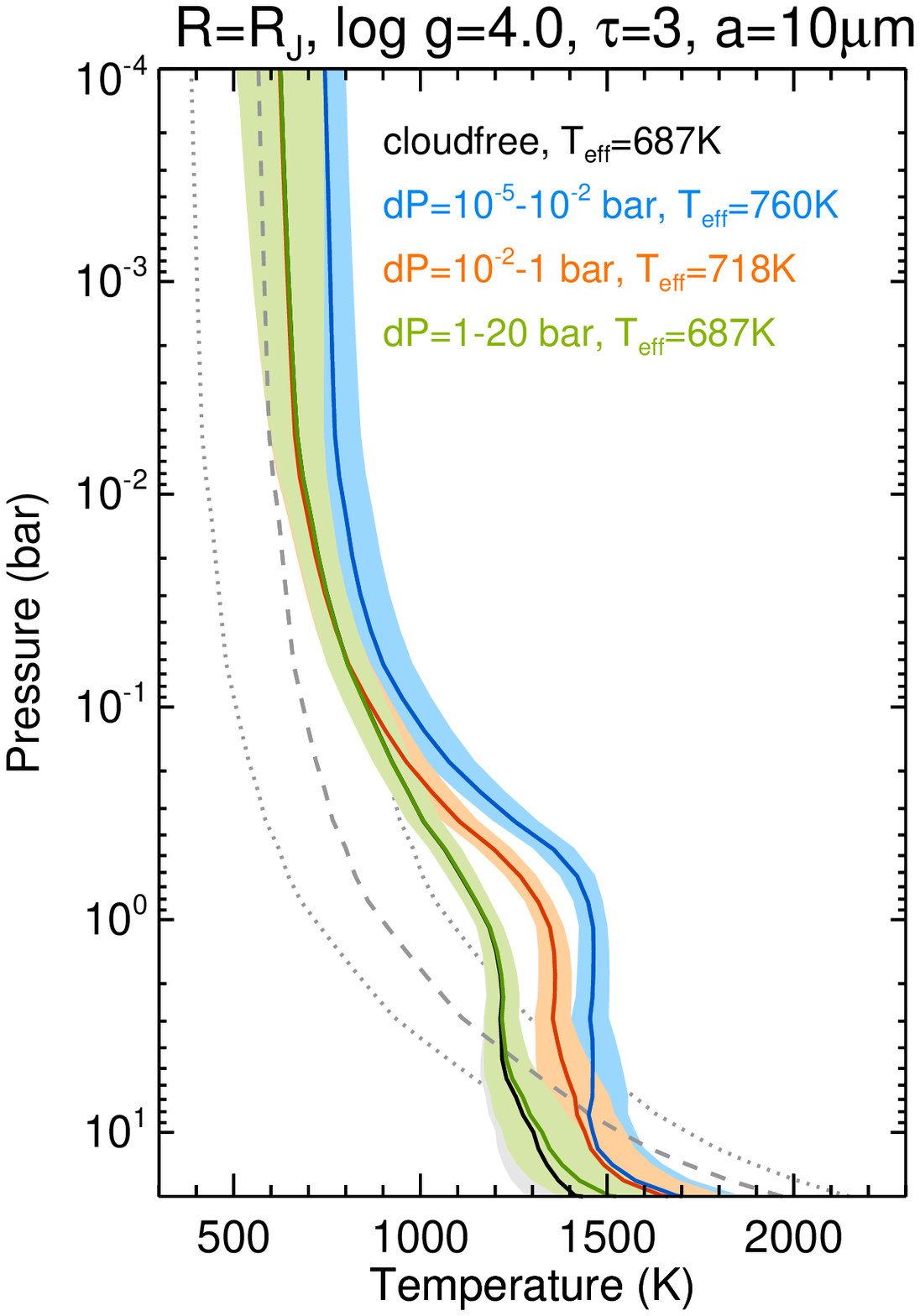}
\includegraphics[width=0.65\columnwidth]{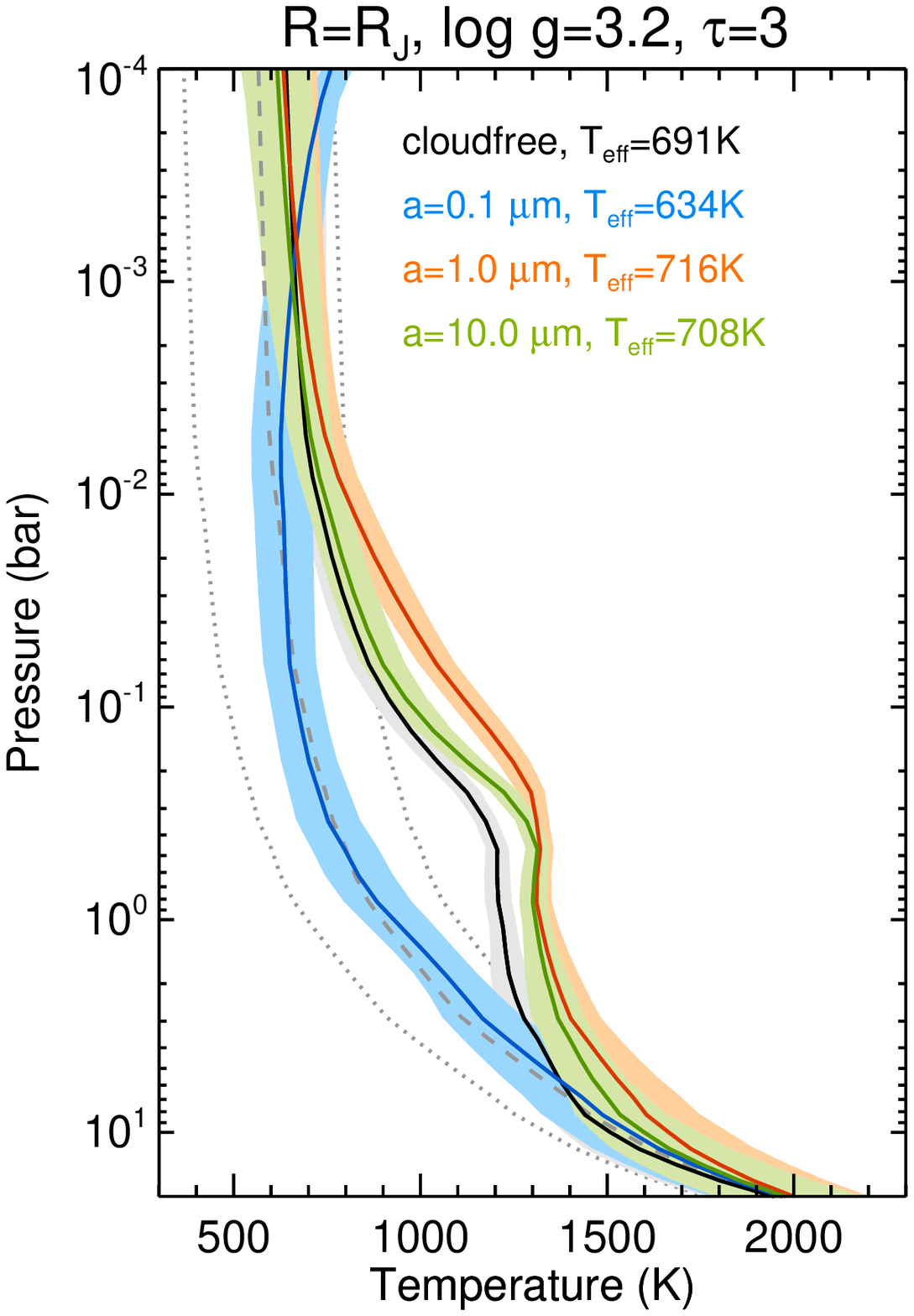}
\caption{Examining the effects of cloud properties (optical depth $\tau$ and particle radius $a$) on the computed temperature-pressure profiles.  Left panel: varying the cloud opacity (with $a=10$ $\mu$m) for the UC models.  Middle panel: varying the cloud deck location.  Right panel: varying the cloud particle radius for the UC models.  In all of the panels, we assume $R=R_{\rm J}$; for the left and middle panels, we use $\log{g}=4.0$.  For the right panel, we use $\log{g}=3.2$ as the $\log{g}=4.0$ suite of models did not produce a reasonable fit for $a=0.1$ $\mu$m.  The dashed curve is the initial temperature-pressure profile used for the retrieval, while the dotted curves represent its uncertainties.}
\label{fig:tp}
\end{figure*}

\begin{figure}
\centering
\includegraphics[width=\columnwidth]{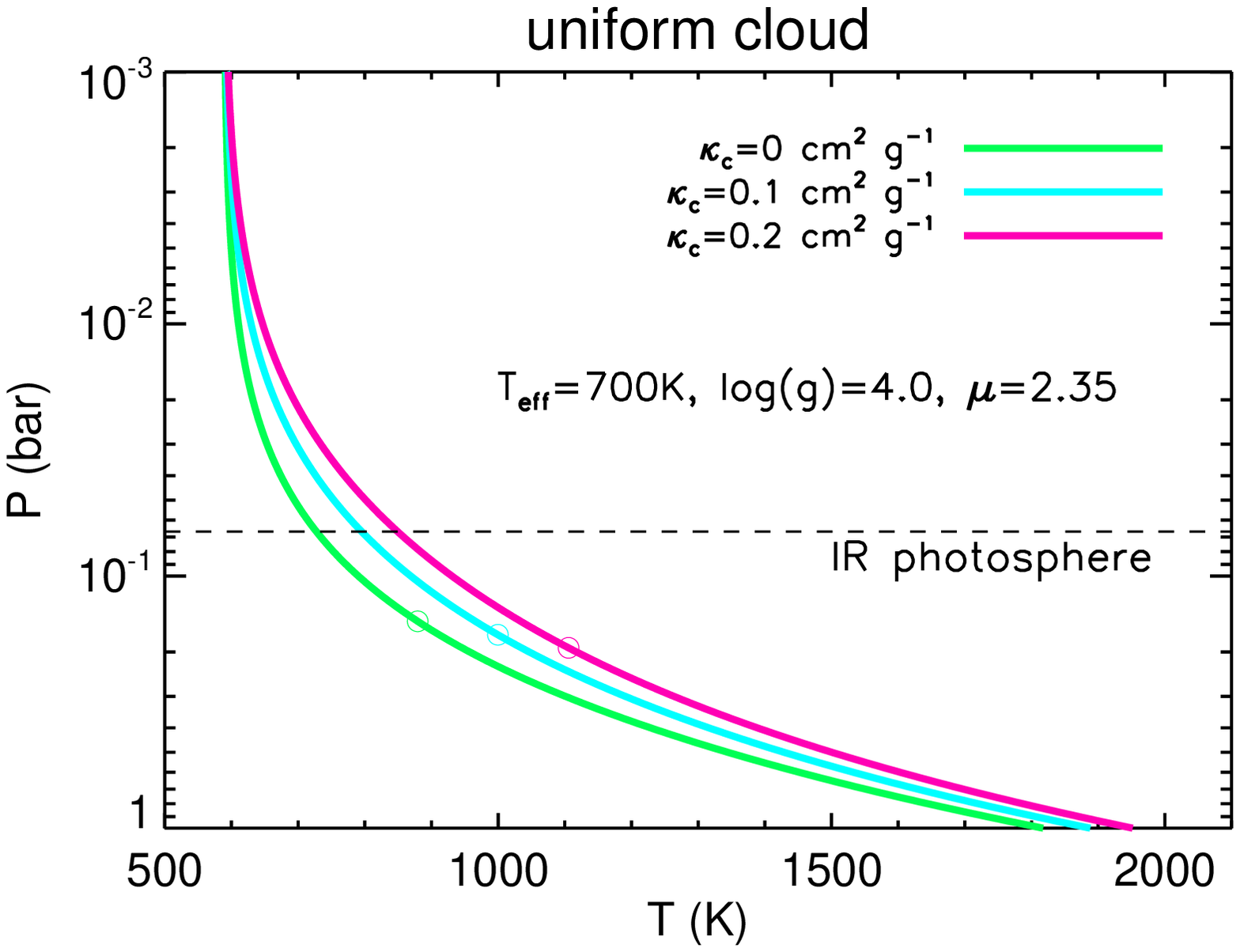}
\includegraphics[width=\columnwidth]{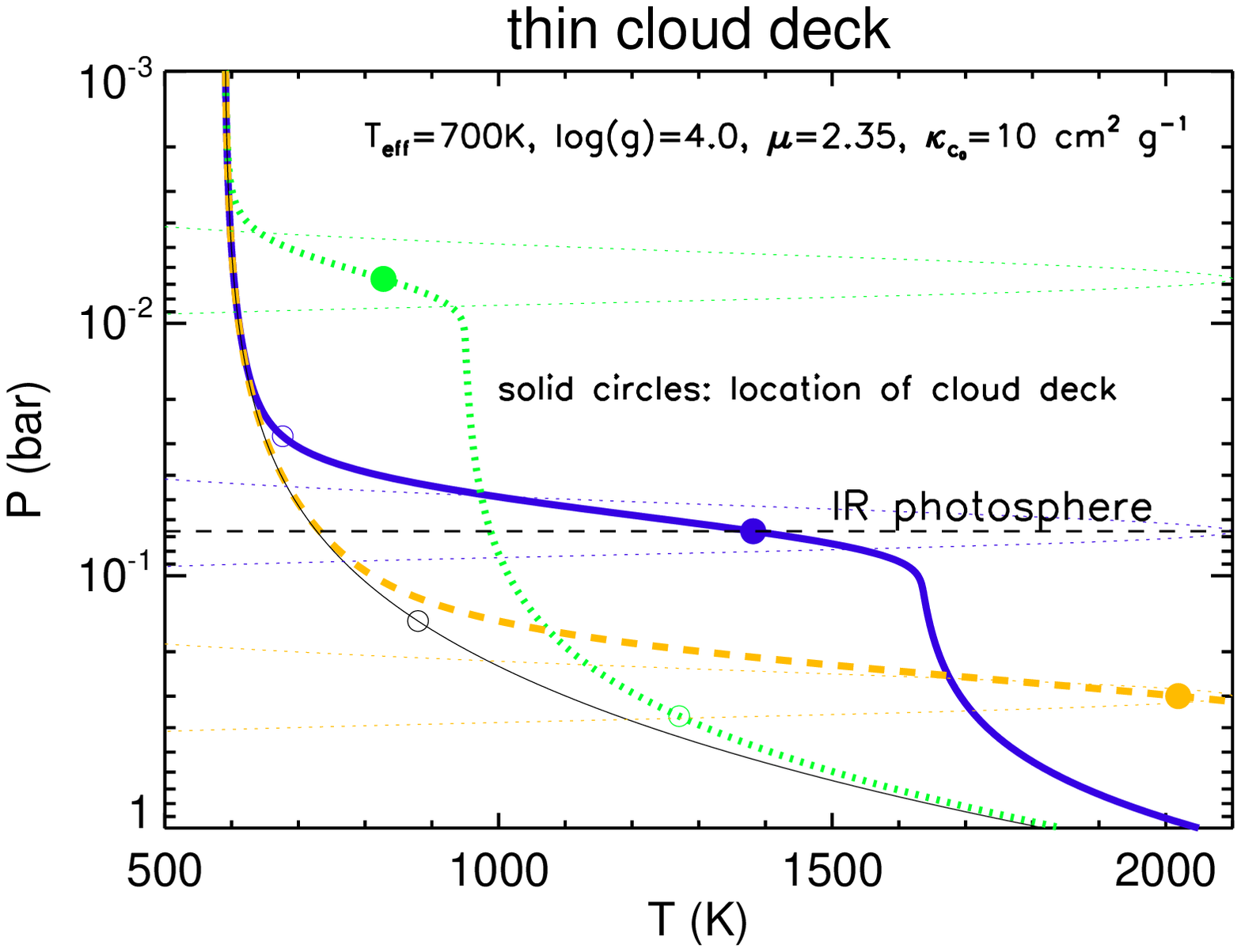}
\caption{Examples of temperature-pressure profiles computed using the analytical models of Heng et al. (2012) for a non-irradiated, self-luminous atmosphere.  The open circles indicate the locations where the lapse rates become super-adiabatic; at atltitudes below this point, the model profiles are unphysical.  Top panel: uniform cloud layer.  Bottom panel: thin cloud deck.  The thin, dotted curves show the Gaussian shape of the cloud deck; it is shown with an arbitrary normalization in the plot but is normalized to $\kappa_{\rm c_0}=10$ cm$^2$ g$^{-1}$ in the actual calculations.  The thin, solid curve is for the cloudfree case.}
\label{fig:hhps}
\end{figure}

The novel technical aspect of our study is the addition of a simple, phenomenological cloud model to our retrieval method.  We assume our cloud particles to be spherical and made of enstatite (MgSiO$_3$).  The latter assumption is reasonable and illustrative, because it is well known that the refractive indices for many materials are fairly similar (Figure \ref{fig:refract}; Table 5.1 of \citealt{p10}).  In principle, the refractive index describing the surface layers of a given material may be different from that of the bulk, but it is currently unknown how to compute these differences (B.T. Draine, 2013, private communication).  In practice, the refractive index is assumed to be the same throughout a given material and thus to be independent of the particle radius.  Examples of refractive indices, including those of enstatite, are shown in Figure \ref{fig:refract} for completeness.

With these assumptions, our cloud model has 4 free parameters.
\begin{enumerate}

\item \textbf{Particle radius ($a$).}

\item \textbf{Cloud deck boundaries ($P_{\rm up}$ and $P_{\rm down}$): } For the intermediate (IN) model, the thickness of the cloud deck is $\Delta P = P_{\rm down} - P_{\rm up}$.  We assume $P_{\rm down} = 1$ bar and $P_{\rm up} = 0.01$ bar, based on the sensitivity of the L$^\prime$-band flux to water and methane (Figure \ref{fig:L_band}).  Specifying $\Delta P$ allows us to compute the spatial extent of the cloud ($\Delta z$).

\item \textbf{Cloud optical depth ($\tau$): } This is described by the expression, $\tau = Q_{\rm ext} \pi a^2 n \Delta z$, where $Q_{\rm ext}$ is the extinction efficiency defined at L$^\prime$ band (3.78 $\mu$m). By specifying the values of $\tau$ and $a$, we determine the value of $n$, the number density of cloud particles.  Varying $\tau$ and keeping $a$ fixed is akin to changing the column mass of enstatite present in the atmosphere.

\end{enumerate}

We have used the extinction efficiency $Q_{\rm ext}$, which is the sum of the absorption ($Q_{\rm abs}$) and scattering ($Q_{\rm scat}$) efficiencies, in our expression for the optical depth of the cloud.  This circumvents the technical difficulty that $Q_{\rm abs}=0$ in the L$^\prime$ band for certain materials (such as enstatite).  Physically, the approximation being taken is that the cloud not only scatters a photon back towards the deep interior, but that it is absorbed locally, which mimics the scattering greenhouse effect in a maximal manner \citep{p10}.  While this approach allows us to account for infrared scattering \citep{dekok11}, it misses the intricacies of multiple scattering events, which allow some fraction of the scattered radiation to be absorbed deeper in the atmosphere and the remaining fraction to depart it.  Thus, from a technical standpoint, our clouds behave like pure absorbers with a first-order correction for scattering.  This view is consistent with the fact that if we generalize the analytical formalism of \cite{hhps12} to include infrared scattering, the term involving the internal heat retains the same mathematical form, except that the infrared absorption opacity is now replaced by the infrared extinction (absorption plus scattering) opacity.

A key advantage of our simple cloud model is that its behavior is easy to decipher.  The particle-radius dependence of extinction is captured within $Q_{\rm ext}$ (Figure \ref{fig:qext}).  Essentially, different particle sizes can be understood in terms of the behavior of $Q_{\rm ext}$ with wavelength ($\lambda$) and the size parameter ($2 \pi a / \lambda$).
\begin{itemize}

\item \textbf{Small particles ($a=0.1$ $\mu$m): } When $a/2\pi\lambda \ll 1$, one is sampling the steep, Rayleigh slope of the $Q_{\rm ext}$ function.  Since we are defining $\tau$ at the L$^\prime$ band, the optical depth at shorter wavelengths becomes much larger than unity.  The implication is that in an attempt to suppress the L$^\prime$-band flux with small grains, the fluxes at shorter wavelengths become even more suppressed, thus yielding a bad overall fit to the data (Figure \ref{fig:spectra}).  For this reason, small grains are generally not good candidates for modeling clouds in the atmosphere of HR 8799b, consistent with previous studies \citep{madhu11b,barman11,marley12}.

\item \textbf{Medium-sized particles ($a=1$ $\mu$m): } When $a/2\pi\lambda \sim 1$, one is sampling the peak of the $Q_{\rm ext}$ function, meaning that the extinction efficiency is comparable across wavelength but with non-negligible variations.  Specifically, an optically thick cloud defined at the L$^\prime$ band tends to suppress the K-band flux by a larger amount, producing a poor fit to the data.

\item \textbf{Large particles ($a=10$ $\mu$m): } When $a/2\pi\lambda \gg 1$, one is sampling the flat portion of the $Q_{\rm ext}$ function.  Flux suppression by clouds containing large grains is essentially grey or flat across wavelength.  For this reason, large grains are good candidates, again consistent with previous studies \citep{madhu11b,barman11,marley12}.  Particles larger than 10 $\mu$m will sample the $Q_{\rm ext}$ curve in a very similar way.  For example, we expect $a=60$ $\mu$m models to yield essentially the same results as the $a=10$ $\mu$m ones.

\item \textbf{Ideal particles ($a=1.5$ $\mu$m): } Understanding the behavior of small, medium-sized and large particles allows us to specify the ideal particle radius.  Specifically, one wishes to maximally suppress the L$^\prime$-band flux, while minimally suppressing the H- and K-band fluxes.  In other words, the ideal particle will sample the $Q_{\rm ext}$ curve in such a manner that its peak sits above the L$^\prime$ band, while the trough between the peak and the first harmonic sits between the H and K bands.  Such a particle has a radius of $a \sim \lambda/\pi \sim 1$ $\mu$m.  A more careful calculation and examination of Figure \ref{fig:qext} reveals that $a=1.5$ $\mu$m.  It is worth noting that this selection of a specific particle size is only meaningful when a monodisperse population of particles is present.  If a size distribution exists, then the clear correspondence between particle size and wavelength-dependent extinction washes out.

\end{itemize}

As a precursor to discussing our results, we need to gain confidence in our cloud model by elucidating its basic properties.  The effect of $\tau$ on the computed spectra was already demonstrated in Figure \ref{fig:spectra}: more optically thick clouds generally suppress the flux more.  The behavior with $a$ is non-monotonic because of the way that the $Q_{\rm ext}$ curve is being sampled (Figure \ref{fig:qext}): $a=1$ $\mu$m particles tend to suppress more flux at 1.5--4.0 $\mu$m, than $a=10$ $\mu$m ones, for a given value of $\tau$, because the value of $Q_{\rm ext}$ is higher in this wavelength range.  The $a=0.1$ $\mu$m particle yields a bad fit to the measured SED (Figure \ref{fig:spectra}), because of the steepness of the $Q_{\rm ext}$ curve sampled at $2\pi a/\lambda < 1$; for example, if $\tau=1$ in the L$^\prime$ band, then we have $\tau \gg 1$ in the H band.  This bad fit translates into a somewhat odd temperature-pressure profile in Figure \ref{fig:tp}.

Figure \ref{fig:tp} shows the effect of varying $a$ and $\tau$ on the temperature-pressure profile.  We expect only a greenhouse warming effect to occur for self-luminous, non-irradiated gas giants like HR 8799b.\footnote{For irradiated exoplanets, scattering in the optical/shortwave produces an anti-greenhouse effect.}  Larger values of the cloud optical depth ($\tau \sim 10$) produce warmer temperature-pressure profiles, but at the expense of yielding a poorer fit to the observed SED.  The trend of the warming with $a$ is again non-monotonic for the reason previously described.  The warming effect of a cloud optical depth may be reproduced using the analytical models of \cite{hhps12}, which we adapt to model a non-irradiated, self-luminous atmosphere (top panel of Figure \ref{fig:hhps}).  The cloud opacity, denoted by $\kappa_{\rm c}$, includes both absorption and scattering.

The same analytical models can also be used to elucidate the effects of a thin cloud deck ($\Delta_{\rm c} = 10$ in the terminology of \citealt{hhps12}).  Essentially, a warming effect is produced only for the atmospheric layers residing below the cloud deck (bottom panel of Figure \ref{fig:hhps}).  If the cloud deck is placed below the infrared photosphere, then the temperature-pressure profile is identical to its cloudfree analogue, at least at the pressure levels probed by the observations.  If the cloud deck is placed at or above the infrared photosphere, then all of the layers below it are warmed.  We have exaggerated the cloud opacity in Figure \ref{fig:hhps} to more clearly illustrate the warming effect of a thin cloud deck.  These qualitative trends are seen in our full retrieval calculations as well (Figure \ref{fig:tp}) when we vary the values of $P_{\rm up}$ and $P_{\rm down}$.  When the cloud deck is placed deep within the atmosphere ($P_{\rm up}=1$ bar, $P_{\rm down}=20$ bar), there is a negligible warming effect on the temperature-pressure profile because it sits below the photosphere across all of the wavelengths considered.  Placing the cloud deck between $P_{\rm up}=10^{-5}$ bar and $P_{\rm down}=0.01$ bar produces the greatest warming effect, while setting $P_{\rm up}=0.01$ bar and $P_{\rm down}=1$ bar (as in the IN model) produces a warming effect that is intermediate between the two extremes.

The disadvantage of our cloud model is that it is not grounded in first-principles condensation physics and chemistry.  But this feature may also be viewed as an advantage: we are attempting to constrain the cloud/haze properties empirically, so that we may further infer additional properties after the fact, free from any preconceived ideas based on brown dwarfs.  Another advantage is that we do not need to prescribe an eddy diffusion coefficient ($K_{\rm zz}$) as a proxy for vertical mixing induced by convection.  The range of values of $K_{\rm zz}$ considered by forward modelers typically span several orders of magnitude and it is used as a physically plausible parameter that sets the conditions for non-equilibrium chemistry and particle lofting to occur \citep{madhu11b,barman11,marley12}.

\section{Results}
\label{sect:results}

\begin{figure*}
\centering
\includegraphics[width=\columnwidth]{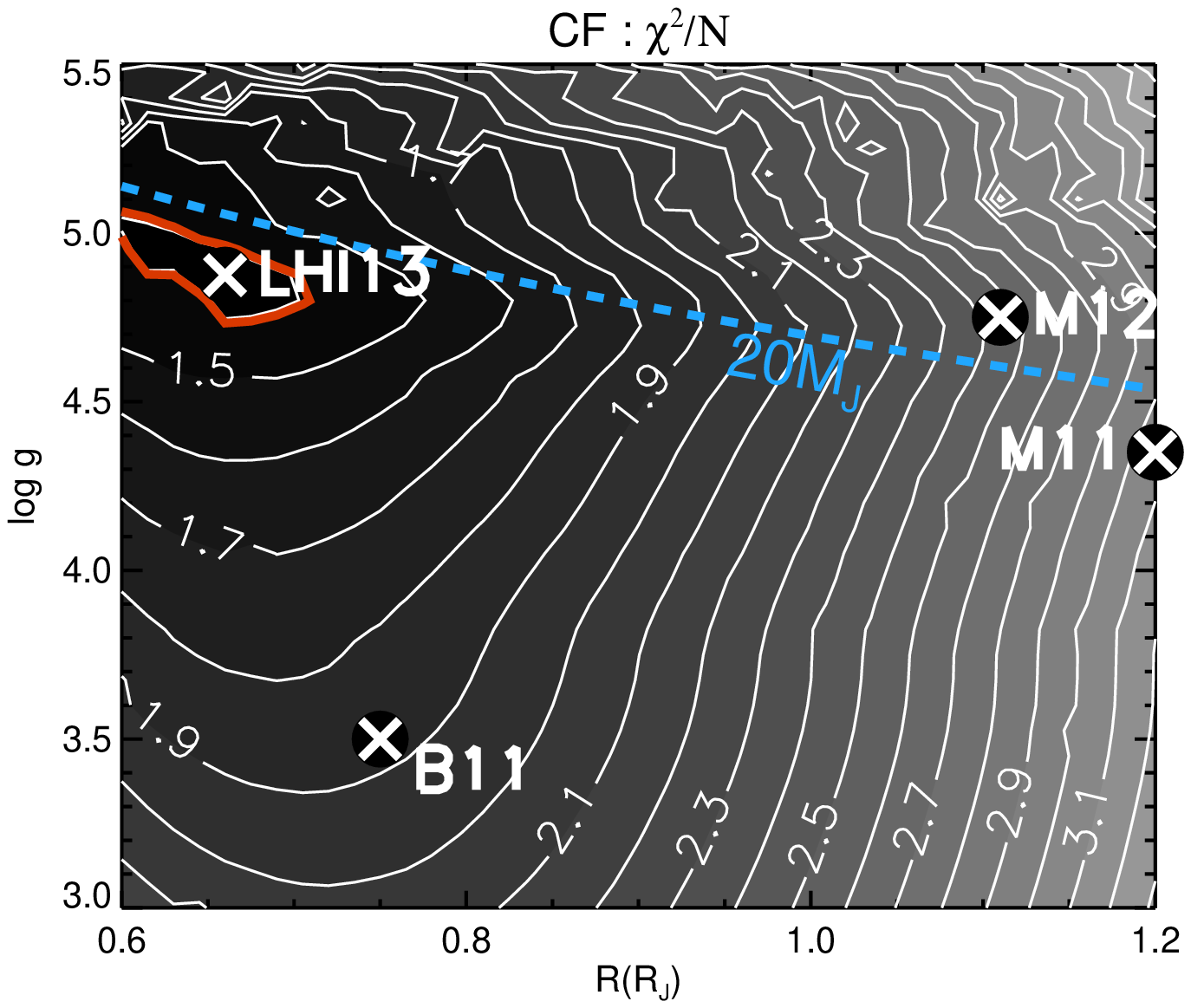}
\includegraphics[width=\columnwidth]{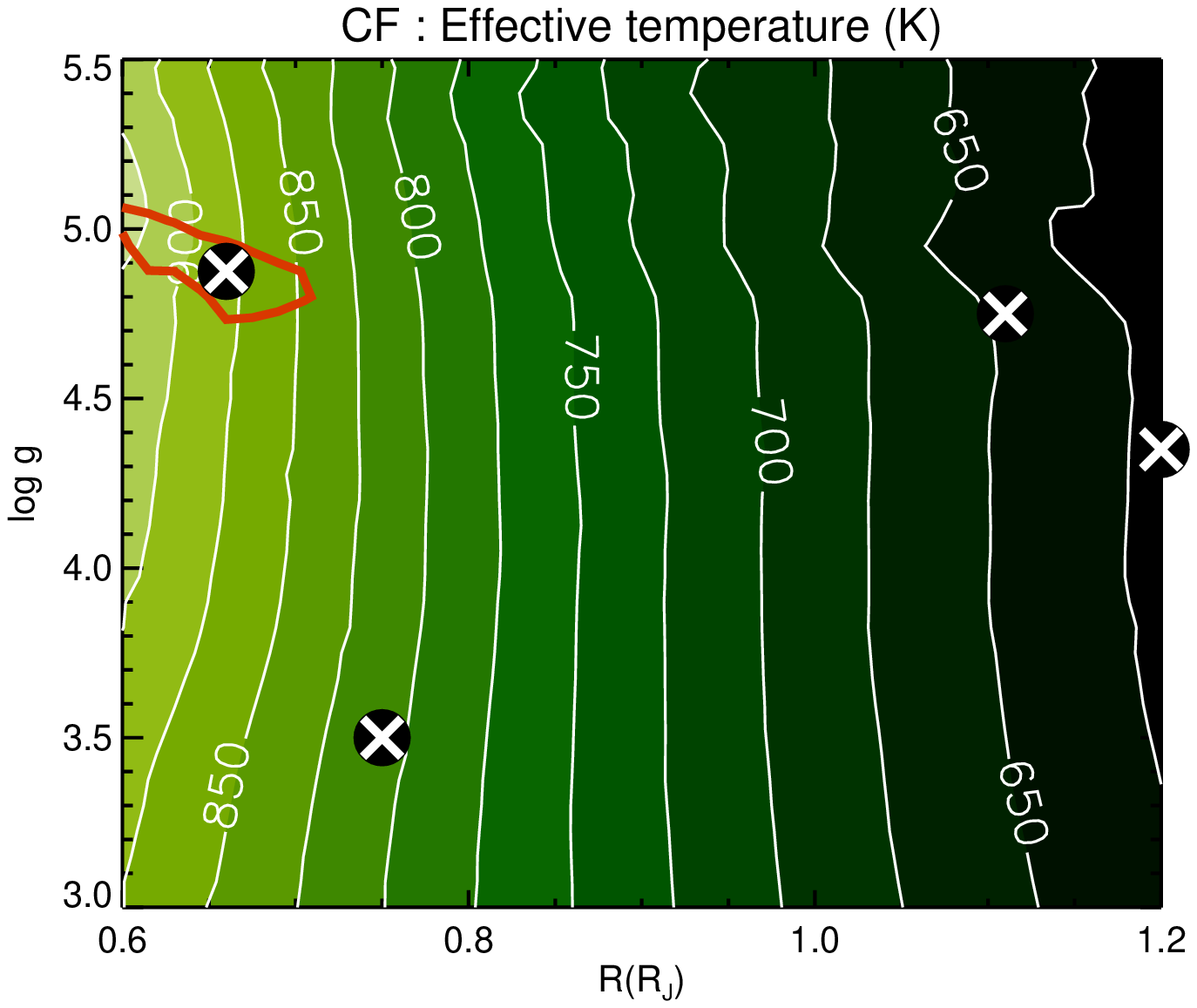}
\includegraphics[width=\columnwidth]{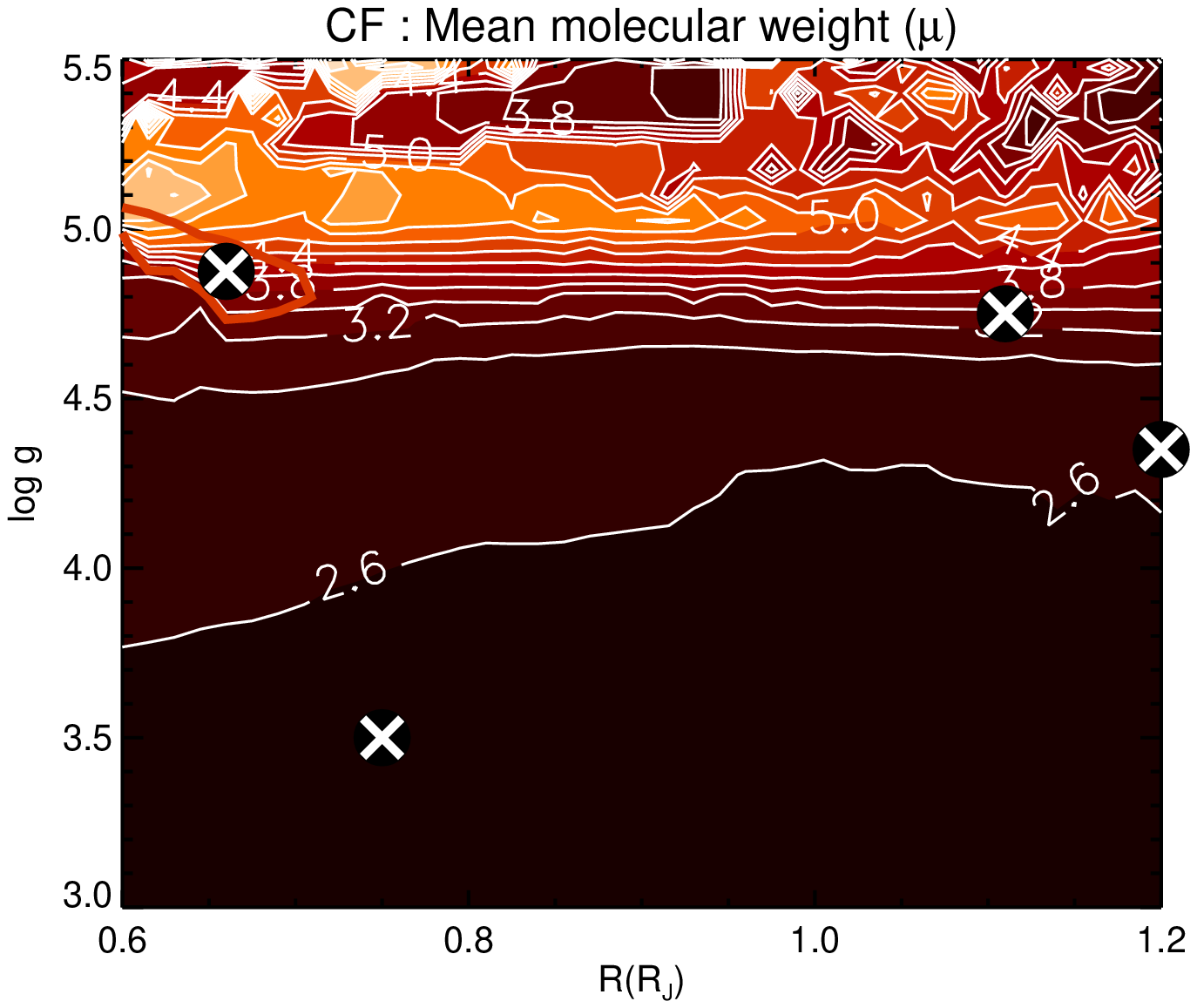}
\includegraphics[width=\columnwidth]{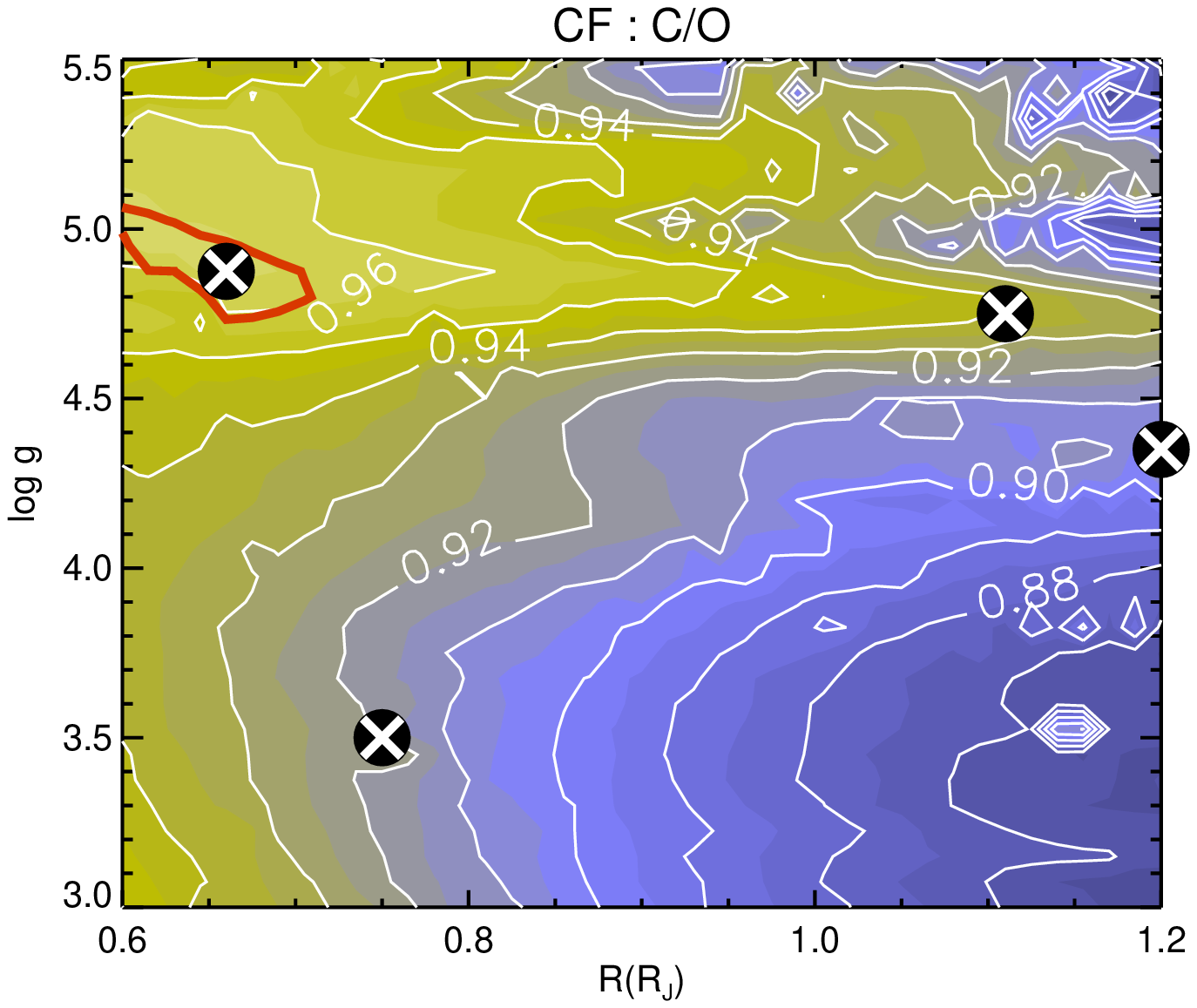}
\includegraphics[width=\columnwidth]{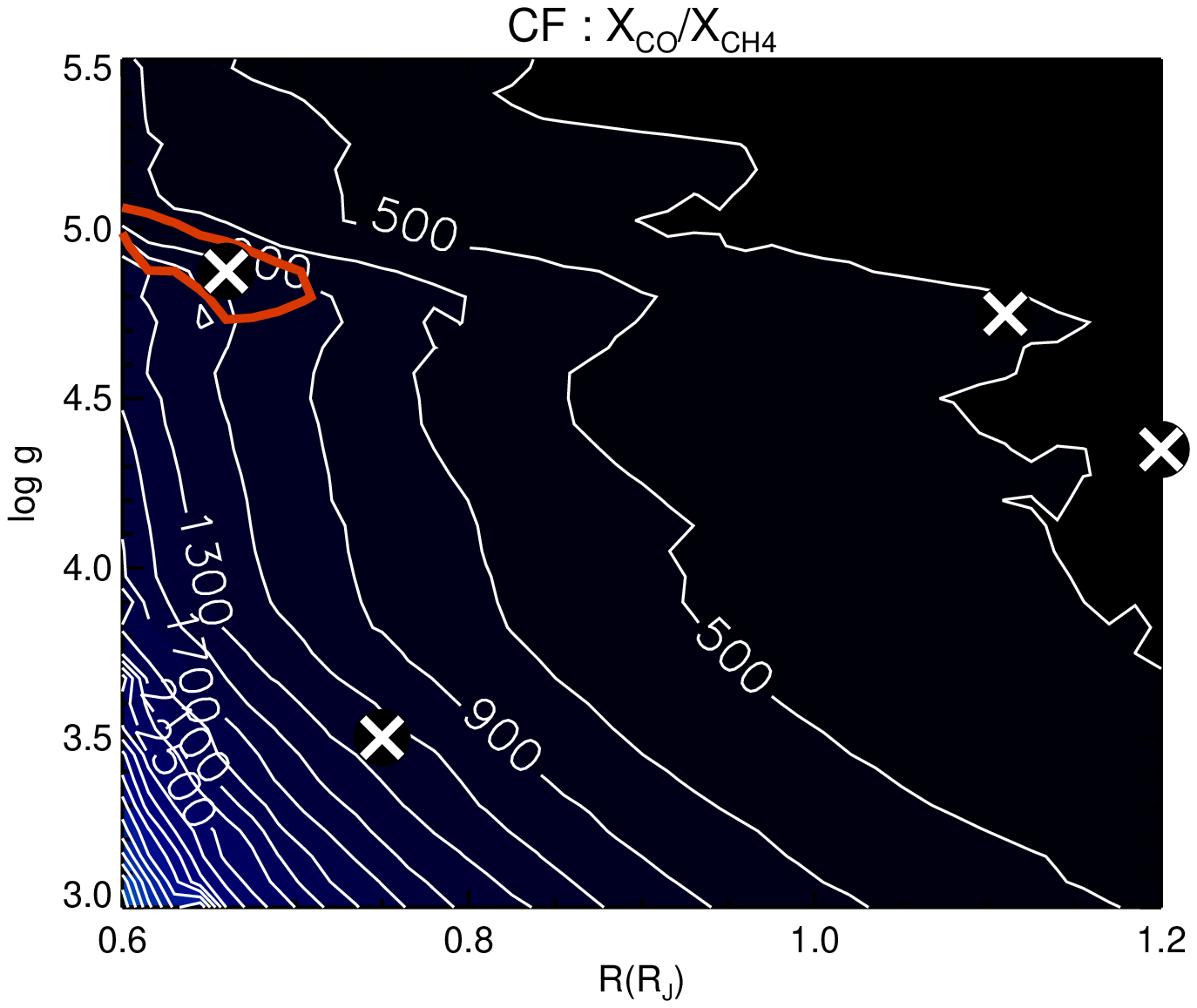}
\includegraphics[width=\columnwidth]{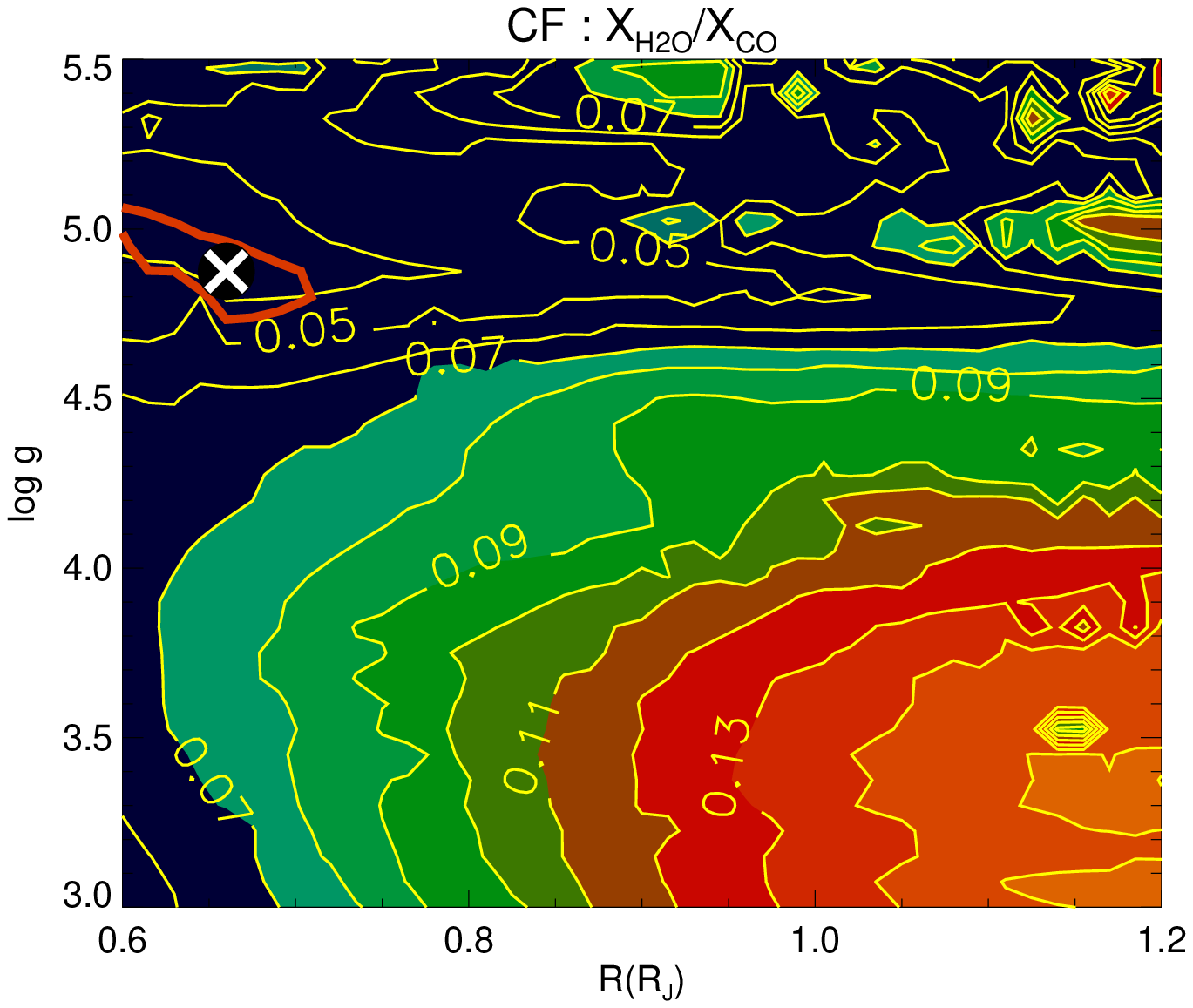}
\caption{Contour plots of various quantities, from the cloudfree (CF) suite of models, as functions of the radius and surface gravity: goodness of fit (top left panel); effective temperature (top right panel); mean molecular weight (middle left panel); carbon-to-oxygen ratio (middle right panel); ratio of carbon monoxide to methane abundances (bottom left panel); ratio of water to carbon monoxide abundances (bottom right panel).  The reported values of the radius and surface gravity from previous studies are marked with crosses: Madhusudhan et al. (M11), Barman et al. (B11) and Marley et al. (M12).}
\label{fig:cf}
\end{figure*}

\begin{figure*}
\centering
\includegraphics[width=\columnwidth]{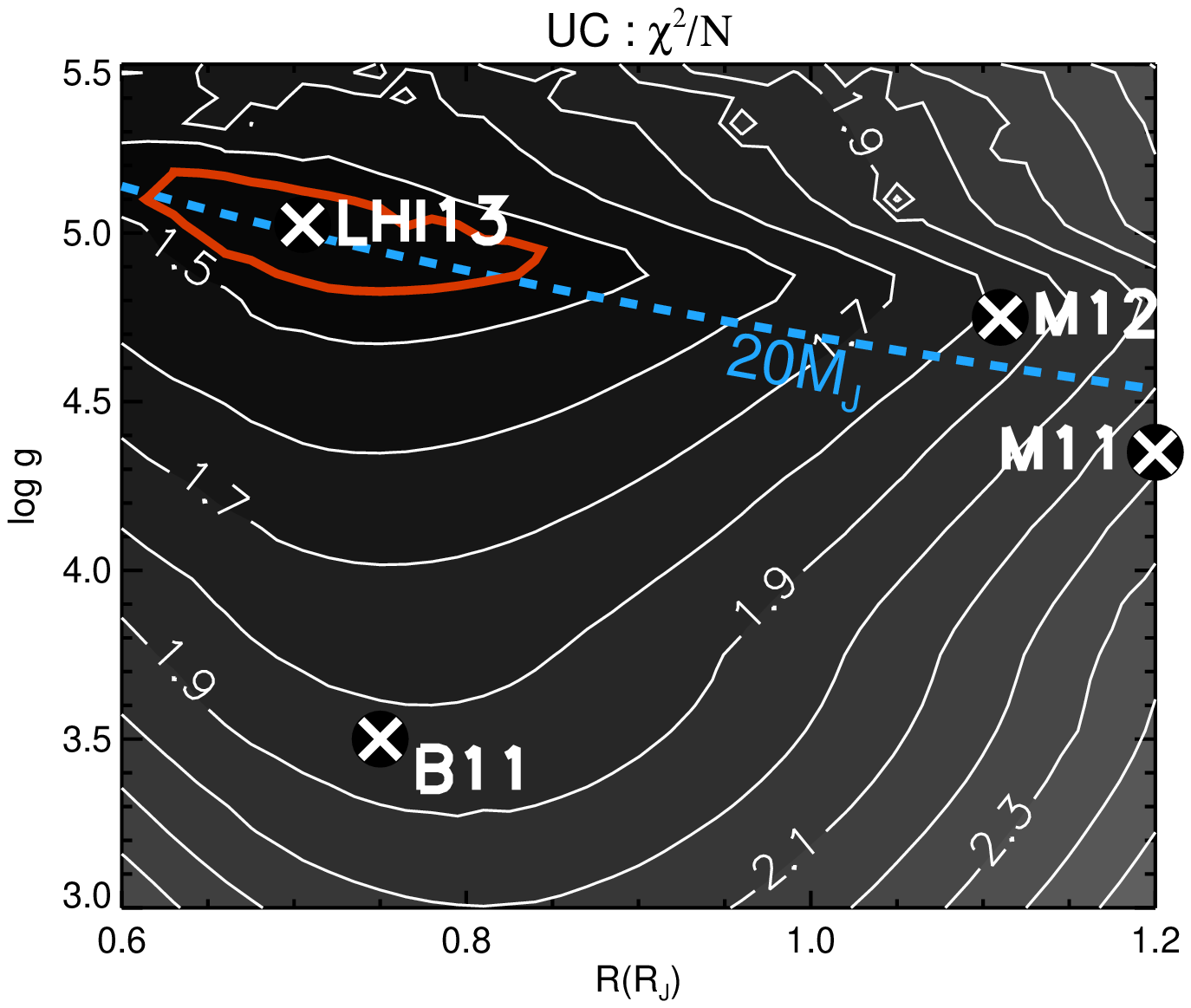}
\includegraphics[width=\columnwidth]{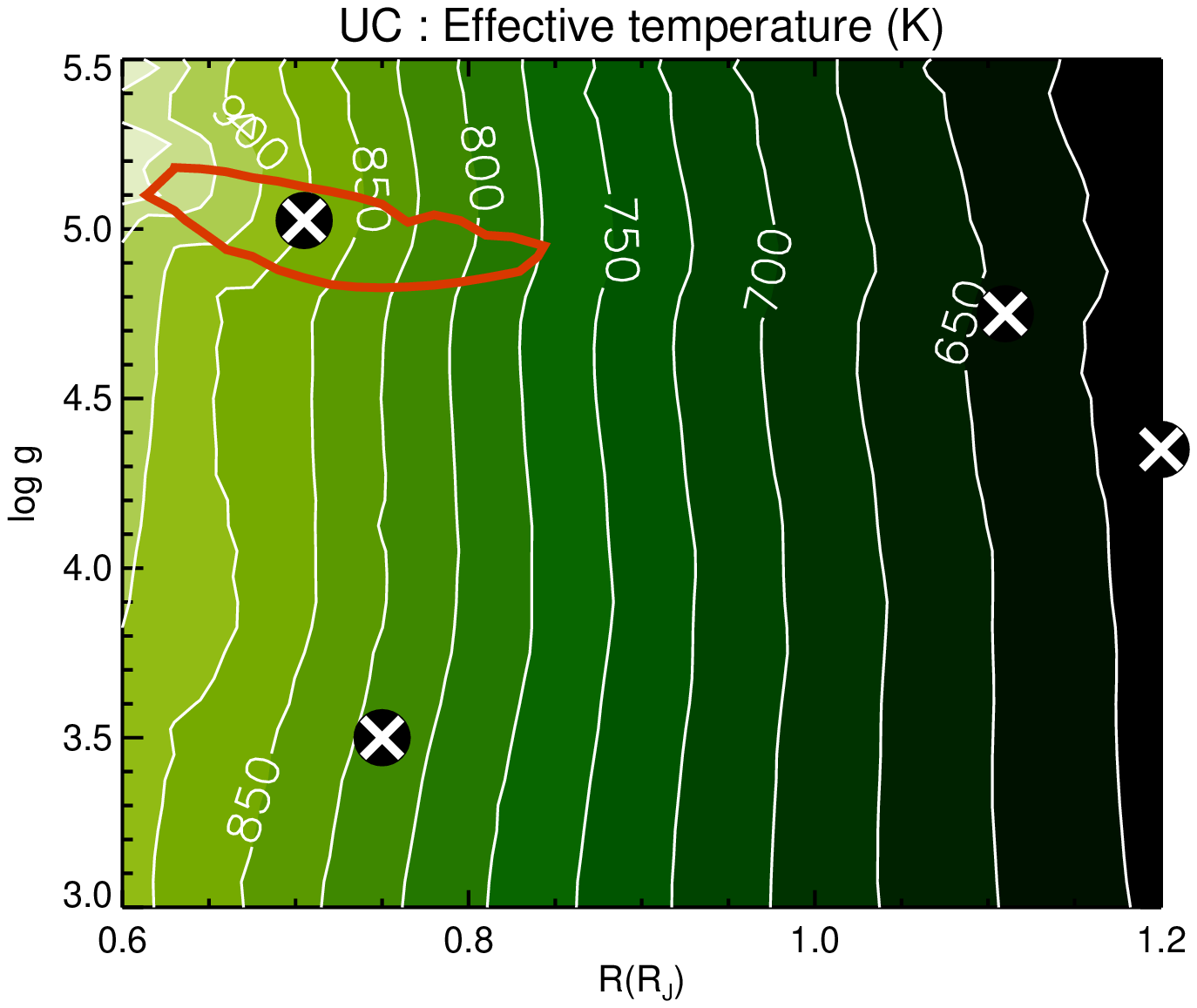}
\includegraphics[width=\columnwidth]{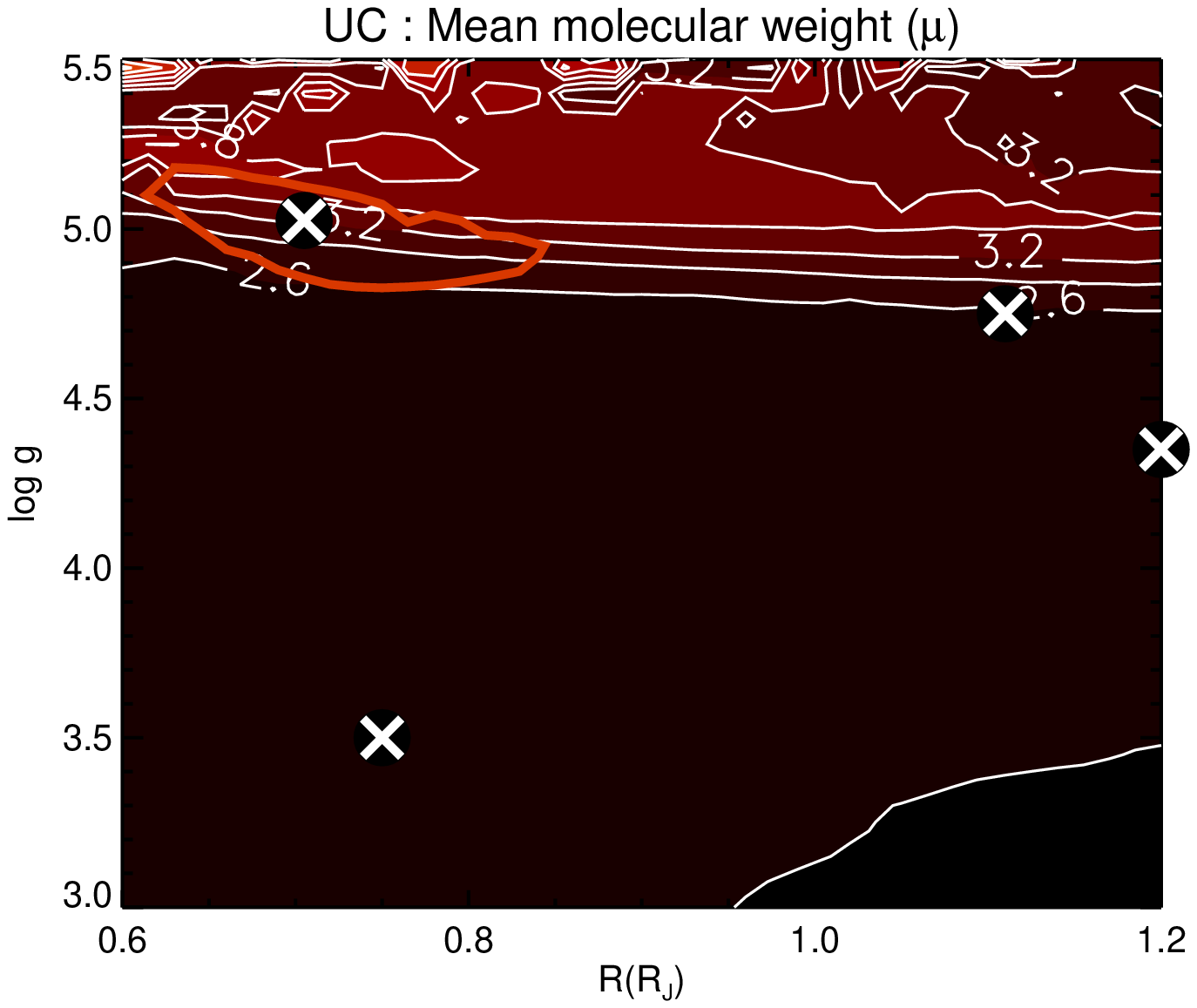}
\includegraphics[width=\columnwidth]{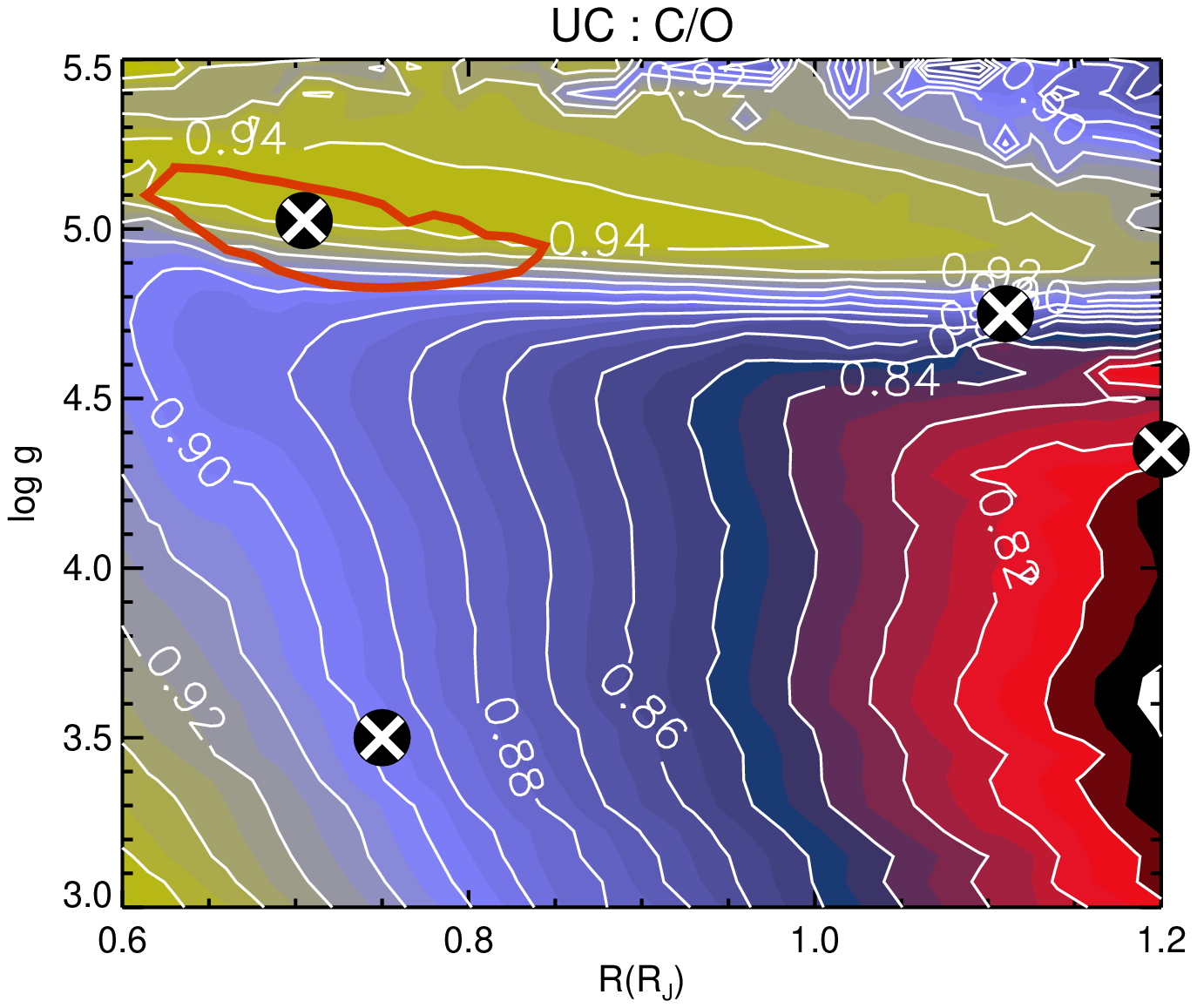}
\includegraphics[width=\columnwidth]{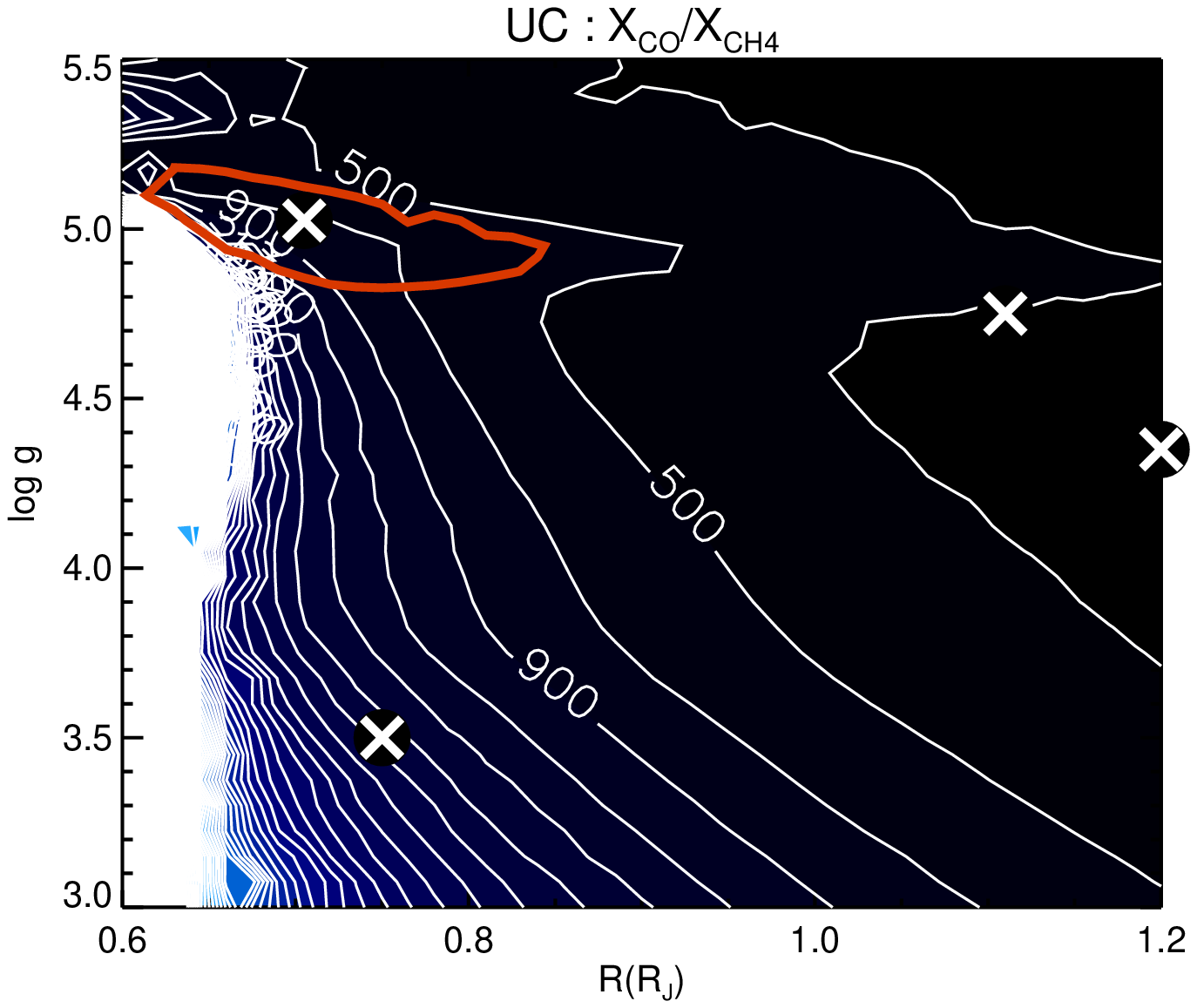}
\includegraphics[width=\columnwidth]{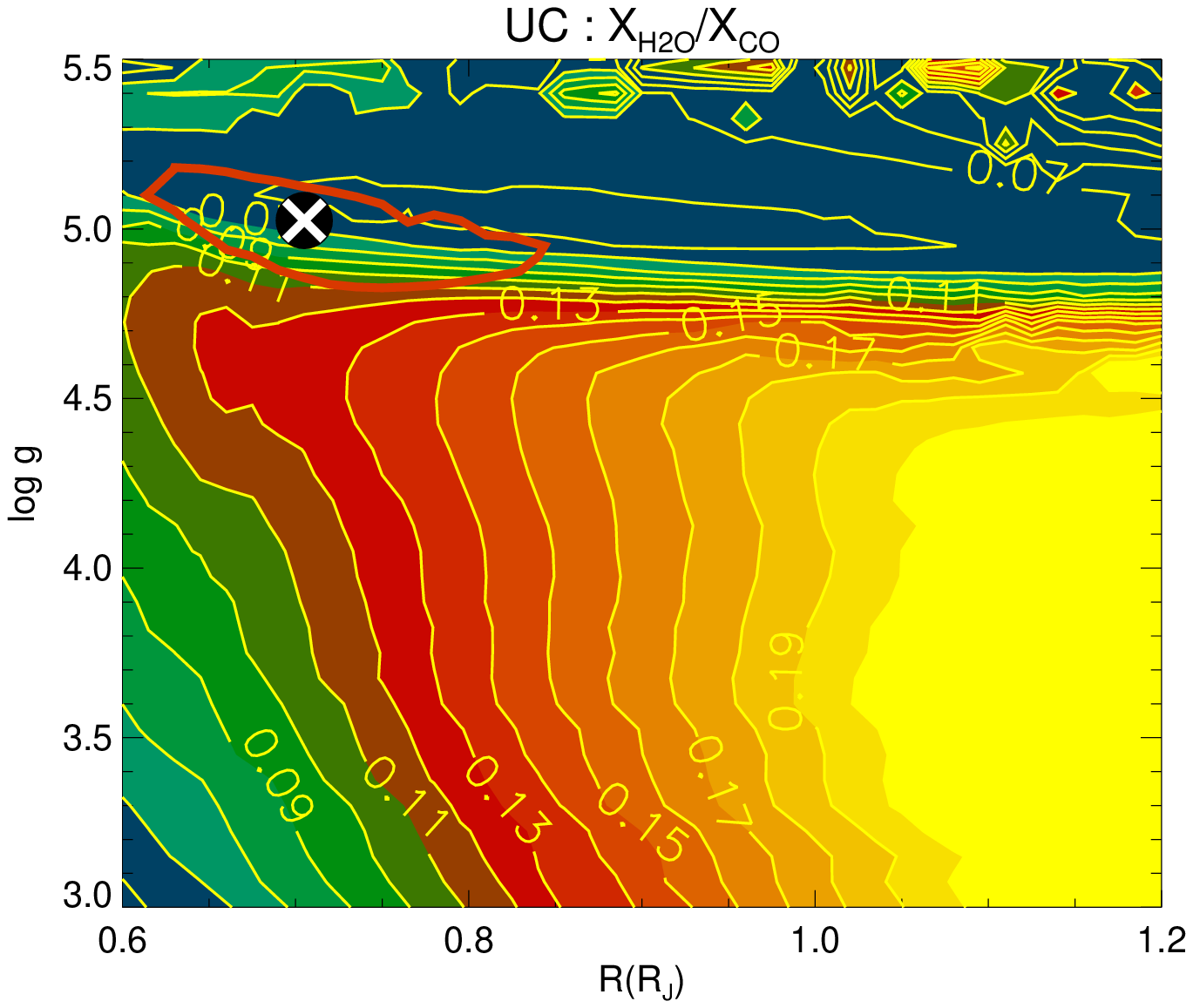}
\caption{Same as Figure \ref{fig:cf}, but for a suite of uniformly cloudy (UC) models with $\tau=2$ and $a=1.5$ $\mu$m.}
\label{fig:uc}
\end{figure*}

\begin{figure*}
\centering
\includegraphics[width=\columnwidth]{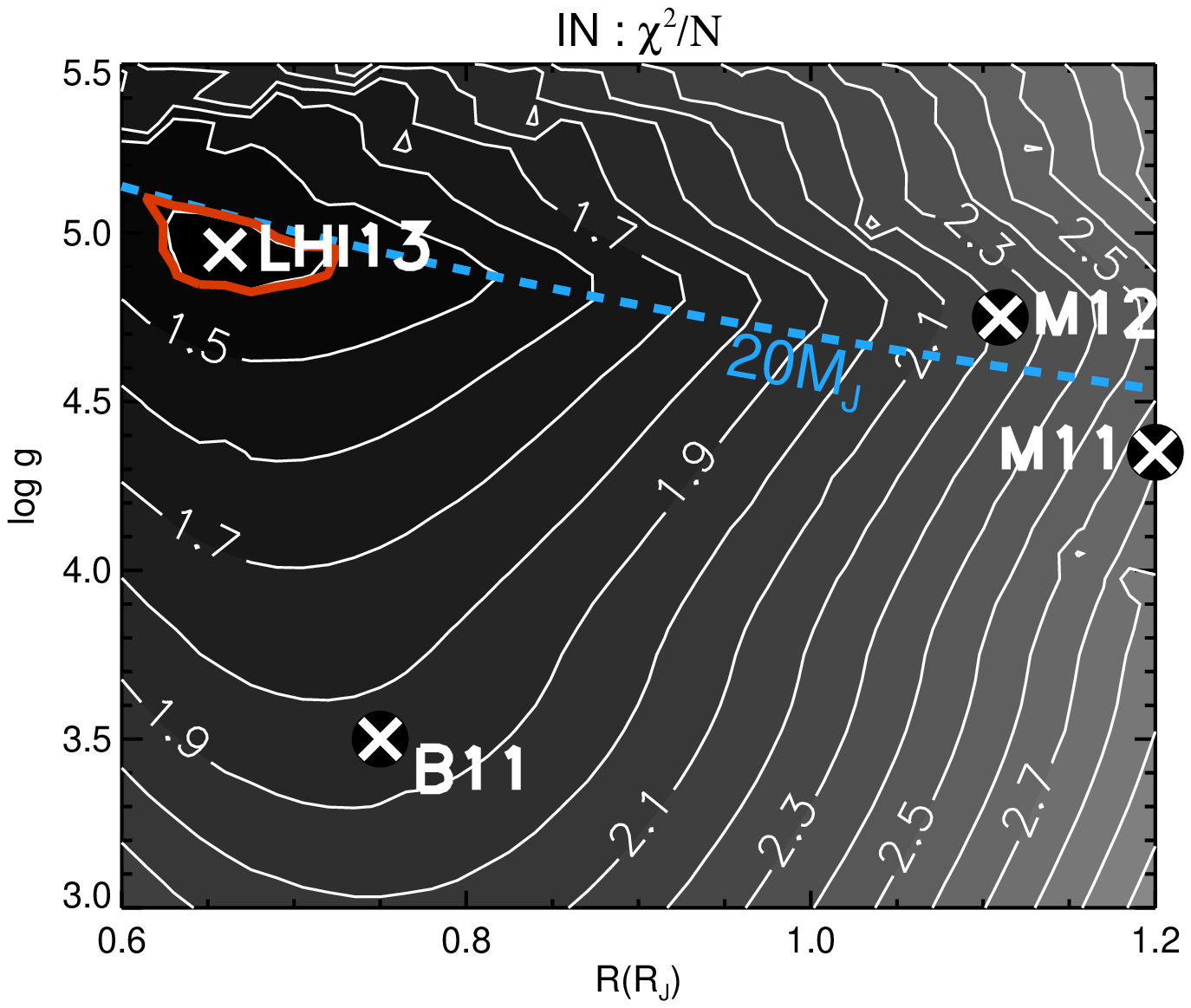}
\includegraphics[width=\columnwidth]{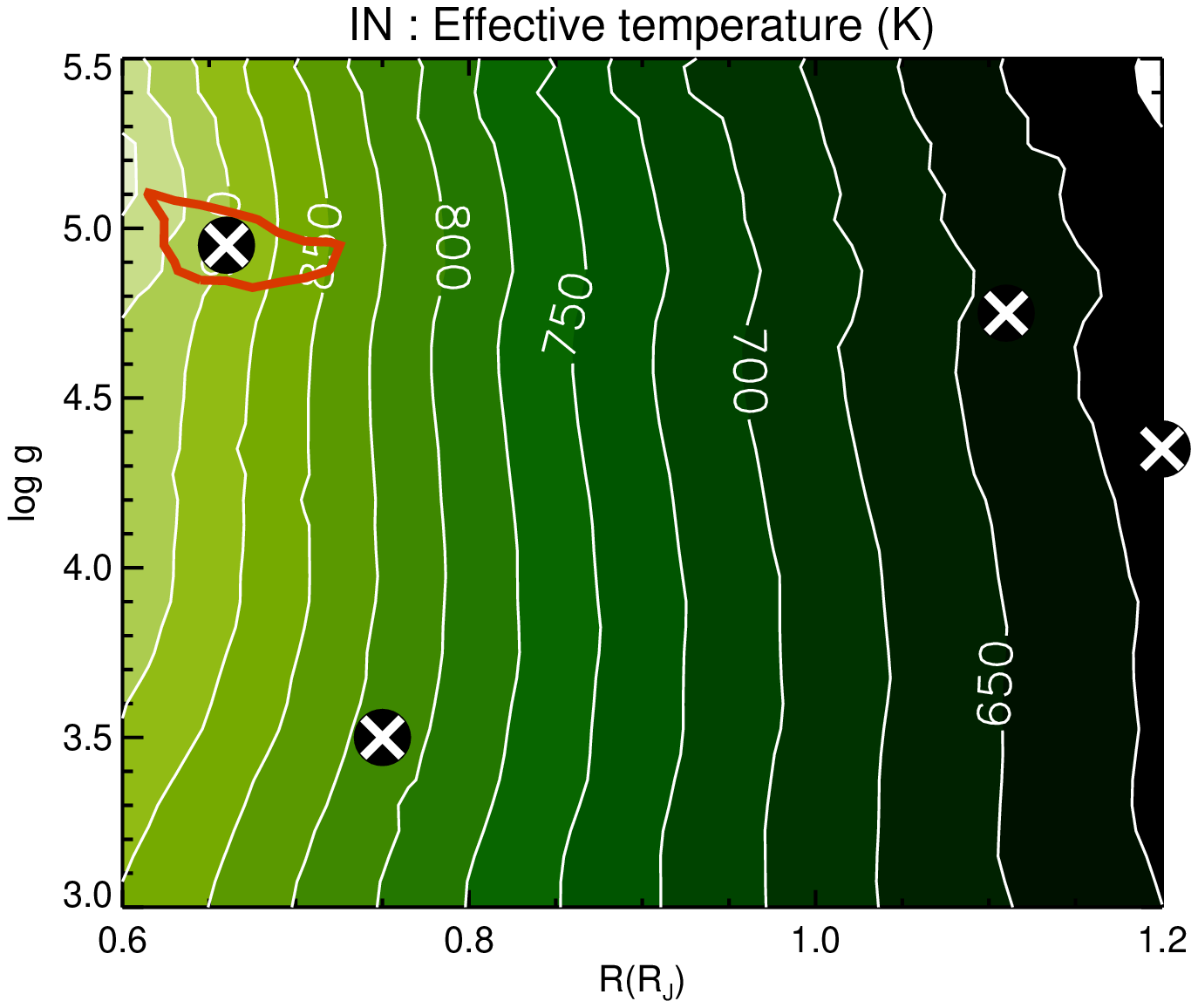}
\includegraphics[width=\columnwidth]{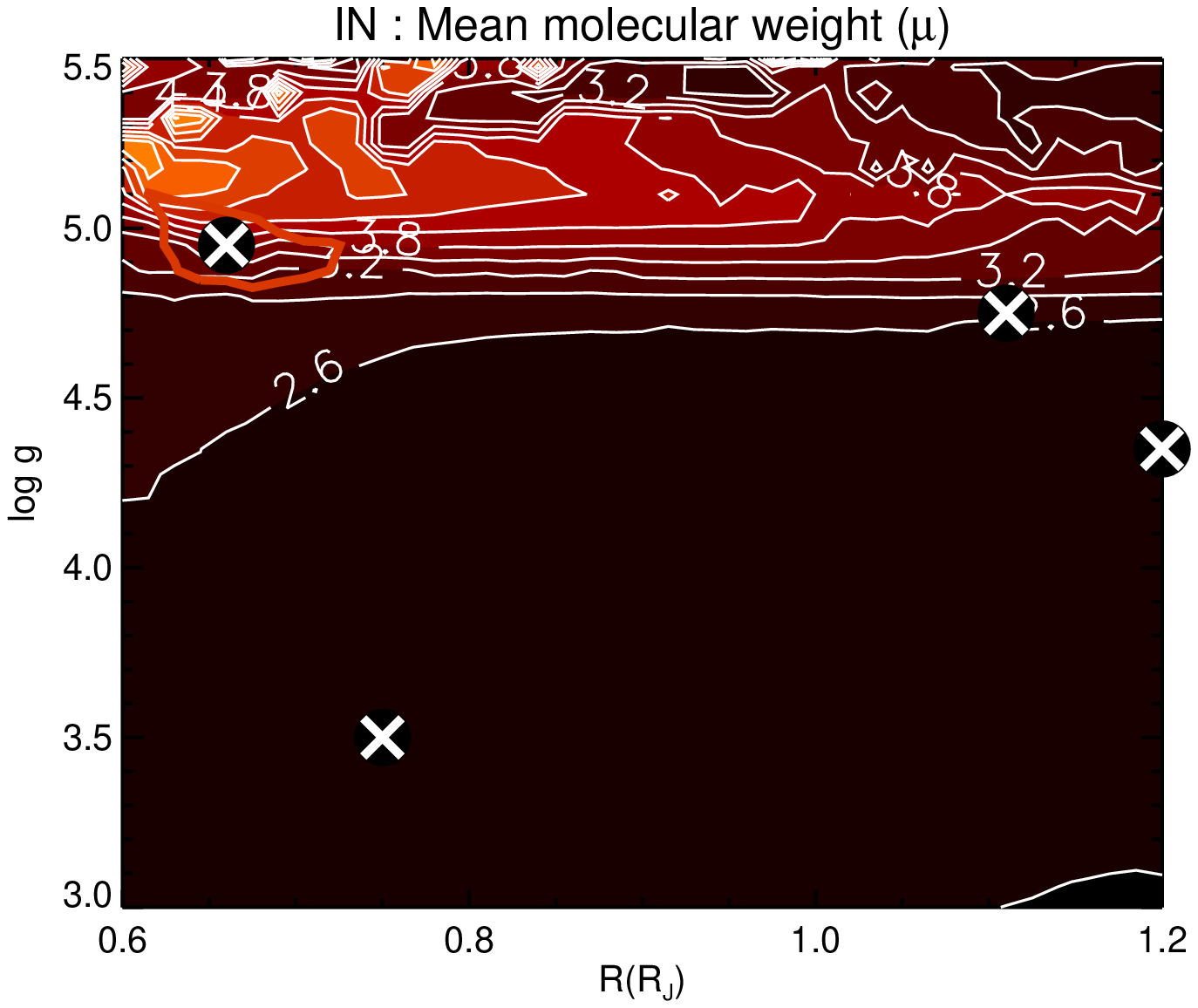}
\includegraphics[width=\columnwidth]{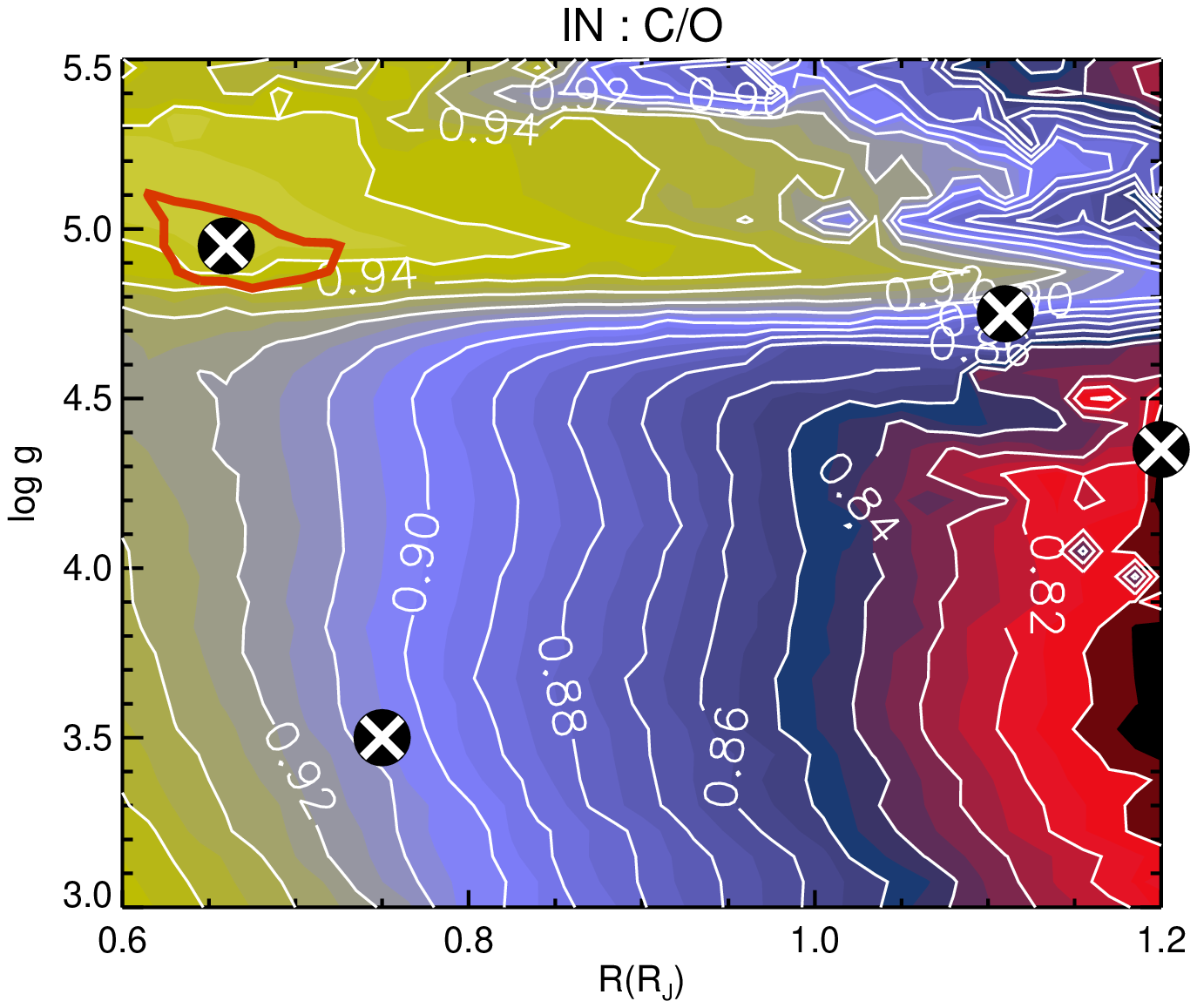}
\includegraphics[width=\columnwidth]{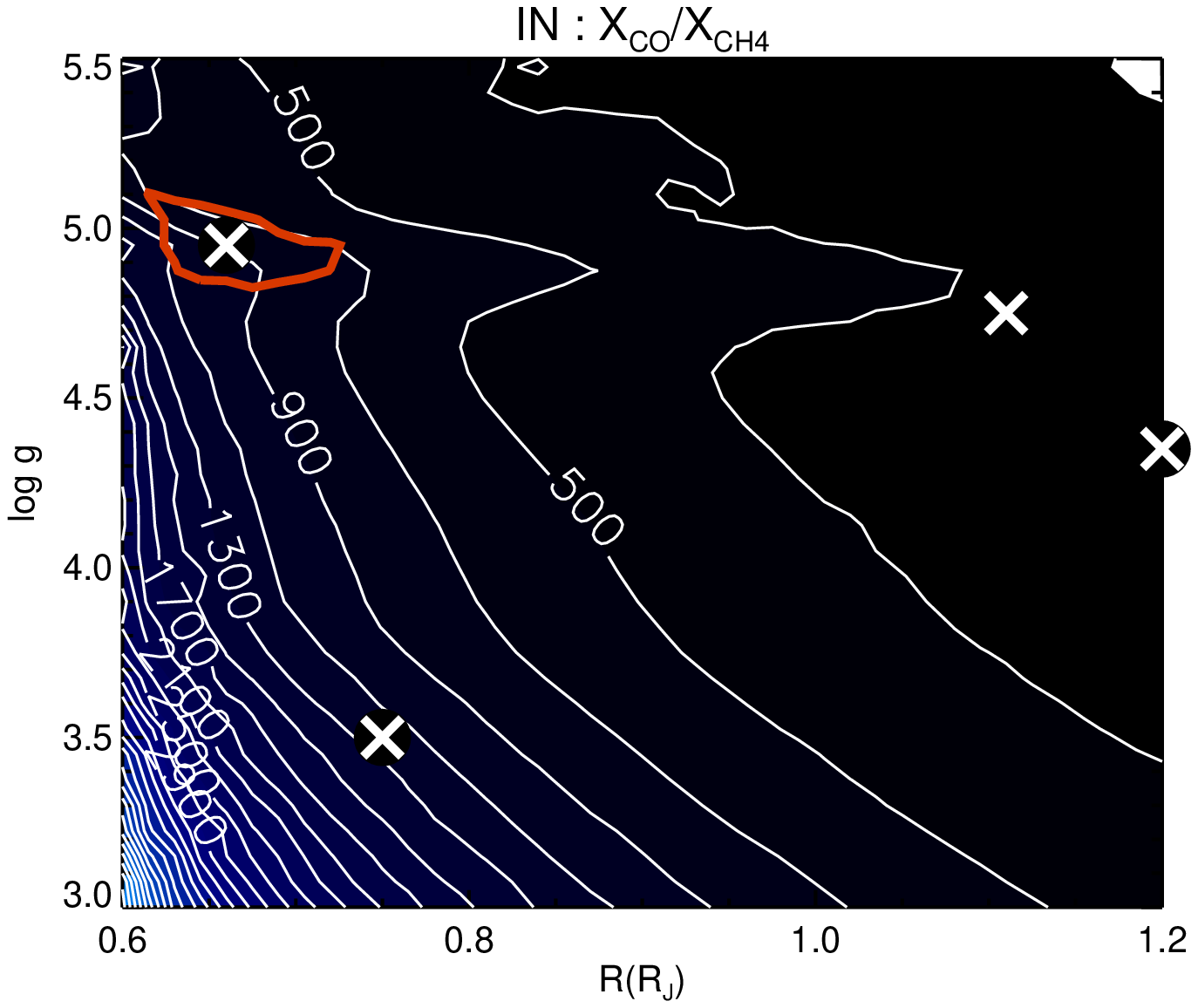}
\includegraphics[width=\columnwidth]{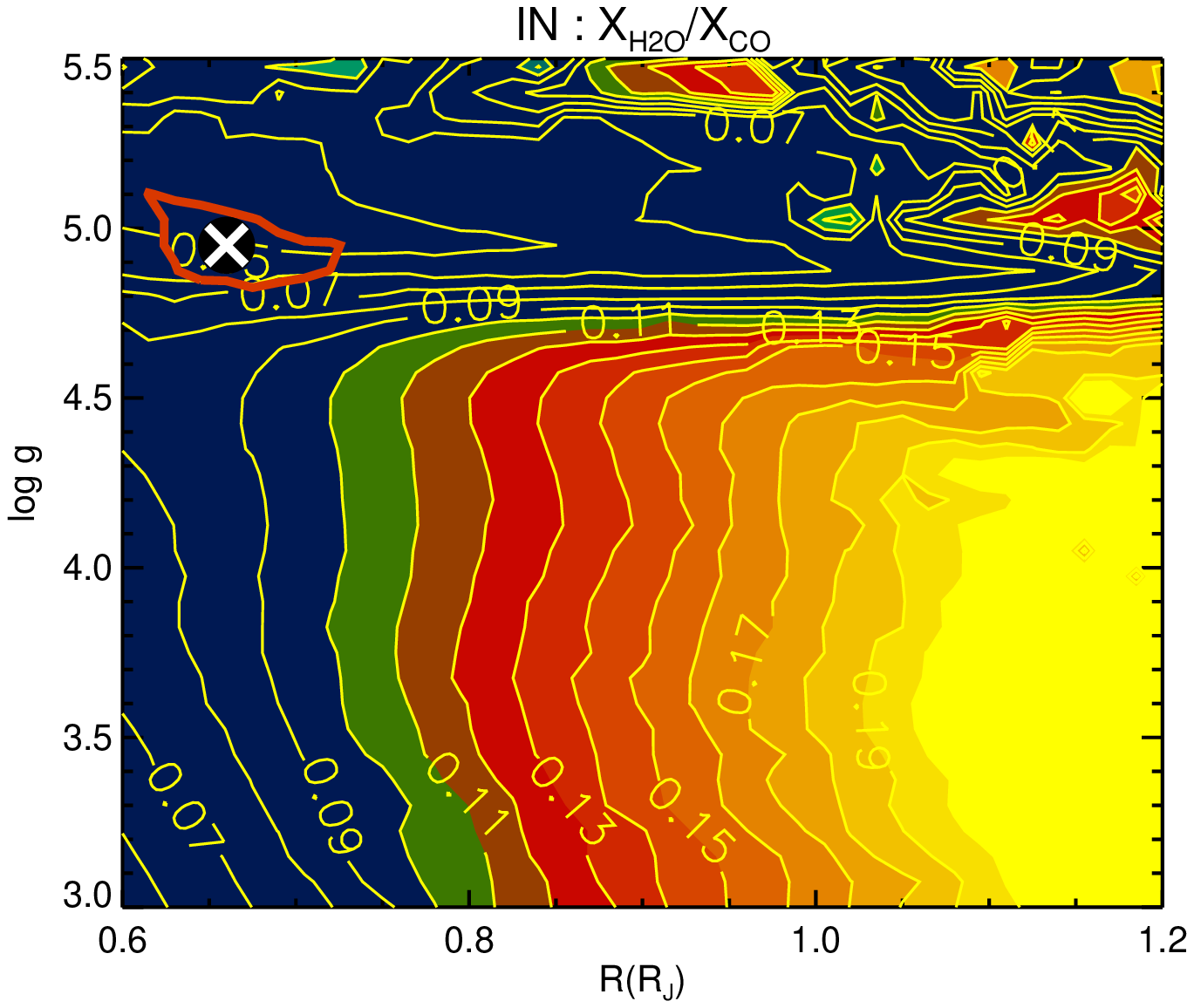}
\caption{Same as Figure \ref{fig:cf}, but for a suite of intermediate (IN) models with $\tau=2$ and $a=1.5$ $\mu$m.}
\label{fig:in}
\end{figure*}

\begin{figure}
\centering
\includegraphics[width=\columnwidth]{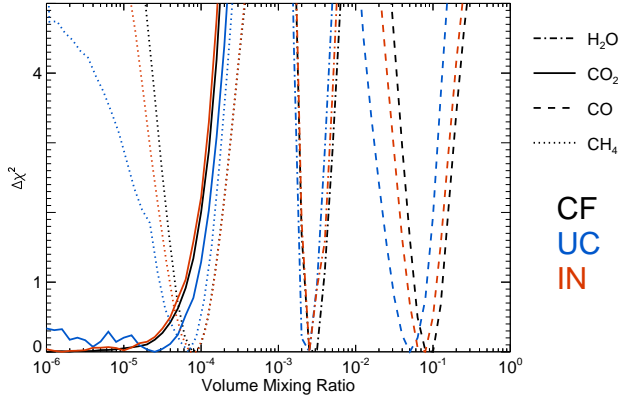}
\caption{Posterior distributions of the molecular abundances of water, carbon dioxide, carbon monoxide and methane for the best-fit CF, UC and IN models.  The distribution of carbon dioxide is unbounded from below due to the relatively small contribution of CO$_2$ to the spectrum.}
\label{fig:abund_var}
\end{figure}

\begin{figure}
\centering
\includegraphics[width=\columnwidth]{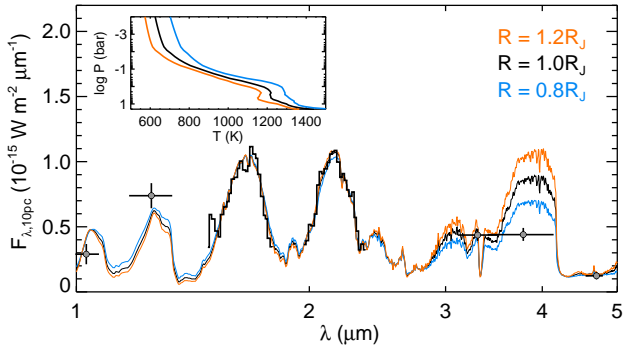}
\caption{Testing the sensitivity of the retrieved, synthetic spectrum to temperature.  Smaller radii correspond to warmer atmospheres.  The baseline, cloudfree model has $\log{g}=4.0$ and $R=R_{\rm J}$.  Using the retrieved solution, we fix the elemental abundances, vary only the radius and perform the fitting again.  It is apparent that it is easier to obtain a lower flux in the L$^\prime$ band when smaller radii (and warmer atmospheres) are assumed.}
\label{fig:spectra_sens1}
\end{figure}

\begin{figure}
\centering
\begin{minipage}{10cm}
\begin{minipage}{10cm}
\hspace{-0.7cm}
\includegraphics[width=9.5cm]{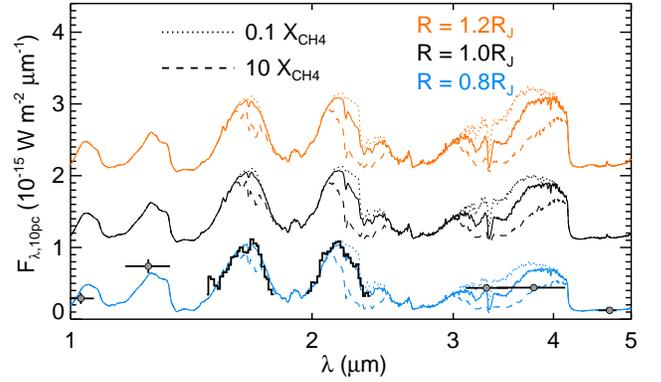}
\vspace{-0.4cm}
\end{minipage}
\begin{minipage}{10cm}
\hspace{-0.7cm}
\includegraphics[width=9.5cm]{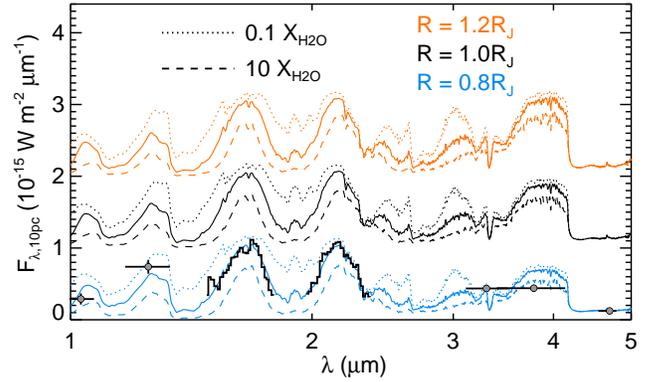}
\vspace{-0.4cm}
\end{minipage}
\begin{minipage}{10cm}
\hspace{-0.7cm}
\includegraphics[width=9.5cm]{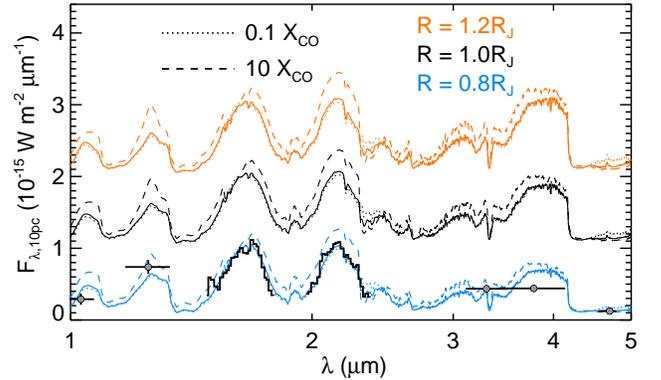}
\vspace{-0.4cm}
\end{minipage}
\begin{minipage}{10cm}
\hspace{-0.7cm}
\includegraphics[width=9.5cm]{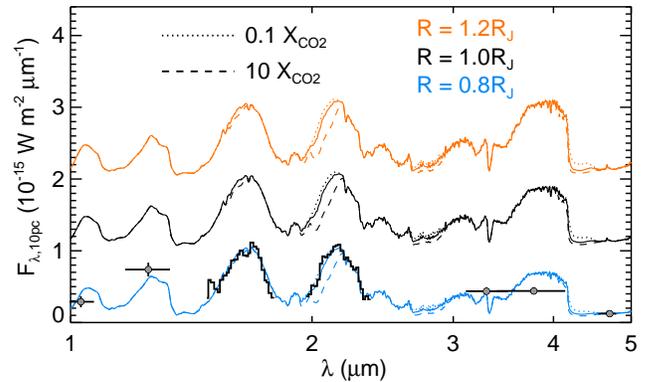}
\end{minipage}
\end{minipage}
\caption{Testing the sensitivity of the retrieved, synthetic spectrum to variations in temperature and specific elemental abundances.  We identify the main molecules determining the L$^\prime$-band flux as being methane and water.  For carbon monoxide, increasing its abundance ($\sim$0.1 in volume mixing ratio) leads to higher synthetic fluxes because the effect of collision-induced absorption (by hydrogen and helium) is reduced.  The actual fluxes computed are shown for the $R=0.8 R_{\rm J}$ case, while the other cases ($R=1, 1.2 R_{\rm J}$) are successively shifted upwards by $10^{-15}$ W m$^{-2}$ $\mu$m$^{-1}$.}
\label{fig:spectra_sens2}
\end{figure}

\begin{figure}
\centering
\includegraphics[width=\columnwidth]{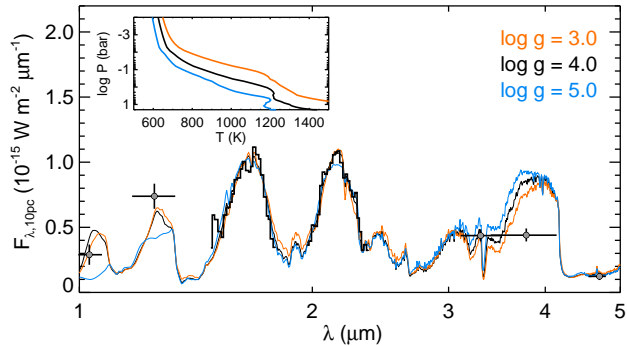}
\caption{Testing the sensitivity of the retrieved, synthetic spectrum to surface gravity.  Lower surface gravities correspond to warmer atmospheres.  The baseline, CF model has $\log{g}=4.0$ and $R=R_{\rm J}$.  Using the retrieved solution, we fix the elemental abundances and vary only the radius and perform the fitting again.}
\label{fig:spectra_grav}
\end{figure}

\begin{figure}
\hspace{-0.5cm}
\begin{minipage}{9.2cm}
\includegraphics[width=\columnwidth]{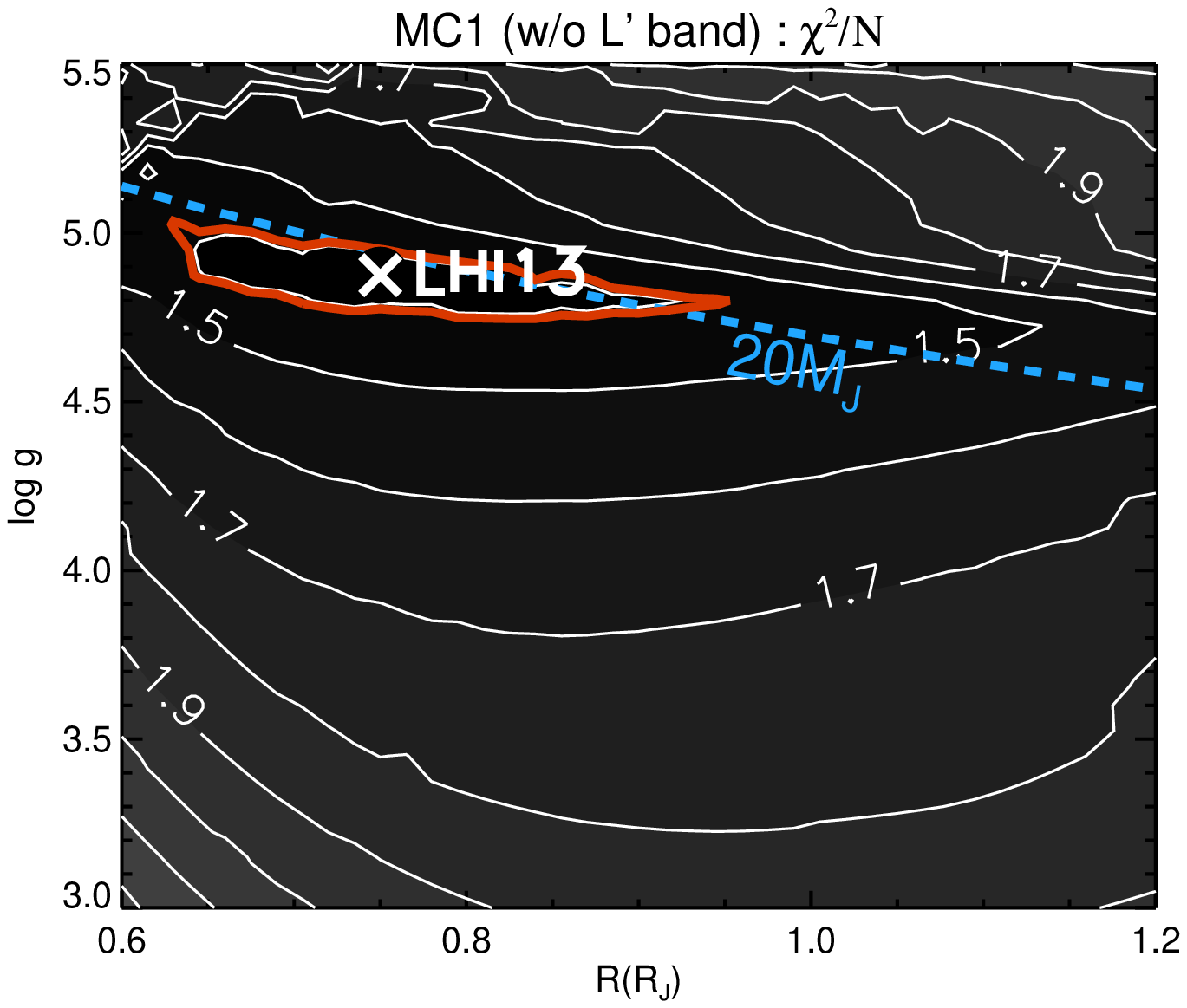}
\includegraphics[width=\columnwidth]{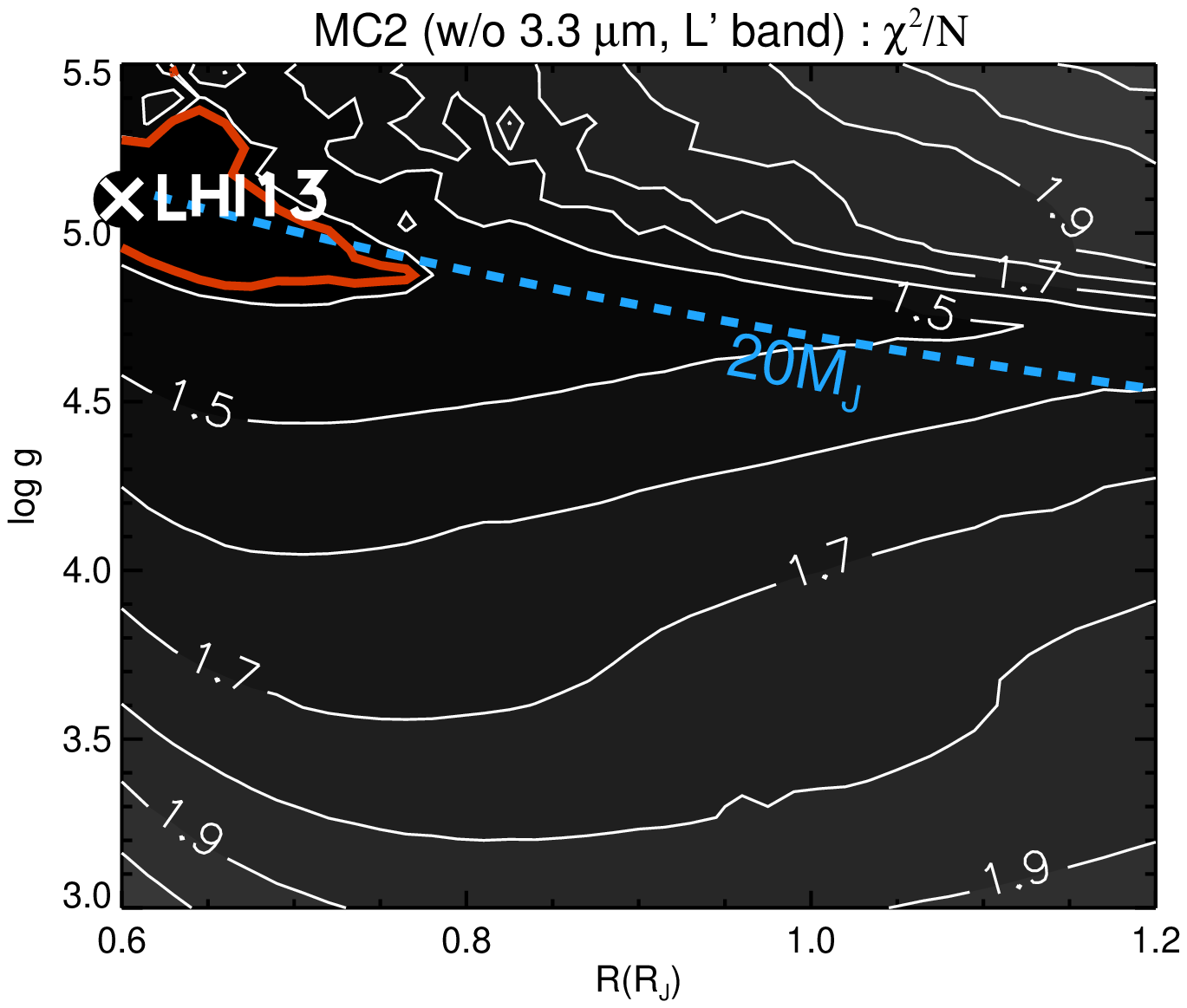}
\includegraphics[width=\columnwidth]{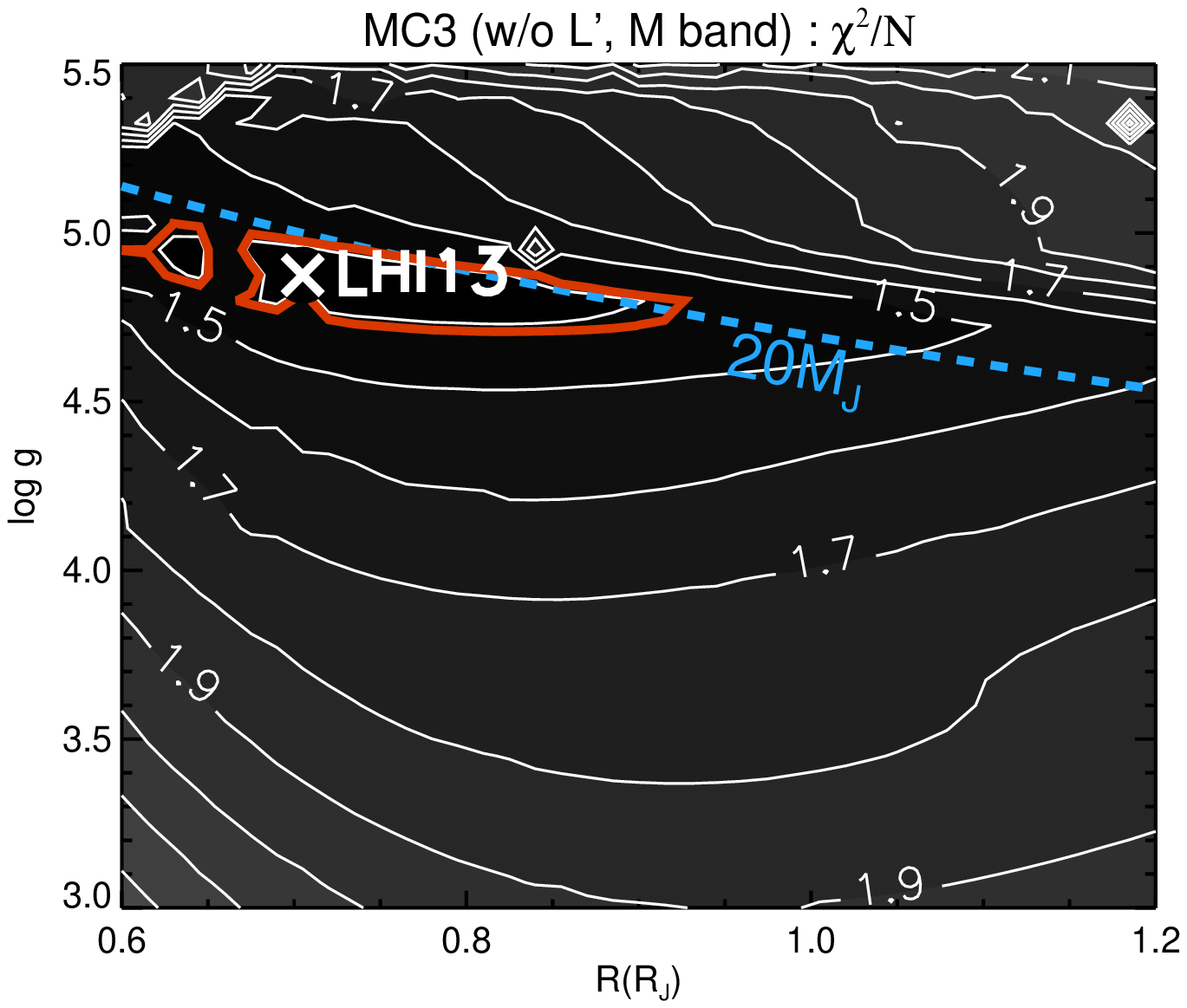}
\end{minipage}
\caption{Contour plots of the goodness of fit, as a function of radius and surface gravity, for the ``Magic Cloud" (MC) suites of models.  The MC1, MC2 and MC3 suites exclude the L$^\prime$ band, 3.3 $\mu$m and L$^\prime$ bands, and L and M bands, in turn.}
\label{fig:tests}
\end{figure}

\subsection{Cloudfree (CF) Models}

The results from the suite of CF models are shown in Figure \ref{fig:cf}.  In the parameter space of radius ($R$) and surface gravity ($g$), we perform retrievals for each pair of parameter values.  We explore 1435 models (41 $R$ values, 35 $\log{g}$ values) within these parameter ranges: $R = 0.6$--$1.2 R_{\rm J}$ and $\log{g} = 3.0$--5.5 (cgs units).  For each model in this suite, we compute the goodness of fit ($\chi^2/N$, the {\it weighted least mean square error}, where $N$ is the number of data measurements), the effective temperature ($T_{\rm eff}$), the mean molecular weight ($\mu$), the carbon-to-oxygen ratio (C/O), the ratio of carbon monoxide to methane abundances ($X_{\rm CO}/X_{\rm CH_4}$) and the ratio of water to carbon monoxide abundances ($X_{\rm H_2O}/X_{\rm CO}$).  The best-fit model in this suite is based on evaluating the $\chi^2/N$.  The uncertainties on the best-fit model are derived by assuming 1-$\sigma$ (68\%) confidence intervals and two interesting parameters ($R$ and $\log{g}$), implying that one computes the range of values of $R$ and $\log{g}$ over $\Delta \chi^2=2.30$ \citep{avni76}.

For the suite of CF models, the goodness-of-fit for the best-fit model is surprisingly reasonable: $\chi^2/N = 1.36$ (stated to two decimal places).  The mean molecular weight departs from the standard solar-abundance value of $\mu=2.35$: $\mu= 4.2^{+1.5}_{-0.5}$.  The carbon-to-oxygen ratio is high: C/O$=$ 0.97$^{+0.00}_{-0.01}$, compared to a value of about 0.5--0.6 for the Sun.  The relative abundance of carbon monoxide to methane increases towards smaller radii and higher temperatures, as it should.  All of the best-fit values of the properties of HR 8799b are reported in Table 1. 

The inadequacy of cloudfree models to match the observed spectra of HR 8799b has previously been claimed by \cite{madhu11b}, who demonstrated that $\log{g} \approx 4$ models over-predict fluxes in the z, J, H and K bands.  They further showed that varying the non-equilibrium chemistry and metallicity alone do not improve the matching substantially.  Their approach is to prescribe thick-enough clouds to suppress the fluxes shortward of 2.2 $\mu$m.  Our conclusion is more subdued: CF models produce reasonable fits to the data if the assumptions of solar abundances and metallicity are relaxed.

\subsection{Uniformly Cloudy (UC) Models}

The second simplest suites of models one can construct are those where the atmospheres are uniformly populated by spherical cloud particles.  Generally, we have examined multiple suites of cloudy models: $a=0.1$--10 $\mu$m, $\tau=0.1$--10.  In this subsection, we will only present results from the $a=1.5$ $\mu$m and $\tau=2$ suite as this produces the best fit among all of the IN suites examined (for the optimal $\tau$ value, see Figure \ref{fig:compare}; results for selecting the optimal $a$ value are not shown).  Although the $\tau=1$ UC suite produces a better fit than the $\tau=2$ one, we pick $\tau=2$ in order to perform a fair comparision between the UC and IN suites.

The same set of contour plots for the UC, rather than the CF, suite of models are presented in Figure \ref{fig:uc}.  The goodness of fit actually \emph{worsens} to $\chi^2/N = 1.41$ (again stated to two decimal places).  The mean molecular weight is now closer to its solar abundance value ($\mu = 3.3^{+1.1}_{-0.3}$), but the carbon-to-oxygen ratio remains high (C/O$= 0.94^{+0.02}_{-0.01}$).  The UC suite of models do not provide an improvement over the CF suite.  Furthermore, part of the allowed UC solution is at conflict with the dynamical stability constraint of \cite{fm10}, who estimated that HR 8799b needs to have a mass of at most $\sim 20 M_{\rm J}$.  This constraint may be even more stringent if the debris disk of HR 8799 is taken into account \citep{mq13}.

\subsection{Intermediate (IN) Models}

The third simplest suite of models adds two parameters to the analysis: the finite boundaries of the cloud deck, which we take to be $P_{\rm up}=0.01$ bar and $P_{\rm down}=1$ bar based on our previously described analysis of the sensitivity function in the L$^\prime$ band (Figure \ref{fig:L_band}).  The goodness-of-fit is improved over the UC best-fit model, but is identical to the CF best-fit one ($\chi^2/N = 1.36$).  The preferred mean molecular weight is $\mu = 3.8^{+1.5}_{-0.4}$, intermediate between the CF and UC models, while the carbon-to-oxygen ratio remains close to unity.  The IN suite of models makes two improvements over the UC suite: the best-fit region of parameter space is smaller and is no longer at conflict with the dynamical constraint of \cite{fm10}.  That an intermediately cloudy model is an improvement over a uniform one has been noted in previous studies of brown dwarfs (e.g., \citealt{tn03}) and HR 8799b (e.g., \citealt{marois08,madhu11b}).

\subsection{General Theoretical Trends}

An interesting feature of the $\chi^2/N$ contours is that the radius solution is driven towards low values.  One may naturally ask why lower values of the radius provide generally better fits.  As previously explained, lower radii correspond to warmer model atmospheres.  Thus, if all of the other model quantities are held constant and only the radius is changed, one may expect to identify the spectral features which tend towards a better fit for higher temperatures.  In Figure \ref{fig:spectra_sens1}, we perform such a sensitivity test.  For $\log{g}=4.0$ and $R=R_{\rm J}$, we first retrieve a best-fit model.  We then fix the retrieved elemental abundances, vary the radius and perform the fit again.  This re-fitting procedure involves changing the best-fit temperature-pressure profile.  It is apparent that the L$^\prime$ band displays the highest sensitivity to variations in temperature.  A natural, follow-up question to ask is: what are the major molecules that influence the L$^\prime$-band flux the most?  Figure \ref{fig:spectra_sens2} shows examples of $\log{g}=4.0$ synthetic spectra where we vary a specific chemical abundance by an order of magnitude, while keeping all of the other model quantities fixed.  From performing such a test, we demonstrate that the L$^\prime$-band flux is mostly determined by the abundance of methane and water.  Higher temperatures activate higher opacities of methane and water \emph{for a fixed set of elemental abundances}.  In an attempt to minimize the L$^\prime$-band flux obtained, the best-fit solution is driven towards higher temperatures and thus lower radii.

The main intention behind introducing clouds to the atmospheric models of HR 8799b is to suppress the high L$^\prime$-band fluxes obtained in the model solutions.  Reducing the L$^\prime$-band flux allows the tension between attempting to fit the H-, K- and L$^\prime$-band fluxes to come into play.  If the cloud optical depth is too high, the predicted shape of the H and K bandheads become distorted (not shown).  To compensate for the over-suppression of the z-band flux, the temperature at higher altitudes is artificially increased, hence producing a retrieved temperature-pressure profile that possesses a high-altitude inversion (not shown).  For this reason, $\tau=10$ models have cloud optical depths that are too high.  For $\tau=2$, the optical depth is sufficiently high to ``beat down" the L$^\prime$-band flux to a level where the fitting tension between it and the other wavebands is better allowed to come into play.  This fitting tension is controlled by the temperature and hence the assumed radius.

Another feature to explore is the sensitivity of our retrieved spectra to variations in the surface gravity (Figure \ref{fig:spectra_grav}).  Consistent with the study of \cite{marley12}, lower surface gravities correspond to warmer atmospheres.  When the cloud optical depth is too high and the L$^\prime$-band flux is overly suppressed, the model spectrum can now be scaled up and down by either varying the radius or surface gravity.  This leads to the $\chi^2/N$ contours in the parameter space of $\log{g}$ versus $R$ to flatten out, due to the degeneracy in the surface gravity and radius.

Finally, we note that in all of the presented suites of models, we derive $X_{\rm CO}/X_{\rm CH_4} \gg 1$.  The prevalence of CO is the reason why we have C/O $\approx 1$.

\subsection{Additional Tests: the ``Magic Cloud"}

As a final check on our results, we consider additional suites of ``magic cloud" (MC)\footnote{Our enstatite cloud model with $a=1.5$ $\mu$m is a physically sound and less extreme version of such a ``magic cloud".  For the currently known SED of HR 8799b, the ideal cloud composition will have an extinction efficiency curve with a deeper trough between its peak and first harmonic.} models (Figure \ref{fig:tests}), which are analogous to the situation where some idealized cloud species is able to modify the L$^\prime$-band flux to any value desired.  We examine additional suites of CF models that in turn exclude the L$^\prime$ band (MC1), the 3.3 $\mu$m and L$^\prime$ bands (MC2), and the L$^\prime$ and M bands (MC3).  By excluding the 3.3 $\mu$m, L$^\prime$ and M bands in turn, we examine the sensitivity of our results to these various bands that are believed to be most affected by the presence of clouds.  We see that our retrieved solutions for the radius are driven to slightly larger values for the MC models, consistent with our previous discussion and understanding of the theoretical trends.  The fragmented best-fit region of the MC3 suite is due to the non-monotonic behavior of the $\chi^2/N$ contours close to the best-fit solution.

\section{Discussion}
\label{sect:discussion}

\begin{figure}
\centering
\includegraphics[width=\columnwidth]{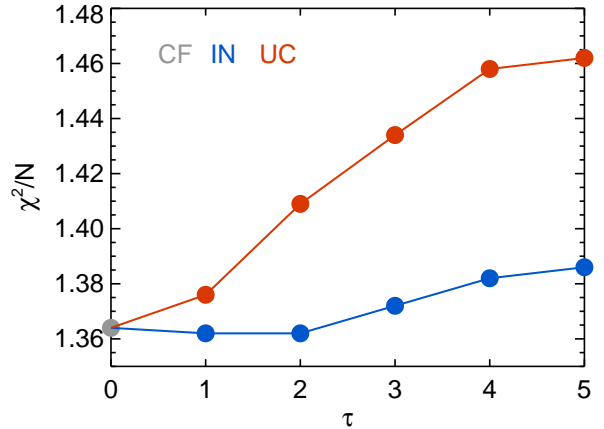}
\caption{Goodness of fit ($\chi^2/N$) as a function of the cloud optical depth ($\tau$) for the UC and IN suites of models with $a=1.5$ $\mu$m.}
\label{fig:compare}
\end{figure}

\begin{figure*}
\centering
\includegraphics[width=1.2\columnwidth]{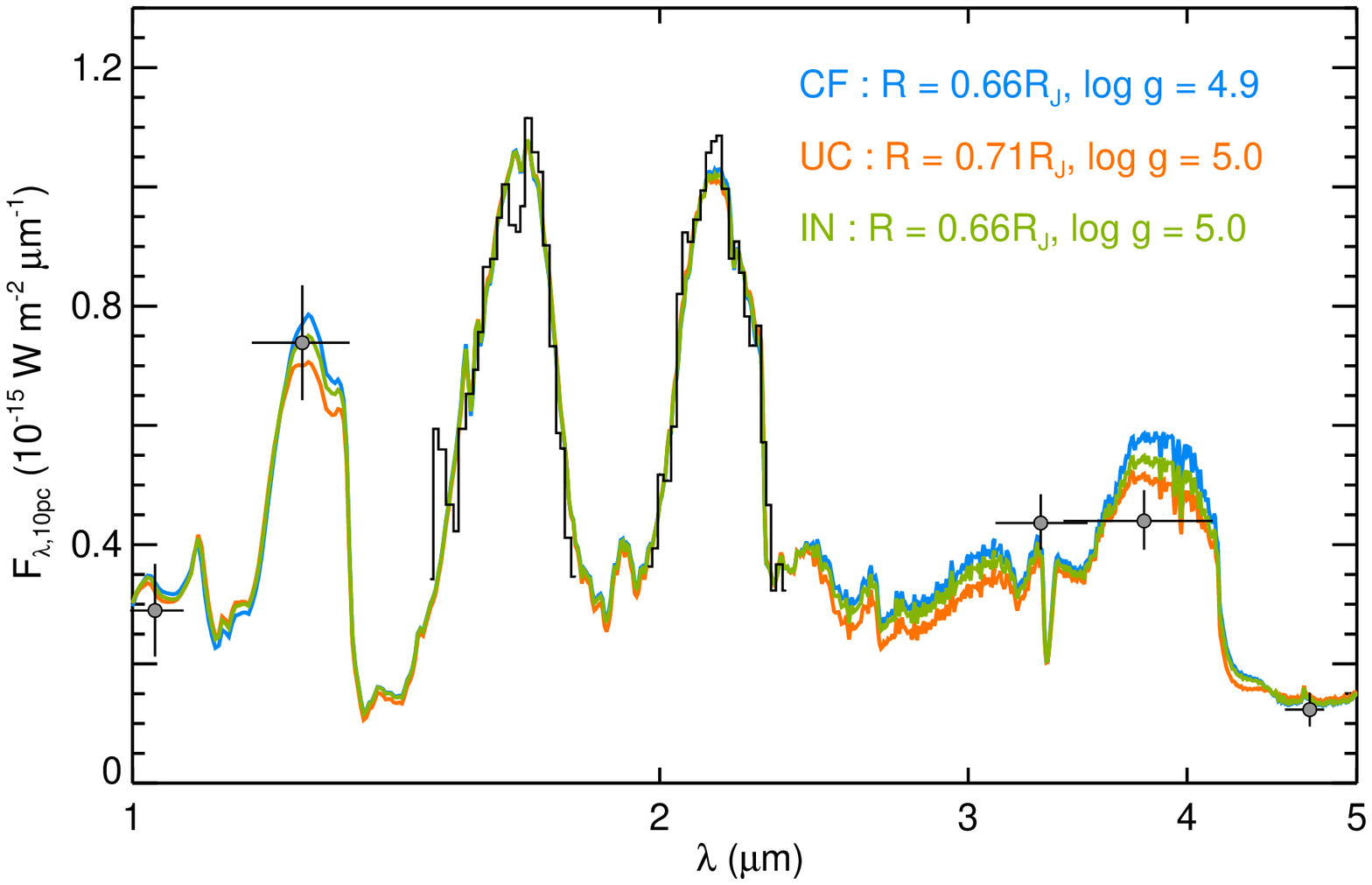}
\includegraphics[width=0.61\columnwidth]{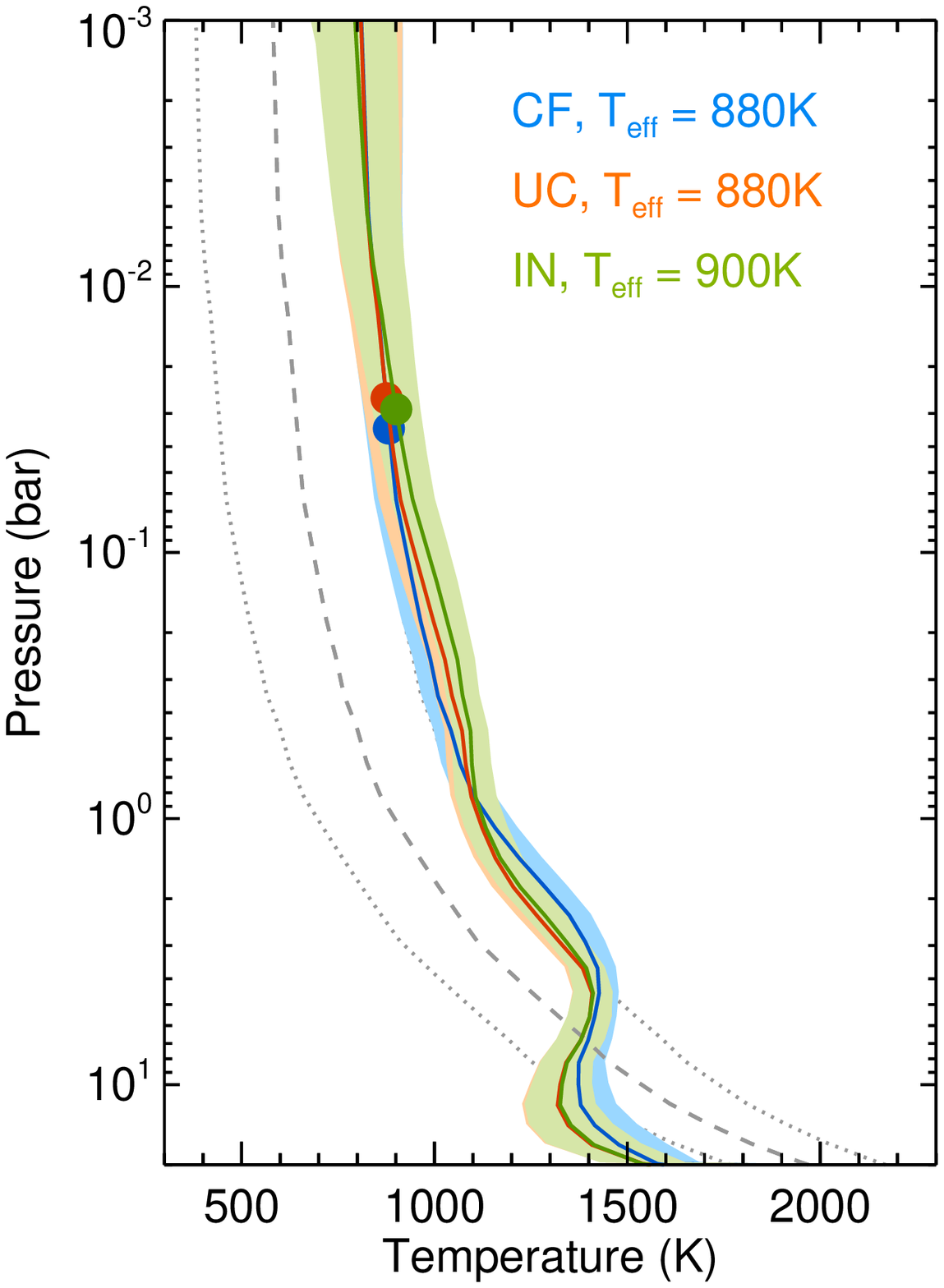}
\caption{Comparison of the best-fit spectra (left panel) and temperature-pressure profiles (right panel) from our cloudfree (CF), uniformly cloudy (UC) and intermediate (IN) suites of models.  Our cloudy models have $a=1.5$ $\mu$m and $\tau=2$. The dots indicate the locations of the photospheres ($\tau_{\rm total}=2/3$).}
\label{fig:spectra_compare}
\end{figure*}

\subsection{Summary}

We have performed an atmospheric retrieval analysis of the directly imaged exoplanet HR 8799b, aiming to infer its properties solely from considering its SED.  The salient points of our study may be summarized as follows.
\begin{itemize}

\item We have augmented the retrieval method of \cite{lee12} to include a simple, phenomenological model of clouds with two essential parameters.  We used our improved retrieval method to perform the first inverse-modeling study of the atmosphere of the directly imaged exoplanet HR 8799b.

\item By analyzing the photometry and spectra of HR 8799b from the published literature, we obtain best-fit solutions for a broad range of radii and surface gravities: $R=0.6$--$1.2 R_{\rm J}$, $\log{g}=3.0$--5.5.  We demonstrate that cloudfree and intermediately cloudy models (with the clouds being concentrated within a finite deck) produce comparable fits to the SED of HR 8799b, although the former requires a higher mean molecular weight.  Both scenarios require super-solar metallicities and carbon-to-oxygen ratios; we have not assessed if these unusual compositions imply non-equilibrium chemistry.

\item If we assume a monodisperse cloud particle size, we find that IN models with $a=1.5$ $\mu$m and $\tau=2$ provide the best match to the data.  The specific cloud particle radius derives from the specific nature in which the extinction efficiency curve is being sampled---the L$^\prime$-band flux is being suppressed at the expense of the H and K bands.  We report the best-fit values for the radius ($R = 0.66^{+0.07}_{-0.04} R_{\rm J}$), mass ($M = 16^{+5}_{-4} M_{\rm J}$), surface gravity ($\log{g} = 5.0^{+0.1}_{-0.2}$), effective temperature ($T_{\rm eff} = 900^{+30}_{-60}$K) and mean molecular weight ($\mu = 3.8^{+1.5}_{-0.4}$) of HR 8799b.

\end{itemize}
Figure \ref{fig:compare} shows the goodness-of-fit versus the cloud optical depth, which justifies our choice of $\tau=2$ for a comparison of our cloudy (UC versus IN) models. Our best-fit values for the properties of HR 8799b are stated in Table 1.  A comparison of our best-fit spectra is presented in Figure \ref{fig:spectra_compare}.

\subsection{Cloudy, Substellar Objects: an Infinity \\ of Mass-Radius Relationships}

In the study of the terrestrial climate, the Solar System and brown dwarfs, a fundamental obstacle is our lack of understanding of clouds from first principles.  Being free of clouds, main sequence stars are relatively simple objects and obey a unique mass-radius relationship.  In cooler, substellar objects, the presence of clouds destroys this uniqueness.  For example, \cite{burrows11} have shown that the measured masses and radii of brown dwarfs (found in binary systems) may be matched by several mass-radius relationships that depend upon metallicity, helium content and clouds.  Essentially, each cloud configuration (composition, size distribution, geometry) yields its own mass-radius relationship.

In the present study, we run into the same obstacle.  If we consider the goodness-of-fit to just two significant digits, we are unable to distinguish between the three different interpretations of the atmosphere of HR 8799b (Table 1).  Even with exquisite data in the future, we anticipate some of this degeneracy to persist, because it stems from an ignorance of basic physics and chemistry, rather than a lack of quality in the data.

\subsection{Are the Inferred Properties of HR 8799b Consistent \\ with Planet Formation Theories?}

In the present study, we are agnostic about the formation and evolution of HR 8799b, instead choosing to infer its properties based solely upon analyzing its atmosphere.  Nevertheless, it is instructive to discuss these properties within the context of current ideas about planet formation.  The small radii ($R \approx 0.7R_{\rm J}$) and high surface gravities ($\log{g} \approx 5$) retrieved imply masses that are either at the edge of or beyond the deuterium-burning mass limit ($\sim 13 M_{\rm J}$).  However, it is worth noting that the deuterium-burning mass limit itself is not a sharp boundary, but rather depends upon the metallicity and may be as low as $11 M_{\rm J}$ for metal-rich objects \citep{spiegel11}.  Even the use of the deuterium-burning mass limit as a demarcation between exoplanets and brown dwarfs is not universally accepted \citep{baraffe08,spiegel11}.  It is worth noting that \cite{baraffe08} have already demonstrated the plausibility of forming exoplanets, via core accretion, with masses $\sim 10 M_{\rm J}$ (and core masses $\sim 100 M_\oplus$) and sub-Jupiter radii that are able to sustain deuterium burning.  Our inferred properties of HR 8799b should be confronted by future models of planet formation and evolution.  If our findings are confirmed, HR 8799b may be the first clear-cut example of a deuterium-burning exoplanet.

\subsection{Vertical Mixing: the Inferred, Minimum Value of $K_{\rm zz}$}

For clouds to be present in the atmosphere of HR 8799b requires that a vertical flow of the atmosphere exists and that the cloud particles are held aloft by it.  While our retrieval method does not require $K_{\rm zz}$ to be specified, the particle radius inferred from the application of our phenomenological cloud model allows for the order-of-magnitude, minimum value of $K_{\rm zz}$ to be estimated after the fact.  We note that there are two ways of specifying $K_{\rm zz}$---either that associated with the cloud particles or with chemistry; here, we examine the former.  Particles with radii $ \sim 1$ $\mu$m typically have associated Knudsen numbers that are greater than unity,
\begin{equation}
N_k = \frac{k_{\rm B}T}{P \sigma_m a} \sim 10 \left( \frac{T}{900 \mbox{ K}} \right) \left( \frac{P}{0.1 \mbox{ bar}} \frac{a}{1 \mu\mbox{m}} \right)^{-1},
\end{equation}
where $k_{\rm B}$ is the Boltzmann constant, $T$ is the temperature, $P$ is the pressure and $\sigma_m \sim 10^{-15}$ cm$^2$ is the cross section for interactions between molecules.  One may compute the minimum velocity needed to loft a cloud particle of radius $a$ and mass density $\rho_c$ \citep{spiegel09}.  In the limit of $N_k \gg 1$, we have
\begin{equation}
v_{\rm z} \gtrsim \frac{\rho_c a g}{2.7 P} \left( \frac{k_{\rm B}T}{\gamma m} \right)^{1/2},
\end{equation}
where $\gamma$ is the adiabatic gas index, $m = \mu m_{\rm H}$ is the mean molecular mass and $m_{\rm H}$ is the mass of the hydrogen atom.  If we approximate the eddy diffusion coefficient by $K_{\rm zz} \sim v_{\rm z} H$, with $H = k_{\rm B} T/mg$ being the pressure scale height, then we obtain
\begin{equation}
K_{\rm zz} \gtrsim \frac{\rho_c a}{2.7 P \gamma^{1/2}} \left( \frac{k_{\rm B}T}{m} \right)^{3/2}.
\end{equation}
If we plug in $\rho_c = 3$ g cm$^{-3}$, $a=1.5$ $\mu$m, $P=0.1$ bar, $\gamma=7/5$, $T=900$ K and $\mu=4$, we obtain $K_{\rm zz} \gtrsim 4 \times 10^6$ cm$^2$ s$^{-1}$.  Values of $K_{\rm zz}$ in the literature are usually associated with chemistry and stated as a lower limit: e.g., \cite{madhu11b} adopt $K_{\rm zz} = 10^2$--$10^6$ cm$^2$ s$^{-1}$. \cite{barman11} and \cite{marley12} both use particle radii $\sim 1$--100 $\mu$m and assume $K_{\rm zz}=10^4$ cm$^2$ s$^{-1}$.  \cite{barman11} remark that the studies of Jupiter typically infer a range of $K_{\rm zz} \sim 10^2$--$10^8$ cm$^2$ s$^{-1}$.

\subsection{How Robust Are Current Retrieval Analyses \\ of Hot Jupiters?}

A lesson we have learned from the present study is that for claims about the nature of an exoplanetary atmosphere to be robust, they need to be based on analyses involving both photometry and spectroscopy and inferred from beyond a few photometric data points.  In the case study of HR 8799b, as presented in this paper, it is the fitting tension between the H, K and L$^\prime$ bands that determine the best-fit values of its basic properties and elemental abundances.  If some of the data points are removed, the retrieved properties of HR 8799b differ rather significantly.  Even with this relative wealth of data, compared to hot Jupiters, our interpretation of the atmosphere of HR 8799b is non-unique due to our ignorance of first-principles cloud physics and chemistry.

This train of thought leads us to be concerned about the existing analyses of the atmospheres of hot Jupiters, many of which are based on the analysis of a small number of photometric data points.  (See \citealt{line13} for a more quantitative version of this opinion.)  Without transit or eclipse spectra, it is difficult to gauge the robustness of these studies.  A notable exception is the prototypical hot Jupiter HD 189733b, for which a relatively extensive measurement of its SED exists, although the robustness of the infrared SED remains controversial \citep{pont13}.  Prudence suggests that we view these results tentatively until robust spectra are obtained using future infrared observatories such as the \textit{James Webb Space Telescope (JWST)}.

\vspace{0.2in}
\textit{JL and KH acknowledge financial and logistical support from the Swiss-based MERAC Foundation, the University of Bern and the University of Z\"{u}rich.  PGJI acknowledges the support of the United Kingdom Science and Technology Facilities Council.  This work was conducted as part of the activities of the Exoplanets and Exoclimes Group based at the Universities of Bern and Z\"{u}rich.  The calculations were performed using the \texttt{zBox4} computing cluster at the University of Z\"{u}rich thanks to support from Doug Potter, Simon Grimm and Joachim Stadel.  We thank Bruce Draine, Thayne Currie, Sascha Quanz and Michael Meyer for useful conversations during the initial stages of this work.  We acknowledge a gracious, thoughtful and constructive report from the anonymous referee, which has improved the clarity and quality of the manuscript.  Michael Line is credited with an in-depth discussion on computing the posterior distributions of molecular abundances.}

\begin{table*}
\label{tab:data}
\begin{center}
\caption{Previously published data for HR 8799b from the literature, selected for the present study}
\begin{tabular}{cccccc}
\hline
\hline
Band & Central Wavelength ($\mu$m) & F$_{\lambda,{\rm 10 pc}}$ (mJy) & F$_{\lambda,{\rm 10 pc}}$ (10$^{-15}$ W cm$^{-2}$ um$^{-1}$) & Reference \\
\hline
z & 1.03 & 0.10$\pm$0.03 & 0.29$\pm$0.08 & \cite{currie11} \\
J & 1.25 & 0.38$\pm$0.05 & 0.74$\pm$0.10 & \cite{currie11} \\
\hline
& 1.48 & 0.25$\pm$0.08 & 0.34$\pm$0.11 \\
& 1.49 & 0.44$\pm$0.04 & 0.59$\pm$0.05 \\
& 1.50 & 0.42$\pm$0.06 & 0.56$\pm$0.08 \\
& 1.52 & 0.36$\pm$0.03 & 0.47$\pm$0.04 \\
& 1.53 & 0.33$\pm $0.08 & 0.42$\pm$0.10 \\
& 1.54 & 0.47$\pm$0.05 & 0.59$\pm$0.06 \\
& 1.56 & 0.53$\pm$0.08 & 0.65$\pm$0.10 \\
& 1.57 & 0.57$\pm$0.06 & 0.69$\pm$0.07 \\
& 1.58 & 0.63$\pm$0.06 & 0.76$\pm$0.07 \\
& 1.59 & 0.73$\pm$0.04 & 0.87$\pm$0.05 \\
& 1.61 & 0.76$\pm$0.04 & 0.88$\pm$0.05 \\
OSIRIS H & 1.62 & 0.83$\pm$0.05 & 0.95$\pm$0.06 & \cite{barman11}\\
& 1.63 & 0.89$\pm$0.03 & 1.00$\pm$0.03 \\
& 1.65 & 0.85$\pm$0.05 & 0.94$\pm$0.06 \\
& 1.66 &  0.85$\pm$0.07 & 0.92$\pm$0.08 \\
& 1.67 & 0.90$\pm$0.04 & 0.97$\pm$0.04 \\
& 1.68 & 1.05$\pm$0.04 & 1.12$\pm$0.04 \\
& 1.70 & 1.02$\pm$0.06 & 1.06$\pm$0.06 \\
& 1.71 & 1.00$\pm$0.04 & 1.03$\pm$0.04 \\
& 1.72 & 0.89$\pm$0.04 & 0.90$\pm$0.04 \\
& 1.74 & 0.74$\pm$0.04 & 0.73$\pm$0.04 \\
& 1.75 & 0.60$\pm$0.06 & 0.59$\pm$0.06 \\
& 1.76 & 0.58$\pm$0.05 & 0.56$\pm$0.05 \\
& 1.77 & 0.43$\pm$0.05 & 0.41$\pm$0.05 \\
& 1.79 & 0.37$\pm$0.05 & 0.35$\pm$0.05 \\
\hline
& 1.97  & 0.47$\pm$0.10 & 0.36$\pm$0.08 \\
& 1.99  & 0.52$\pm$0.07 & 0.39$\pm$0.05 \\
& 2.00  & 0.69$\pm$0.12 & 0.52$\pm$0.09 \\
& 2.02  & 0.69$\pm$0.11 & 0.51$\pm$0.08 \\
& 2.04  & 0.83$\pm$0.10 & 0.60$\pm$0.07 \\
& 2.05  & 1.15$\pm$0.12 & 0.82$\pm$0.09 \\
& 2.07  & 1.32$\pm$0.13 & 0.92$\pm$0.09 \\
& 2.08  & 1.31$\pm$0.12 & 0.91$\pm$0.08 \\
& 2.10  & 1.39$\pm$0.12 & 0.94$\pm$0.08 \\
& 2.12  & 1.49$\pm$0.13 & 0.99$\pm$0.09 \\
& 2.13  & 1.60$\pm$0.15 & 1.06$\pm$0.10 \\
OSIRIS K & 2.15  & 1.66$\pm$0.15 & 1.08$\pm$0.10 &  \cite{barman11} \\
& 2.16  & 1.69$\pm$0.15 & 1.09$\pm$0.10 \\
& 2.18  & 1.58$\pm$0.15 & 1.00$\pm$0.09 \\
& 2.20  & 1.42$\pm$0.14 & 0.88$\pm$0.09 \\
& 2.21  & 1.48$\pm$0.14 & 0.91$\pm$0.09 \\
& 2.23  & 1.42$\pm$0.14 & 0.86$\pm$0.08 \\
& 2.24  & 1.31$\pm$0.13 & 0.78$\pm$0.08 \\
& 2.26  & 1.25$\pm$0.12 & 0.73$\pm$0.07 \\
& 2.28  & 1.33$\pm$0.12 & 0.77$\pm$0.07 \\
& 2.29  & 1.00$\pm$0.10 & 0.57$\pm$0.06 \\
& 2.31  & 0.83$\pm$0.08 & 0.47$\pm$0.04 \\
& 2.32  & 0.58$\pm$0.06 & 0.32$\pm$0.03 \\
& 2.34  & 0.67$\pm$0.08 & 0.37$\pm$0.04 \\
& 2.36  & 0.60$\pm$0.07 & 0.32$\pm$0.04 \\
\hline
3.3 $\mu$m & 3.30 & 1.59$\pm$0.17 & 0.44$\pm$0.05 & \cite{skemer12} \\
L' & 3.78 & 2.10$\pm$0.23 & 0.44$\pm$0.05 & \cite{currie11} \\
M & 4.70 & 0.91$\pm$0.21 & 0.12$\pm$0.03 & \cite{galicher11} \\
\hline
\end{tabular}
\end{center}
\end{table*}

\appendix

\section{The Nearly Universal Shape of the Extinction Efficiency Curve}
\label{append:qext}

\begin{figure}
\centering
\includegraphics[width=\columnwidth]{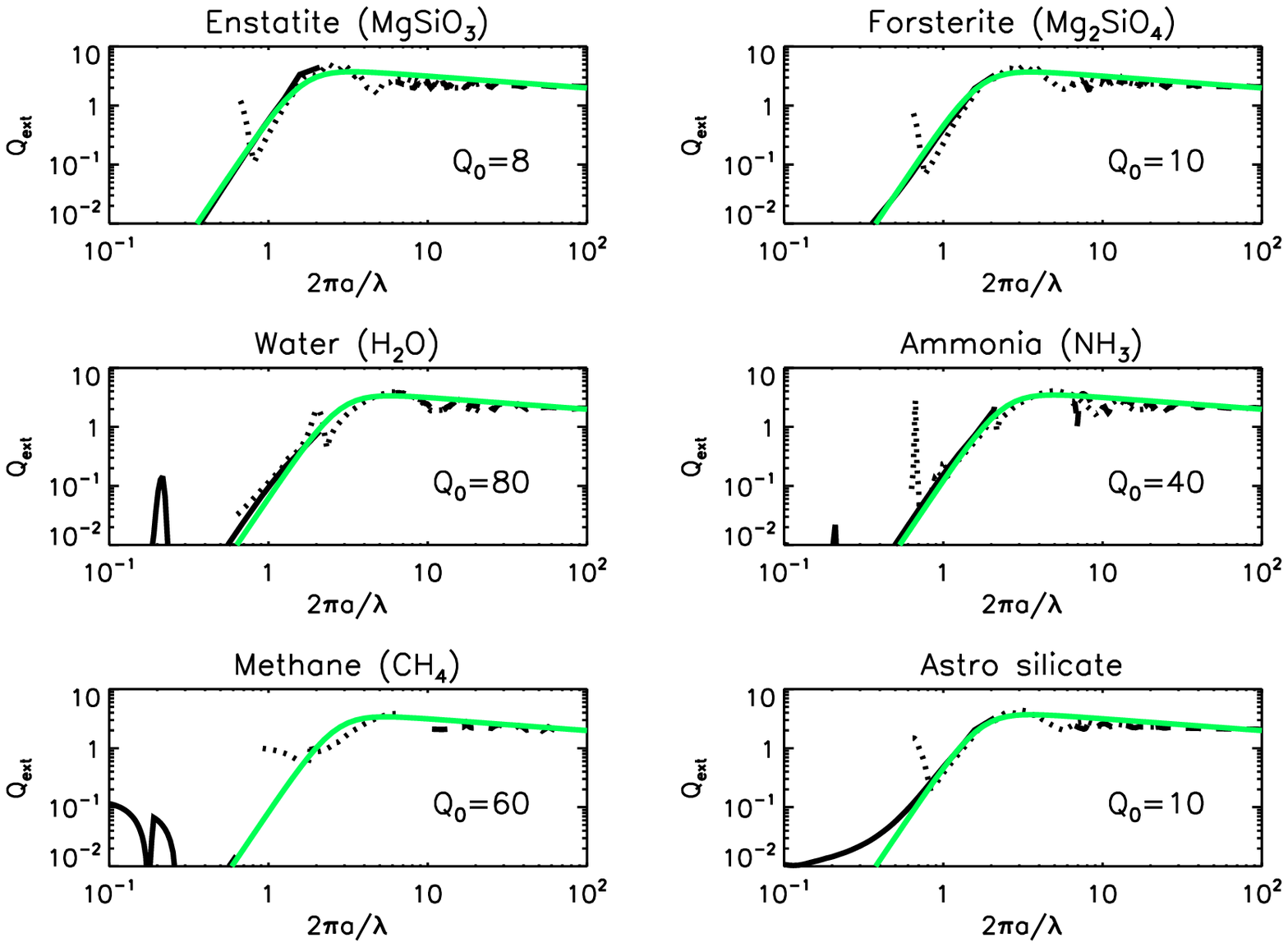}
\caption{Extinction efficiency curve for various chemical species as a function of the size parameter.  It may be approximately described by a one-parameter family ($Q_0$; see text for details).  The black solid, dotted and dashed curves are for $a=0.1, 1$ and 10 $\mu$m calculations, while the green solid curve is the empirical fitting function.  The data for astronomical silicate are taken from \cite{draine84} and \cite{ld93}.  Note that these fits to the $Q_{\rm ext}$ data are for the purpose of illustration and are not used in our analysis.}
\label{fig:qext_univ}
\end{figure}

As stated in \S\ref{subsect:cloudmodel}, we use the extinction efficiency $Q_{\rm ext}$ in our definition of the cloud optical depth.  While the $Q_{\rm ext}$ curve may contain features that are specific to a given composition, its main functional form may be approximately described by a one-parameter, empirical fitting function (Figure \ref{fig:qext_univ}),
\begin{equation}
Q_{\rm ext} = \frac{5}{Q_0 x^{-4} + x^{0.2}},
\end{equation}
where $x \equiv 2 \pi a/\lambda$ and $a$ is the radius of our spherical aerosol or cloud particle.  The $x^{-4}$ term represents the contribution from Rayleigh scattering when $x \ll 1$, while the $x^{0.2}$ term mimics the peaking of the $Q_{\rm ext}$ curve when $x \sim 1$, followed by a gentle decline when $x \gg 1$.  The quantity $Q_0$ determines the exact $x$-value where $Q_{\rm ext}$ peaks and is determined by comparing the fitting function to detailed calculations of the extinction efficiency.  Details such as resonant behavior are not captured by such a simple description.  It is apparent that more volatile material has $Q_0 \sim 100$, while silicates are better described by $Q_0 \sim 10$.  Note that these fits to the $Q_{\rm ext}$ data are for the purpose of illustration and are not used in our analysis.

\section{The Radius Ratio Non-Problem}
\label{append:radius_ratio}

In our earlier analyses of the atmosphere of HR 8799b, we defined two radii: the model radius ($R$) and the photospheric radius ($R_{\rm ph}$).  The latter is defined by the relation,
\begin{equation}
R_{\rm ph} = \left(\frac{{\cal L}}{\pi \sigma_{\rm SB}}\right)^{1/2} \frac{1}{2 T^2_{\rm eff}}
\approx 1.1 ~R_{\rm J} \left( \frac{{\cal L}}{10^{-5.1}{\cal L}_\odot} \right)^{1/2} \left( \frac{T_{\rm eff}}{900 \mbox{ K}} \right)^{-2},
\end{equation}
where ${\cal L} = 10^{-5.1 \pm 0.1} {\cal L}_\odot$ is the inferred bolometric luminosity of HR 8799b \citep{marois08} and ${\cal L}_\odot$ is the solar bolometric luminosity.  By inferring the effective temperature from the retrieved temperature-pressure profile, one can compute $R_{\rm ph}$.  Our original reasoning was that $R_{\rm ph}/R = 1$ should serve as an additional figure of merit for judging whether our fits to the SED were physically reasonable.  We discovered that all of our models produced $R_{\rm ph}/R \ne 1$, which we termed the ``radius ratio problem".

However, it is important to note that this value of the bolometric luminosity was inferred by \cite{marois08} based on comparing photometric data points with forward spectral models.  Thus, using it to compute $R_{\rm ph}$, for comparison with $R$, is not self-consistent with our own retrieval models.  Self-consistency is achieved by integrating the retrieved SED of a model (for fixed values of $R$ and $g$) over wavelength to obtain the bolometric flux $F$.  The bolometric luminosity is then given by ${\cal L} = 4 \pi R^2 F$.  Only one radius remains.

Thus, the radius ratio problem is an artefact of not using a self-consistent definition of the bolometric luminosity.

\newpage

\label{lastpage}

\end{document}